\pdfoutput=1
\documentclass[12pt,preprint]{aastex}
\usepackage{graphicx,apjfonts,emulateapj5,onecolfloat5,dblfloatfix}
\usepackage{amssymb,mathrsfs,color,gensymb}
\usepackage{ulem}
\usepackage{subfigure}
\usepackage{amsmath} 
\newfont{\rsfsten}{rsfs10 scaled 1200}
\newfont{\rsfsseven}{rsfs7 scaled 1200}
\newfont{\rsfsfive}{rsfs5 scaled 1200}

\slugcomment{Draft: \today}

\shorttitle{}
\shortauthors{Zhao, Morris, Goss}
\begin{document}
\twocolumn[
\title{A procedure for making high dynamic-range radio images:
Deep imaging of the kiloparsec-scale radio
structures of a distant blazar, NRAO 530, with JVLA data} 
\author{
Jun-Hui Zhao\altaffilmark{1}, 
Mark R. Morris\altaffilmark{2} \& 
W. M. Goss\altaffilmark{3}
} 
\affil{$^1$Center for Astrophysics | Harvard-Smithsonian, 60
Garden Street, Cambridge, MA 02138, USA; jzhao@cfa.harvard.edu}
\affil{$^2$Department of Physics and Astronomy, University of California Los Angeles, Los Angeles, CA 90095}
\affil{$^3$NRAO, P.O. Box O, Socorro, NM 87801, USA}
\begin{abstract}
Using JVLA data obtained from A-, B- and C-array observations of Sgr A* and 
NRAO 530 (as a calibrator) at 5.5, 9 and 33 GHz during the period 
between 2012 and 2015, we developed a procedure for the 
reduction of wideband data at high-angular
resolution. We have demonstrated that, correcting for 
residual interferometer errors such as antenna-based 
errors caused by residual delays as well as baseline-based closure errors, 
radio astronomers can now achieve high-fidelity radio images 
with a dynamic range (peak:rms) exceeding 1,000,000:1.
We outline the procedure in detail, noting that it can have 
broad application to the analysis of broadband continuum 
observations. Here, we apply this procedure to observations 
of a distant blazar, NRAO 530, revealing the radio structures 
surrounding the core in unprecedented detail.   
Our wideband JVLA image of this source at 5.5 GHz 
(C band) shows that the kiloparsec radio 
structure of NRAO 530 is prominently characterized by a
moderately curved western jet terminating at a hot spot where the 
radiation is stretched further out into a diffuse radio lobe or plume.
Close to the radio core, an abrupt bending of the jet is
revealed in the high-resolution (< 100 milliarcsec) images at 33
GHz (Ka band), showing an evolution of the position angle of the 
jet from the north at the VLBI scale (50 mas, or $\sim$400 pc \textcolor{black}{projected}), 
increasing toward the west at the larger VLA scales 
(1 arcsec, or $\sim$10 kpc).  The continuation
of the jet axis drift forms the curved western jet extending out to 200 kpc.
In contrast to the main radio structure, a faint and
broad counter-jet is present on the eastern side
with a curvature antisymmetric to the western jet. 
The eastern jet terminates at a bright hotspot, forming 
an edge-brightened diffuse lobe. A newly recognized compact component,
located 0.6 arcsec east of the radio core, is detected in the A-array 
images at 9 GHz (X band), suggesting that a more recent ejection has 
taken place towards the east. In addition, a lower-brightness 
emission extended N-S from the core is detected at a level 
of 1-10 K in brightness temperature in our 5.5 GHz C-array data.
The observed contrast in surface brightness between the
western and eastern jet components suggests that the 
jets on the VLA scale are mildly relativistic. The radiation from
the western jet is boosted while the radiation from the 
receding eastern jet is plausibly suppressed owing to the relativistic
Doppler effect.
  
\end{abstract}

]

\section{Introduction}
As a number of radio interferometer systems increase sensitivity
by employing large increases in bandwidth, several important corrections
must be applied due to time variations of various instrumental 
parameters during the observation. An example is antenna-based residual
delays that vary across the band and that change with time.  
For the wideband data distorted by these  time-variable issues, 
it becomes difficult to restore the true information in the process 
of imaging  the continuum structure of a radio source. Consequently, 
without corrections for such residual effects, the dynamic range of
the resulting images is limited.

In previous narrow-band observations, the dynamic range of continuum 
observations was typically sensitivity-limited, so it was challenging 
at best to learn about the full spatial structure of the brightest 
radio sources such as blazars, distant radio galaxies and QSOs characterized 
with dominant bright radio cores. Such objects have often served the purpose 
of being calibrator sources for interferometer data inasmuch as they 
appear largely as bright point sources. Historically, the emission 
from their extended environs, as well as background radio sources, 
were neglected, except at the highest frequencies, where the bright 
emission components might be partially resolved.

When extended-source signals from such high-contrast regions are sampled 
with recent wideband techniques, the {\it residual errors} (RE) left over from 
standard calibration procedures from previous eras challenge the imaging 
reliability, particularly for the faint features in the vicinity of bright 
radio sources, features that are often confused with artifacts resulting 
from the uncorrected RE. The {\it dynamic range} (DR) of the image is therefore 
limited, and the image fidelity is degraded owing to the residual errors 
in phase and amplitude. DR is a good indicator of {\it image fidelity} --
the difference between any produced image and the true image \citep{perl99},
so DR can be used as a measure of image quality. In the presence of bright 
radio sources, DR is particularly important for reliably capturing the 
distribution of much fainter extended emission. For example, NRAO 530 was 
listed in the VLA calibrator catalog accompanied  by images showing little 
more than a dominant radio core at 14.9 and 8.46 GHz (see Fig. A1 of 
this paper). Faint extended structure was revealed by  later VLA observations 
at 1.46 GHz with the A-array, showing east-west radio lobes and jets \citep{kha2010}.
With an angular resolution of 1.5" and noise level of 0.34 mJy beam$^{-1}$,
this observation achieved a  DR of$\sim$18,000:1. 

Motivated by our JVLA observations of the Galactic center source, Sgr A 
\citep{zmg2016, mzg2017}, we developed a procedure for restoring the emission 
structure of the calibrators used for our programs to a substantially greater 
depth than had previously been accomplished. The image DR was improved by 
iteratively converging the reference model to the true structure by applying 
various nonstandard corrections, in addition to eliminating {\it radio frequency 
interference} (RFI). In particular, we corrected for the {\it time-variable 
residual delays}.

In this paper, we illustrate our detailed procedure by applying it to 
the radio structure of the complex-gain (phase and amplitude) 
calibrator NRAO 530 -- a distant radio blazar -- using
the JVLA data obtained in support of the observations of Sgr A. 
Using the best DR image model of the calibrator, the solutions for 
correcting RE were derived and applied to the Sgr A data \citep{mzg2017}.  
As by-products, an unprecedentedly deep image of NRAO 530
was also obtained. So while we use this source to demonstrate the 
wideband capability for RE correction to maximize the DR of the image, 
we also report a new scientific perspective on the blazar itself.

NRAO 530 is a gamma-ray blazar, associated with an optically 
violent variable source at z=0.902 \citep{heal08} or $D_{\rm L}= 5.8$ Gpc 
(1\arcsec = 7.8 kpc)\footnote{Calculated using the cosmology 
calculator of Ned Wright, UCLA
http://www.astro.ucla.edu/~wright/CosmoCalc.html assuming
$H_0=71$ km s$^{-1}$ Mpc$^{-1}$, $\Omega_{\rm M} = 0.27$, 
$\Omega_{\rm vac} = 0.73$ \citep{kom09}.}. 
The radio source was first discovered with the 300-foot telescope of
the NRAO \citep{paul66}. The VLBI observations show a 
jet structure to the north of its compact core, suggesting a prominent
example of  a periodic oscillation of 
the jet axis \citep{list13} or a helical motion \citep{lu2011} that the jet 
traces on scales of milli-arcsec or parsec. The northern VLBI jet is superluminal 
\citep{lu2011,list16,list18}. On kpc scales, 
NRAO 530 is characterized as having double radio lobes in 
the E-W direction. The western lobe shows relatively higher 
intensity, showing a curved jet connecting 
the western hot spot to the core. NRAO 530 is  one of the 
MOJAVE\footnote{Monitoring Of Jets Active galactic nuclei with
VLBA Experiments: http://www.physics.purdue.edu/MOJAVE/.} 
blazar sample mapped with VLA data by \cite{kha2010}.
Based on their kpc-scale radio study of the MOJAVE blazar sample, 
those authors have questioned the simple radio-loud unified scheme linking 
BL Lac objects ({\it i.e.} blazars) to FR-I galaxies and quasars to FR II sources 
\citep{FR1974, urry1995} and have also pointed out that a significant fraction
of MOJAVE blazars show parsec-to-kiloparsec scale jet misalignment. 
NRAO 530 seems to fall into the FR I/II category
because of the $\sim90${\degree}  misalignment
between the VLBI jet and the VLA structure \citep{kha2010}. 

In general, given the bright
cores of blazars and typically large distances, 
the relatively faint extended structure
of their jets, radio lobes and halos of radio cores are hidden under 
the detection limits of the old-generation narrow-band telescopes. 
As an illustrative example for a distant blazar, deep imaging of NRAO 530 
with the JVLA endowed with  wideband capability is necessary to reveal 
the puzzles concerning the connection
of the morphology on the VLA scale and the powerful core observed with VLBI. 

The rest of the paper is outlined as follows. The details of the 
procedure for correcting RE, and achieving high-DR imaging for wideband radio interferometer 
array data are given and discussed in Appendix A. In the main text, section 2 
describes the data and new images of NRAO 530. Our findings from the blazar 
NRAO 530 are presented in section 3, and the astrophysical implications are
discussed in section 4. Section 5 summarizes our astrophysical results. 

\begin{table*}[ht!]
\tablenum{1}
\setlength{\tabcolsep}{1.7mm}
\caption{Log of dataset}
\begin{tabular}{lcccccccl}
\hline\hline \\
{Project (PI)}&
{Observing date~~}&
{Array}&
{Band$^\dagger$}&
{$\nu$}&
{~~~$\Delta\nu$~~~} &
{HA range}&
{Time}&
\multicolumn{1}{c}{Sources}\\
{} &
{} &
{} &
{} &
{(GHz)} &
{(GHz)} &
{} &
{on source (h)} &
\\
{(1)}&{(2)}&{(3)}&{(4)}&{(5)}&{(6)}&{(7)}&{(8)}&{(9)}\\
\hline \\
\vspace{5pt}
12A-037 (Zhao)&2012-03-29&C &C$^\ddagger$& 5.5&2&$-3^h.0$ --- $+0^h.8$&0.94&J1733$-$1304\\
\vspace{5pt}
12A-037 (Zhao)&2012-04-22&C &C$^\ddagger$& 5.5&2&$-0^h.5$ --- $+2^h.9$&0.82&J1733$-$1304\\
\vspace{5pt}
12A-037 (Zhao)&2012-07-24&B &C$^\ddagger$& 5.5&2&$-0^h.9$ --- $+3^h.0$&0.63&J1733$-$1304\\
\vspace{5pt}
12A-037 (Zhao)&2012-07-27&B &C$^\ddagger$& 5.5&2&$+0^h.4$ --- $+3^h.9$&0.71&J1733$-$1304\\
\vspace{5pt}
14A-346 (Morris)&2014-05-17&A &C$^\ddagger$& 5.5&2&$-3^h.2$ --- $+0^h.7$&0.31&J1733$-$1304\\
\vspace{5pt}
14A-346 (Morris)&2014-05-26&A &C$^\ddagger$& 5.5&2&$-0^h.4$ --- $+3^h.5$&0.47&J1733$-$1304\\
14A-231 (Yusef-Zadeh)&2014-03-01&A &X$^\ddagger$& 9.0&2&$-3^h.0$ --- $+3^h.2$&0.27&J1733$-$1304\\
\vspace{5pt}
                     &       &&&&&$-3^h.0$ --- $+3^h.5$&1.18&J1744$-$3116\\
14A-232 (Yusef-Zadeh)&2014-04-17&A &X$^\ddagger$&9.0&2&$-3^h.3$ --- $+2^h.0$&0.19&J1733$-$1304\\
\vspace{5pt}
                     &       &&&&&$-3^h.4$ --- $+3^h.4$&1.12&J1744$-$3116\\
SF-0853 (Haggard)    &2014-04-28&A &X$^\ddagger$& 9.0&2&$-3^h.0$ --- $+3^h.2$&0.26&J1733$-$1304\\
\vspace{5pt}
                     &       &&&&&$-3^h.0$ --- $+3^h.5$&1.05&J1744$-$3116\\
\vspace{5pt}
15A-293 (Yusef-Zadeh)&2015-09-11& A &Ka$^\sharp$& 33& 8&$-2^h.6$ --- $+0^h.0$&0.17&J1733$-$1304 \\
\\
\hline
\end{tabular}\\
\begin{tabular}{p{0.90\textwidth}}
{\footnotesize
(1) The JVLA program code and PI name.
(2) The calendar dates of the observations.
(3) The Array configurations.
(4) The JVLA band codes$^\dagger$.
(5) The frequencies at the observing band center.
(6) The total bandwidth of the observations.
(7) The coverage of hour angle (HA) range for the data.
(8) The time on sources.
(9) The source name of gain calibrators included in the data.}\\
{\footnotesize
$^\dagger$The band codes "C", "X" and "Ka" used for the JVLA correspond
to the receiver bands in the frequency
ranges of 4.0 - 8.0 GHz, 8.0 - 12.0 GHz and 26.5 - 40.0 GHz, respectively
(https://science.nrao.edu/facilities/vla/docs/manuals/oss2013B/performance/bands).
}\\
{\footnotesize
$^\ddagger$Correlator setup: 64 channels in each of 16 subbands with channel width of 2 MHz. 
 }\\
{\footnotesize
$^\sharp$Correlator setup: 64 channels in each of 64 subbands with channel width of 2 MHz.}
\\
\end{tabular}
\end{table*}

\section{Data sets and imaging}
Along with the dramatic hardware improvements of the JVLA, 
data reduction software packages are being developed emphasizing 
the wideband capability that allows astronomers to make images much deeper 
than have previously been possible. However, many successive
corrections for possible data issues are needed in order
to construct VLA images approaching  theoretical image fidelity
of the equipped broadband instruments (see Appendix A). 
\begin{figure*}[t!]
\centering
\includegraphics[angle=0,width=180.mm]{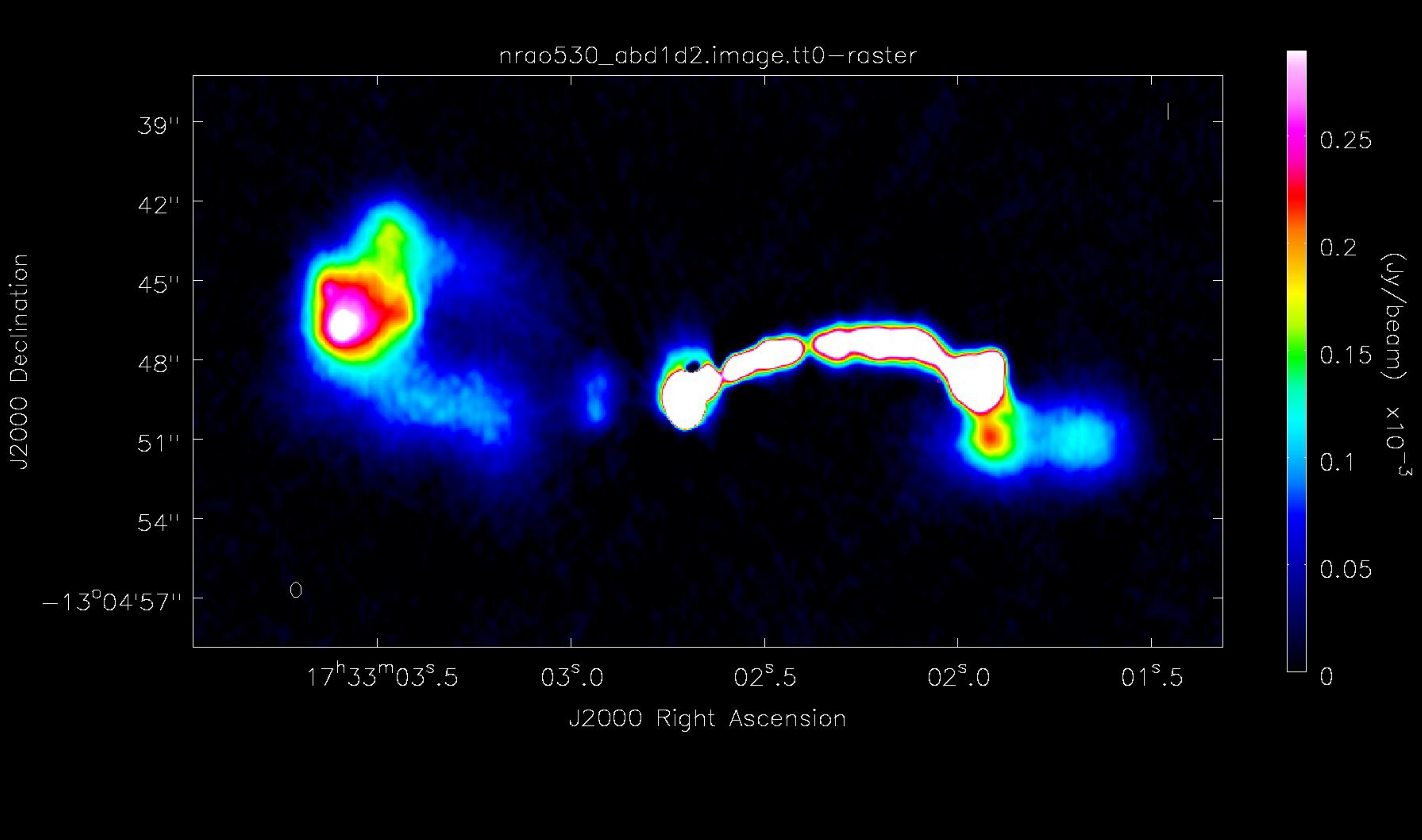}
\caption{The image of NRAO 530 made with natural weighting,
{\it i.e.} weighting all the visibility data equally, by combining
four sets of data observed at 5.5 GHz using the JVLA in the A- and B-arrays
in  July of 2014 and March/April of 2012, respectively (see Table 1). The peak intensity 
and rms are  4.715 Jy beam$^{-1}$ and
 3.4 $\mu$Jy beam$^{-1}$, respectively, corresponding to a dynamic range of 1,300,000:1. 
\textcolor{black}{The wedge (right) shows the color scale (0 $-$ 0.3 mJy beam$^{-1}$) in the display of
the extended emission.}
The synthesized beam is 0.57"x0.41" ($-2.4${\degree}); \textcolor{black}{the ellipse at the left-bottom
indicates FWHM of the beam.}
}
\label{fig1}
\end{figure*}

\begin{figure}[t!]
\centering
\includegraphics[angle=0,width=110mm]{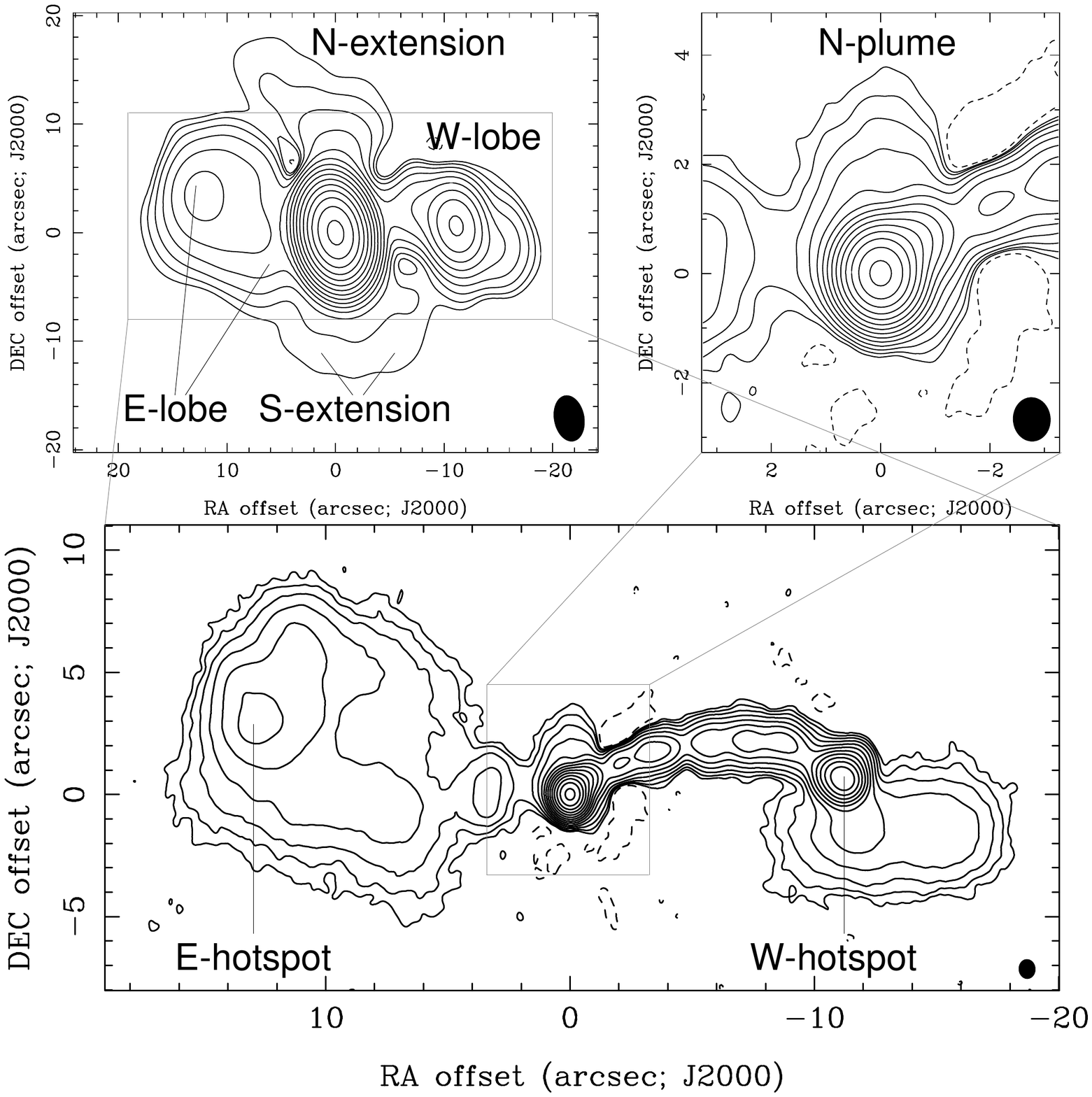}
\includegraphics[angle=0,width=85mm]{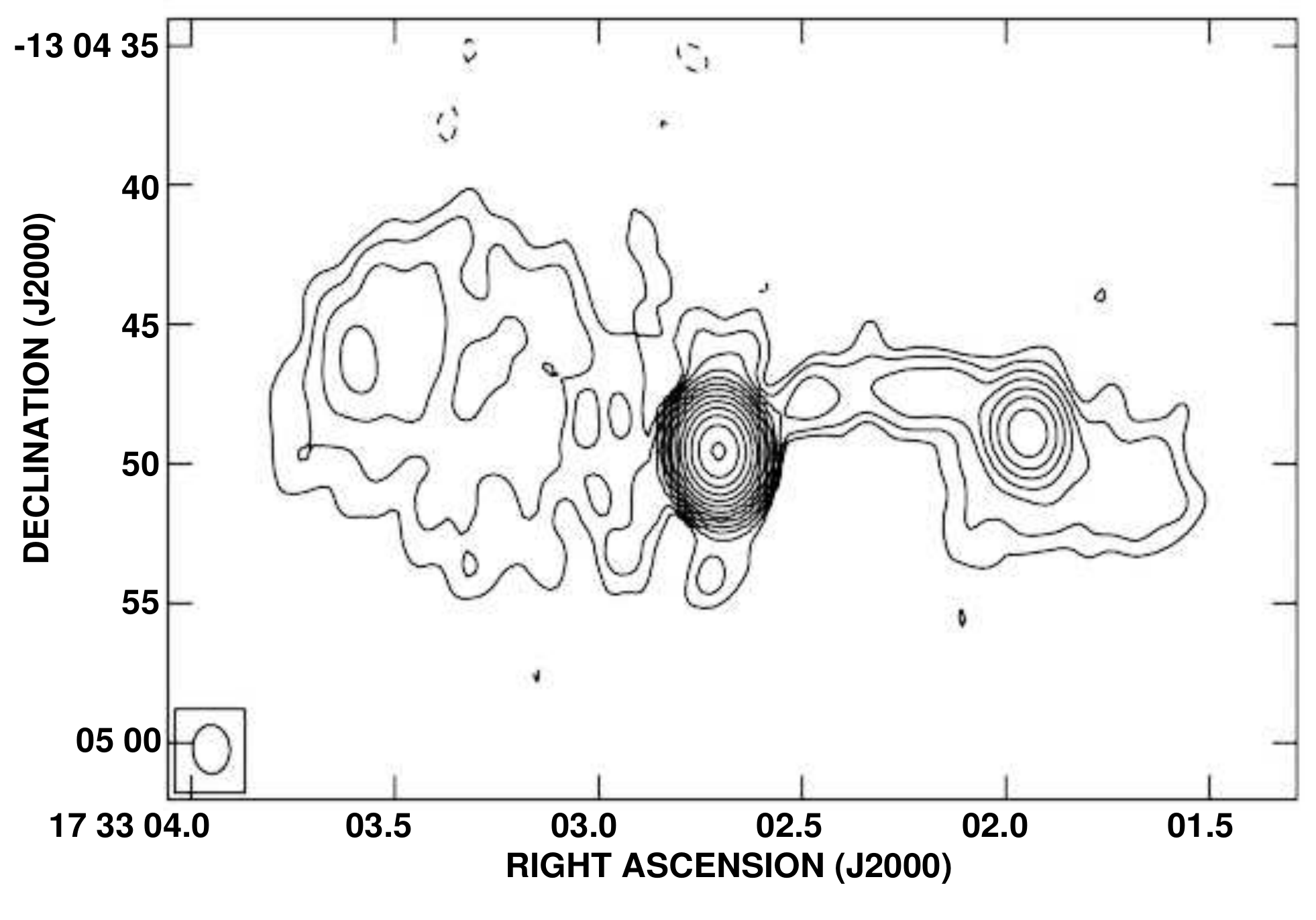}
\caption{
Top-left: Image of NRAO 530 at 5.5 GHz made with robustness 
weight R=0 by combining  data observed in the 
C-array of the JVLA on  2012-03-29 and 2012-04-22. The contours 
are 4.45 Jy beam$^{-1} \times$ ($-$0.00005,0.00005$\times2^{\rm n}$),
where n=0, 1, 2, ..., 14. The rms noise in the region $\sim$10" 
from the core is 6 $\mu$Jy beam$^{-1}$. The synthesized beam is 
4.3" $\times$ 2.8", PA = 11{\degree}. The lowest positive contour 
corresponds to brightness temperature $T_{\rm B}= 0.7$ K.
Middle: The image of NRAO 530 made with six data sets observed at 5.5 GHz
in the A-, B- and C-arrays. A gaussian taper function was applied to the data, 
weighting down the long-baseline visibilities with 
\textit{OUTERTAPER} = 300 k$\lambda$ (see footnote 4 in Section 2.1). 
The final image is convolved to a synthesized beam 0.81"$\times$0.69" 
(4{\degree}), with rms = 5 $\mu$Jy beam$^{-1}$. The inner 6"$\times$8" 
of the core, indicated by a box in the image of the whole source, is 
enlarged at top-right. The contours are S$_{\rm p}\times\left(-10^{-4}, 10^{-4}\times2^n\right)$,
where n = 0, 1, 2, ..., 13 and S$_{\rm p}= 0.18$  Jy beam$^{-1}$. The lowest positive 
contour corresponds to brightness temperature $T_{\rm B}=$ 1.3K.
Bottom: Image of NRAO 530 made with the VLA data observed in 
the A array at 1.46 GHz with the synthesized beam size of $\sim$1.5"
\citep{kha2010}. 
The peak intensity and rms noise are  6.11 Jy beam$^{-1}$ and 0.34 mJy beam$^{-1}$, 
respectively, corresponding to a dynamic range of 18,000:1.}
\label{fig2}
\end{figure}

\subsection{Observations of NRAO 530 and datasets}
In the recent study of Sgr A with the JVLA, we developed procedures 
to demonstrate the possibility of correcting for the data errors and minimizing 
the instrumental and atmospheric effects. In the following text, we discuss 
the data that were used for the images of J1733-1304 (hereafter, NRAO 530) 
and J1744-3116, both used as calibrators for observations of Sgr A. The relevant C-band, 
X-band and Ka-band data sets used in this paper are summarized in Table 1. We 
have applied the technique of RE-correcting and DR-imaging to the radio study 
of the Galactic center; {\it e.g.}, \cite{mzg2017} have published a paper on 
deep imaging the non-thermal radio filament (SgrAWF) near the Galactic black 
hole. More papers on the nonthermal radio filaments and faint
compact radio sources in the radio bright zone (RBZ) at the Galactic center 
will follow.
 
\subsection{Images of NRAO 530}
\subsubsection{C-band images at 5.5 GHz}
Our C-band observations are described in a sequence of papers 
\citep{zmg2013,zmg2016,mzg2017}. NRAO 530 was used as a gain calibrator. 
Using CASA\footnote{the Common Astronomy Software Applications package, 
is being developed with the primary goal of supporting the data post-processing 
needs of the next generation of radio astronomical telescopes such as ALMA 
and VLA. https://casa.nrao.edu/}, we followed the standard calibration for a
continuum source. In order to achieve a high dynamic range (HDR) in a 
 high-resolution image (A-array), in addition to the
VLA standard calibrations, two further corrections must be 
taken: 1) correction for residual delays\footnote{Small phase errors,
present as slopes across sub-bands with a typical value of 
$\sim$2{\degree} per 128 MHz, remain in the data after applying 
JVLA standard calibration. The slopes appear to be caused by residual
delays  with typical values of $\sim$0.04 nsec.
The residual delays are antenna-based and vary with 
time but with unpredictable trend.} across each of
the sub-bands and 2) removal of flux-density variability 
\citep{zhao91,zmg2016, mzg2017}. At radio wavelengths, the 
former is related to antenna-based instrumental issues of
unclear origin (see Appendix A.4. for further discussions), 
and the latter is intrinsic to AGNs. 

We processed the C-band data taken in the A- and B-array at 5.5 GHz
with the JVLA, following the procedure described in  the flow-chart exhibited 
in Fig. A4 of the Appendix. After the first five steps with antenna-based corrections, 
the rms noise of the image with robustness weighting R=0 is 4.5 $\mu$Jy beam$^{-1}$ in 
a region far away from the core ($>15$\arcsec), giving a DR exceeding 1,000,000:1. 
However, within the central 1\arcsec\/ region, the rms noise is about a factor of 10 higher, 
indicating residual errors remaining in the data, including minor baseline-based 
closure issues. Thus with the model produced at step 5, we performed a baseline-based 
correction with a solution interval of 30 seconds. Fig. 1  shows the final image 
made with natural weighting by  combining the A- and B-array data. An rms noise of 
3.5 $\mu$Jy beam$^{-1}$ is achieved \textcolor{black}{in the region without source emission 
($>$15" from the image center) while the rms noise is 4.5 $\mu$Jy beam$^{-1}$ 
near the core ($\sim4"$)\footnote{\textcolor{black}{Note: the regions used for statistics may contain faint extended emission near
the core.}}}, comparable to the expected thermal 
noise of $\sim$3 $\mu$Jy beam$^{-1}$.

Given an observing frequency, the positional errors caused by residual delays
are \textcolor{black}{proportional} to the size of the synthesized beam (see Appendix). For the 
C-array data at 5.5 GHz, contaminated with the mixed issues of larger uncertainties 
of the core position owing to the residual delays, \textcolor{black}{additional flux density} 
from extended emission and lower-level RFI, we initially had a difficulty to converging the 
RE-correction and high-DR imaging simply following the procedure outlined 
in the flow-chart (Fig. A4). A special procedure is developed with more detailed 
sub-steps \textcolor{black}{to complement} each of the major correction cycles. 
A detailed description for 
the reduction of the C-array data is given in Appendix A.5. At step ${\rm M8}$, 
a decent image was obtained from the antenna-based correction process, 
showing a N-S extension. At this stage, it is not clear whether the weak N-S 
feature is a pattern of possible faint residual sidelobes likely caused by 
the postulated baseline-based residual \textcolor{black}{errors}. At the final step of the 
baseline-based correction with the imaging model of the radio core and 
E-W lobes and jets excluding any N-S structure, we solved baseline-based 
solutions with intervals of 120 s, 60 s and 30 s \textcolor{black}{and applied 
them separately to the data.} This final step can effectively eliminate the significant 
\textcolor{black}{basedline-based} errors in comparison to the noise levels corresponding 
to the solution intervals. \textcolor{black}{Therefore, the final image model 
is guatanteed to be produced from antenna-based signals.} The three trials with 
different solution intervals all give the same source structure, revealing the faint 
N-S \textcolor{black}{extended} emission. This necessary test excludes the possibility that 
the N-S pattern is a residual sidelobe caused by the baseline-based closure 
issues on some individual baselines in certain periods, {\it e.g.}, temporary 
issues of some correlator chip. Therefore, we conclude that the N-S extension 
\textcolor{black}{is most likely the true 
\textbf{\textit {weak emission}}  
extended from the core.} Fig. 2 (top-left panel) shows the C-array image 
at 5.5 GHz, revealing a lower brightness feature elongated north-south, labelled 
as N-extension and S-extension, \textcolor{black}{ with an rms noise of 6 $\mu$Jy beam$^{-1}$
in the region no contamination from source emission (20'' away from the core).}
     
Then with the point-source at the core, we aligned the C-array data with 
the A- and B-array data. The flux density of the core was reduced down 
to a common value of 0.18 Jy for all six data sets by subtracting a 
point source model from the phase center of the data to below the contamination level 
spread from the wing of the clean beam. A Gaussian taper function was 
applied to the combined data with {\it OUTERTAPER} \footnote{{\it OUTERTAPER}
is an input parameter of the CASA task \textbf{\textit {clean}},
which sets the boundary for long baselines of the sampled data.} 
$=300$ k$\lambda$ to weight 
down the long-baseline visibilities. The dirty image was cleaned with 
the multi-scale and multi-frequency-synthesis (MS-MFS) algorithm \citep{rau11}  
of CASA and the clean image (Fig. 2, middle) 
was convolved to a synthesized beam 0.81"$\times$0.69" (4{\degree}). The rms 
noise is $\sigma=5\mu$Jy beam$^{-1}$. The top-right panel in Fig. 2 shows 
the central 6"$\times$8" region,  revealing a weak extended emission 
structure stretched out from the core to the north up to 4 synthesized 
beams (3 arcsec) at an intensity level of 4$\sigma$. The combined A+B+C-array 
image at 5.5 GHz shows evidently the detection of the weak component near 
the bright core although the brightness of the southern extension is 
below the detection limit. Our results confirm the N-S extension observed  
in the A-array image at 1.4 GHz (Fig. 2, bottom) taken from \cite{kha2010}.

\begin{figure*}[htp]
\centering
\includegraphics[angle=0,width=160mm]{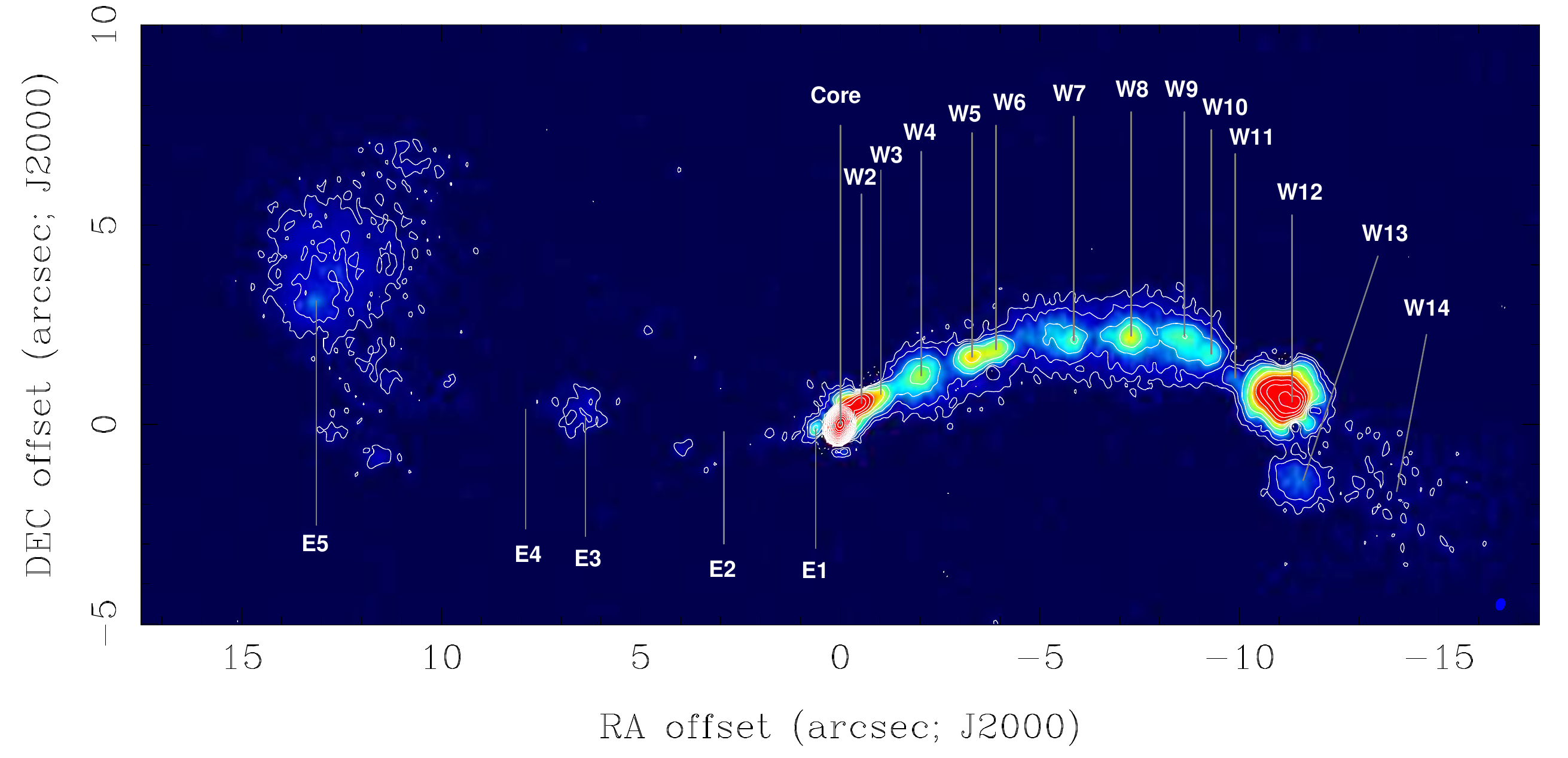}
\caption{Image of NRAO 530 made with robustness weighting R=0 by combining
three data sets from the A-array observations carried out with the JVLA 
at 9 GHz on 2014-03-01, 2014-04-17 and 2014-04-28. The synthesized beam 
is 0.27"x0.19" ($-$17{\degree}). The image peak and rms noise are $S_{\rm p}=$4.915 
Jy beam$^{-1}$ and $\sigma=$3.6 $\mu$Jy beam$^{-1}$, corresponding to a dynamic
range of 1,300,000:1. The compact components with peak intensity $>8\sigma$ are 
labelled. The contours are 2$\sigma\times$(-2, 2$^n$), for $n$ = 1, 2, 3, ..., 19. The 
lower-brightness components W14, E2 and E4 are detected at 5.5 GHz as labelled in 
Fig. 6, top.  
}
\label{fig3}
\end{figure*}

\subsubsection{X-band images at 9 GHz}
Two gain calibrators, NRAO 530 and J1744-3116, were included in 
the X-band observations of Sgr A* with the JVLA in the A-array 
during March and April in 2014 (see Table 1). J1744-3116 is 2.5{\degree}
away from Sgr A while NRAO 530 is 16{\degree} away. Applying the 
calibration solutions derived from closer gain calibrators can
more effectively eliminate the residual errors from the data of
the target sources, in particular for the time-variable residual 
delays (see \cite{mzg2017} and the Appendix below). Following the five-step 
procedure for the RE-correcting and DR-imaging process
discussed in Appendix A.4, we performed further corrections 
for the residual gains, residual delays and flux-density variations 
after the standard calibration for JVLA data, obtaining deep image models
for both calibrators. The final image model of J1744-3116 is 
present in the Appendix (Fig. A3, right), and is utilized to discuss the consequence
of flux-density variation of the core on imaging  and is not  
discussed in the main text. The images of NRAO 530 are discussed 
as follows.

Fig. 3 shows the image of NRAO 530 made with robustness weight 
R=0 by combining the three data sets taken on 2014-3-1, 
2014-4-17 and 2014-4-28. The peak of the image is 4.9 Jy beam$^{-1}$ 
and the rms noise in the outer region is 
 3.6 $\mu$Jy beam$^{-1}$,  giving 
DR of 1,300,000:1 with the synthesized beam of 0.27"$\times$0.19" 
($-17${\degree}).  Twelve compact components \textcolor{black}{in the western jet} (W2-W13) and two 
components \textcolor{black}{in the eastern jet} 
(E1 and E5) are detected at the level of 
$>8 \sigma$ (see Fig. 3). These components are fitted with 2D Gaussian function 
\textcolor{black}{ and the parameters resulting from the fitting are summarized 
in Table 2.} 

Fig. 4 shows uniform weighted X-band images at three individual 
epochs, corresponding to 
the observations at 9 GHz on  March 1, April 17 and April 28 in 2014. A 
comparable rms noise level in the range between 11 - 15 $\mu$Jy beam$^{-1}$ 
is achieved. Each of the epoch data sets has been processed
independently by following the RE-correcting and DR-imaging procedure 
(Fig. A4). The resulting images agree with each other. \textcolor{black}{However, the UV data
were sampled in different hour-angle ranges  at the three epochs (Table 1),
which resulted in different FWHM of the synthesized beams (see  Fig. 4 caption).
The detailed difference between the epochal images,
{\it e.g.}, the apparent peak intensity of
W4, is attributable to the different UV coverage in
the three observations. }
\begin{table*}[ht!]
\center
\tablenum{2}
\setlength{\tabcolsep}{1.7mm}
\caption{Compact components in NRAO 530}
\begin{tabular}{lcccccccccc}
\hline\hline \\
{ID}&
{$S_{\rm p}\pm\sigma$}&
{$T_{\rm B}$} &
{$\nu$}&
{$\alpha$}&
{$S_{\rm t}\pm\sigma$ }&
{$\Delta\theta\pm\sigma$}&
{PA$_\theta$} &
{$\Theta_{\rm maj}\pm\sigma$}&
{$\Theta_{\rm min}\pm\sigma$}&
{PA$_\Theta$} \\
{} &
{(mJy beam$^{-1}$)} &
{(K)}&
{(GHz)} &
{($\nu^\alpha$}) &
{(mJy)} &
{(arcsec)} &
{(deg)}&
{(arcsec)} &
{(arcsec)} &
{(deg)}
\\
(1)&(2)&(3)&(4)&(5)&(6)&(7)&(8)&(9)&(10)&(11)
\\
\hline \\ 
\vspace{3 pt}
Core &${3400\pm 1\over\dots}$&$4.6\times10^5\over\dots$&${33\over\dots}$ &\dots& $3422\pm1\over\dots$&$0.000\over\dots$&$0.0\over\dots$&$0.011\pm0.001\over\dots$&$0.003\pm0.001\over\dots$&$176\pm2\over\dots$\\
\vspace{5 pt}
     &${4915\pm 2\over4715\pm2}$&$1.4\times10^6\over3.7\times10^6$&${9\over5.5}$ &$0.08\pm0.01\over\dots$& $4922\pm2\over\dots$&$0.000\over\dots$&$0.0\over\dots$ &$0.016\pm0.001\over\dots$&$0.007\pm0.001\over\dots$&$174\pm4\over\dots$\\ 
\vspace{5 pt}
W1& $0.23\pm3\over\dots$&$31\over\dots$&${33\over\dots}$&\dots& $0.57\pm0.11\over\dots$ &$0.31\pm0.02\over\dots$&$-27\pm2\over\dots$&$0.21\pm0.05\over\dots$&$0.05\pm0.03\over\dots$&$146\pm7\over\dots$\\
\vspace{3 pt}
W2&${0.31\pm0.02\over\dots}$&$42\over\dots$&${33\over\dots}$& \dots & $1.95\pm0.12\over\dots$ &$0.69\pm0.01\over\dots$&$-42\pm1\over\dots$&$0.31\pm0.02\over\dots$&$0.14\pm0.01\over\dots$&$111\pm3\over\dots$\\
\vspace{5 pt}
   &${3.07\pm 0.24\over4.70\pm 1.70}$&$904\over3700$&${9\over5.5}$&$-0.86\pm0.18\over\dots$&$8.38\pm0.89\over\dots$&$0.69\pm0.01\over\dots$&$-42\pm1\over\dots$&$0.49\pm0.05\over\dots$&$0.30\pm0.02\over\dots$&$125\pm6\over\dots$ \\
\vspace{5 pt}
W3&${0.022\pm0.005\over\dots}$&$3\over\dots$&${33\over\dots}$& \dots & $0.18\pm0.05\over\dots$ &$1.23\pm0.04\over\dots$&$-52\pm1.3\over\dots$&$0.35\pm0.10\over\dots$&$0.16\pm0.05\over\dots$&$128\pm18\over\dots$\\
\vspace{5 pt}
W4   &${0.12\pm 0.01\over0.30\pm0.02}$&$35\over240$&${9\over5.5}$&$-1.8\pm0.17\over\dots$&$0.71\pm0.03\over\dots$&$2.35\pm0.01\over\dots$&$-58\pm0.3\over\dots$&$0.65\pm0.03\over\dots$&$0.49\pm0.02\over\dots$ &$126\pm5\over\dots$\\
\vspace{5 pt}
W5   &${0.29\pm 0.01\over0.46\pm 0.02}$&$85\over360$&${9\over5.5}$&$-0.94\pm0.11\over\dots$ &$1.52\pm0.08\over\dots$ &$3.74\pm0.01\over\dots$&$-63\pm0.2\over\dots$ &$0.64\pm0.03\over\dots$&$0.45\pm0.02\over\dots$ &$106\pm4\over\dots$ \\
\vspace{5 pt}
W6   &${0.21\pm 0.01\over0.32\pm 0.01}$&$62\over250$&${9\over5.5}$&$-0.86\pm0.12\over\dots$ &$1.20\pm0.06\over\dots$&$4.28\pm0.01\over\dots$&$-64\pm0.2\over\dots$&$0.74\pm0.03\over\dots$&$0.41\pm0.02\over\dots$&$112\pm2\over\dots$\\
\vspace{5 pt}
W7   &${0.08\pm 0.01\over0.13\pm 0.01}$&$24\over100$&${9\over5.5}$&$-0.99\pm0.30\over\dots$ &$1.20\pm0.09\over\dots$ &$6.06\pm0.03\over\dots$&$-69\pm0.2\over\dots$ &$1.17\pm0.08\over\dots$&$0.75\pm0.05\over\dots$&$80\pm6\over\dots$ \\
\vspace{5 pt}
W8   &${0.17\pm0.01\over0.32\pm0.02}$&$50\over250$&${9\over5.5}$&$-1.3\pm0.17\over\dots$ &$0.87\pm 0.06\over\dots$&$7.61\pm0.01\over\dots$&$-73\pm0.1\over\dots$&$0.55\pm0.03\over\dots$&$0.49\pm0.03\over\dots$&$98\pm22\over\dots$\\
\vspace{5 pt}
W9   &${0.10\pm0.01\over0.25\pm0.01}$&$29\over200$&${9\over5.5}$&$-1.8\pm0.21\over\dots$ &$1.74\pm0.09\over\dots$&$8.86\pm0.03\over\dots$&$-76\pm0.3\over\dots$ &$1.34\pm0.07\over\dots$&$0.81\pm0.04\over\dots$ &$70\pm4\over\dots$ \\
\vspace{5 pt}
W10   &${0.08\pm0.01\over<0.2}$&$24\over<160$&${9\over5.5}$&$>-1.9\over\dots$&$0.67\pm0.06\over\dots$&$9.40\pm0.04\over\dots$&$-79\pm0.3\over\dots$ &$0.74\pm0.06\over\dots$&$0.59\pm0.05\over\dots$&$24\pm14\over\dots$ \\
\vspace{5 pt}
W11  &${0.06\pm0.01\over0.13\pm0.01}$&$18\over100$&${9\over5.5}$&$-1.6\pm0.37\over\dots$&$0.62\pm0.06\over\dots$&$10.03\pm0.04\over\dots$&$-84\pm0.2\over\dots$&$1.07\pm0.11\over\dots$&$0.55\pm0.05\over\dots$&$56\pm5\over\dots$\\
\vspace{5 pt}
W12  &${5.00\pm0.16\over7.72\pm0.32}$&$1500\over6100$&${9\over5.5}$&$-0.88\pm0.11\over\dots$&$38.1\pm1.3\over\dots$&$11.20\pm0.01\over\dots$&$-87\pm0.1\over\dots$&$0.76\pm0.03\over\dots$&$0.54\pm0.02\over\dots$ &$63\pm4\over\dots$\\
\vspace{5 pt}
W13  &${0.05\pm0.01\over0.08\pm0.02}$&$15\over63$&${9\over5.5}$&$-0.96\pm0.65\over\dots$&$1.35\pm0.06\over\dots$&$11.58\pm0.02\over\dots$&$-97\pm0.1\over\dots$&$1.32\pm0.06\over\dots$&$1.09\pm0.05\over\dots$&$70\pm9\over\dots$\\
\vspace{5 pt}
W14$^\dagger$  &${\dots\over0.124\pm0.002}$&$\dots\over21$&${9\over5.5}$&$\dots$&$\dots\over6.54\pm0.10$&$\dots\over14.74\pm0.04$&$\dots\over-96\pm0.1$&$\dots\over3.99\pm0.06$&$\dots\over3.00\pm0.05$&$\dots\over88\pm2$\\
\vspace{5 pt}
E1   &${0.12\pm0.02\over\dots}$&$35\over\dots$&${9\over5.5}$&\dots&$0.20\pm0.04\over\dots$&$0.62\pm0.02\over\dots$&$97\pm2\over\dots$ &$0.37\pm0.05\over\dots$ &$0.24\pm0.03\over\dots$&$148\pm10\over\dots$\\
\vspace{5 pt}
E2$^\dagger$   &${\dots\over0.10\pm0.4}$&$\dots\over17$&${9\over5.5}$&\dots&$\dots\over2.05\pm0.08$&$\dots\over3.4\pm0.1$&$\dots\over87\pm2$&$\dots\over2.6\pm0.1$&$\dots\over1.7\pm0.1$&$\dots\over170\pm4$ \\
\vspace{5 pt}
E3   &${0.02\pm0.01\over0.03\pm0.01}$&$6\over24$&${9\over5.5}$&\dots&$0.62\pm0.06\over\dots$&$6.5\pm0.1\over\dots$&$89\pm2\over\dots$&$1.27\pm0.11\over\dots$&$1.11\pm0.09\over\dots$&$89\pm26\over\dots$ \\
\vspace{5 pt}
E4$^\dagger$   &$\dots\over{0.083\pm0.002}$&$\dots\over14$&${9\over5.5}$&\dots&$\dots\over4.87\pm0.08$&$\dots\over7.76\pm0.04$&$\dots\over91\pm2$&$\dots\over4.01\pm0.07$&$\dots\over3.30\pm0.06$&$\dots\over38\pm4$ \\
\vspace{5 pt}
E5   &${0.03\pm0.01\over0.06\pm0.01}$&$9\over47$&${9\over5.5}$&$-1.4\pm0.7\over\dots$ &$5.4\pm0.2\over\dots$&$13.2\pm0.1\over\dots$&$73\pm1\over\dots$&$3.4\pm0.2\over\dots$&$2.7\pm0.2\over\dots$&$173\pm6\over\dots$\\
\hline
\end{tabular}\\
\begin{tabular}{p{0.95\textwidth}}
{\footnotesize
Col. 1 is component ID. 
Col. 2 is peak flux densities at 9 GHz / 5.5 GHz.
Col. 3 is the surface brightness temperature.
Col. 4 is the observing frequency.
Col. 5 is spectral index derived from the values listed in col. 2. 
Col. 6 is total flux density. 
Col. 7 is the angular offset from the core,  
RA (J2000) = 17:33:02.706, Dec (J2000) = -13:04:49.55.
Col. 8 is the position angle of the offset vector 
pointing from the core towards the jet components. 
Cols. 9, 10 and 11 are  the sizes 
along the major and minor axis, and the position angle from 2D Gaussian fitting.
The values listed in column 2 and columns 5 to 10 corresponding 
to 9 GHz were determined the A-array image  
with a synthesized beam of 0.27"$\times$0.19" ($-17${\degree}) (Fig. 3). 
The intensities at 5.5 GHz listed in col. 2 
were determined from an image with uniform weighting, convolved to
the same synthesized beam as the 9 GHz image.  
The rms noise of  9 $\mu$Jy beam$^{-1}$ in the 5.5 GHz 
image is larger by a factor of 2.5 than that of the 
9 GHz image. $^\dagger$The quantities for these components 
are determined from the natural weighted A+B image at 5.5 GHz
with the synthesized beam 0.57"$\times$0.41" ($-2.4${\degree}). 
}
\end{tabular}\\
\end{table*}

\begin{figure}[h!]
\centering
\includegraphics[angle=0,width=65mm]{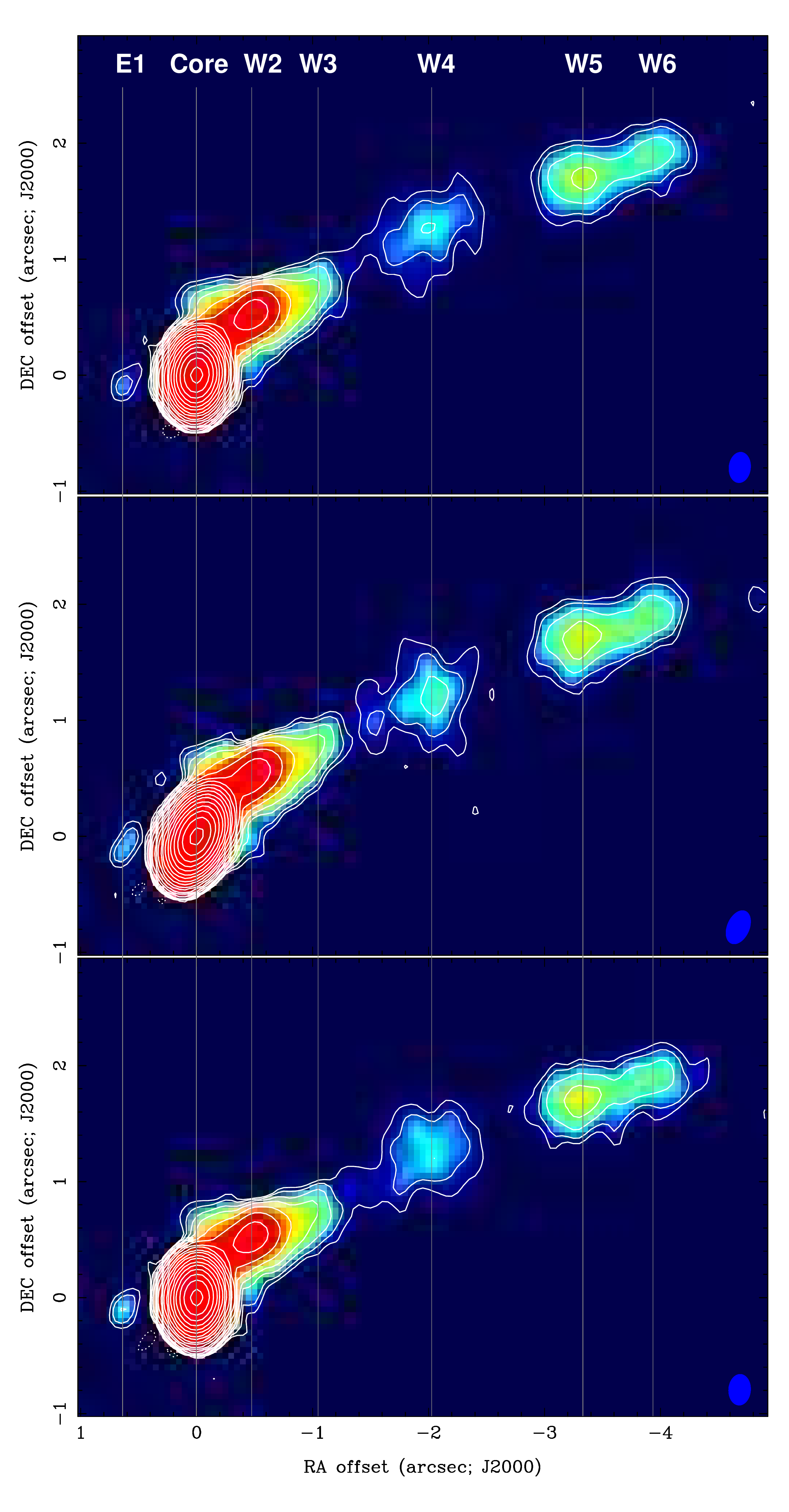}
\caption{The images of NRAO 530 made with uniform weighting, 
corresponding to R=$-2$, from the JVLA observations at 9 GHz 
in the A-array at three epochs. 
Top: March 1, 2014, the rms noise is $\sigma_{\rm d1}= 11 \mu$Jy 
beam$^{-1}$, beam = 0.26"x0.18" ($-$6.1{\degree}). Middle: 
April 17, 2014, $\sigma_{\rm d2} = 15 \mu$Jy beam$^{-1}$, 
beam = 0.30"x0.18" ($-$23{\degree}). Bottom: April 28, 2014, 
$\sigma_{\rm d3} = 12 \mu$Jy beam$^{-1}$, beam = 0.26"x0.18" 
($-$3.8{\degree}). The compact components are labelled with 
the same identifications as in Fig. 3. The contours are 
2$\sigma_{\rm d2}\times$(-2, 2$^n$), for $n$ = 1, 2, 3, ..., 18.
}
\label{fig4}
\end{figure}

\subsubsection{Ka-band images at 33 GHz}
The VLA Ka-band data set 
observed on September 11, 2015, was obtained from the NRAO archive (15A-293).
The correlator setup for the  dataset is described in the footnote of Table 1.
The resulting total bandwidth is 8 GHz.
The central band frequency is 33 GHz.
The data were first calibrated following the standard JVLA procedure.
The corrections for the residual errors follow our new method. 
This data set was used to test and demontrate
the procedure for  RE-correction and high-DR imaging  discussed
in the  Appendix (A.4). 
Fig. 5 shows the image at 33 GHz with
natural weighting. An rms noise of 8$\mu$Jy beam$^{-1}$ is achieved with
a resolution better than 0.1", revealing the bending 
transition from the northern VLBI jet to the  western VLA jet.

\begin{table}[t!]
\tablenum{3}
\setlength{\tabcolsep}{1.7mm}
\caption{Extended components in NRAO 530}
\begin{tabular}{lcccc}
\hline\hline \\
{ID}&
{S$_{\rm t}$}&
{Size} & $\displaystyle {S_{\rm core}\over S_{\rm t}}$ & $P_\nu$\\
{} &
{(mJy)} &
{(arcsec)} &
{} &
{($10^{27}$ W Hz$^{-1}$)}
\\
(1)&(2)&(3)&(4)&(5)
\\
\hline \\
\underline{5.5 GHz}$^\dagger$     &   &   &   & \\
Core        &4450$\pm$5       & $\dots$&   1 &11\\
W-lobe      & 96$\pm$10 & 19    &  46 &0.36\\
E-lobe      & 42$\pm$5  & 18    & 106 &0.16\\
N-extension &  7$\pm$1  & 15    & 636 &0.03 \\
S-extension &  3$\pm$1  & 13    &1480 &0.01\\
Extended    & 148$\pm$13& 37    &  30 &0.56\\
Total       &4598$\pm14$& \dots & 0.968 &11.6\\  
\hline \\
\underline{1.46 GHz}$^\ddagger$     &   &   &   & \\
Core        &6130$\pm$6&$\dots$&     1 & 23 \\
Extended    &518$\pm$46& 37    &    12 & 2  \\
Total       &6648$\pm$47&\dots &  0.922& 25 \\
\hline \\
\underline{Sp. index}$^\sharp$&\multicolumn{2}{r}{\underline{Core}$^\spadesuit$}&\multicolumn{2}{r}{\underline{Extended}} \\
$\alpha_{1.46/5.5}$  & \multicolumn{2}{r}{$-0.23\pm0.12$}& \multicolumn{2}{r}{$-0.94\pm0.13$} \\
\hline
\end{tabular}\\
\begin{tabular}{p{0.45\textwidth}}
{\footnotesize
Col. 1 is ID. Col. 2 is integrated flux density.
Col. 3 is the maximum linear size.
Col. 4 is the ratio of core flux density to the 
integrated flux density for each component.
Col. 5 is radio power $P_\nu = S_t 4\pi D_L^2 / (1+z)^{1+\alpha}$, assuming
$D_L=5.8$Gpc.
}\\
{\footnotesize
$^\dagger$The values are determined from the C-array image at 5.5 GHz (Fig. 2).
}\\
{\footnotesize
$^\ddagger$ \cite{kha2010}.
}\\
{\footnotesize
$^\sharp$The spectral index ${\displaystyle {S_{1.4}\over S_{5.5}}=\left(1.4 \over 5.5\right)^\alpha}$.
}\\
{\footnotesize
$^\spadesuit$The uncertainty in $\alpha$ is dominated by the 6\% variation in flux density
at 5.5 GHz observed in the period between March 29, 2012 and May 26, 2014. 
}
\end{tabular}\\
\end{table}

\begin{table*}[t!]
\centering
\tablenum{4}
\setlength{\tabcolsep}{1.7mm}
\caption{Total flux density of NRAO 530}
\begin{tabular}{ccccccccccc}
\hline\hline \\
{$\nu$ (GHz) }& 
{S$_\nu$ (Jy) }& For fitting & Duration&
 {Ref.} && {$\nu$ (GHz) }&  
{S$_\nu$ (Jy) }& For fitting & Duration&
 {Ref.}\\
(1)&(2) &(3)  &(4) &(5)&&(6)&(7) &(8)  &(9)&(10) \\
\hline \\
\textcolor{black}{0.076}&\textcolor{black}{9.93}$\textcolor{black}{\pm}$\textcolor{black}{ 0.20}&\textcolor{black}{Yes}
&\textcolor{black}{\dots}&\textcolor{black}{[18]}&&0.080 & 7.00&Yes&\textcolor{black}{\dots}&[1] \\
\textcolor{black}{0.084}&\textcolor{black}{9.49}$\textcolor{black}{\pm}$\textcolor{black}{0.15}&\textcolor{black}{Yes}
&\textcolor{black}{\dots}&\textcolor{black}{[18]}&&
\textcolor{black}{ 0.092}& \textcolor{black}{ 9.52}$\textcolor{black}{\pm}$\textcolor{black}{ 0.14}&\textcolor{black}{ Yes}
&\textcolor{black}{\dots}&\textcolor{black}{ [18]}\\
\textcolor{black}{ 0.099}& \textcolor{black}{ 9.09}$\textcolor{black}{\pm}$\textcolor{black}{ 0.12}&\textcolor{black}{ Yes}
&\textcolor{black}{\dots}&\textcolor{black}{ [18]}&&
\textcolor{black}{ 0.107}& \textcolor{black}{ 8.71}$\textcolor{black}{\pm}$\textcolor{black}{ 0.12}&\textcolor{black}{ Yes}
&\textcolor{black}{\dots}&\textcolor{black}{ [18]}\\
\textcolor{black}{ 0.115}& \textcolor{black}{ 8.68}$\textcolor{black}{\pm}$\textcolor{black}{ 0.09}&\textcolor{black}{ Yes}
&\textcolor{black}{\dots}&\textcolor{black}{ [18]}&&
\textcolor{black}{ 0.122}& \textcolor{black}{ 8.60}$\textcolor{black}{\pm}$\textcolor{black}{ 0.08}&\textcolor{black}{ Yes}
&\textcolor{black}{\dots}&\textcolor{black}{ [18]}\\
\textcolor{black}{ 0.130}& \textcolor{black}{ 8.23}$\textcolor{black}{\pm}$\textcolor{black}{ 0.07}&\textcolor{black}{ Yes}
&\textcolor{black}{\dots}&\textcolor{black}{ [18]}&&
\textcolor{black}{ 0.143}& \textcolor{black}{ 8.09}$\textcolor{black}{\pm}$\textcolor{black}{ 0.07}&\textcolor{black}{ Yes}
&\textcolor{black}{\dots}&\textcolor{black}{ [18]}\\
\textcolor{black}{ 0.150}& \textcolor{black}{ 7.65}$\textcolor{black}{\pm}$\textcolor{black}{ 0.77}&\textcolor{black}{ Yes}
&\textcolor{black}{\dots}&\textcolor{black}{ [17]}&&
\textcolor{black}{ 0.151}& \textcolor{black}{ 7.96}$\textcolor{black}{\pm}$\textcolor{black}{ 0.06}&\textcolor{black}{ Yes}
&\textcolor{black}{\dots}&\textcolor{black}{ [18]}\\
\textcolor{black}{ 0.158}& \textcolor{black}{ 7.65}$\textcolor{black}{\pm}$\textcolor{black}{ 0.05}&\textcolor{black}{ Yes}
&\textcolor{black}{\dots}&\textcolor{black}{ [18]}&&
\textcolor{black}{ 0.166}& \textcolor{black}{ 7.61}$\textcolor{black}{\pm}$\textcolor{black}{ 0.05}&\textcolor{black}{ Yes}
&\textcolor{black}{\dots}&\textcolor{black}{ [18]}\\
\textcolor{black}{ 0.174}& \textcolor{black}{ 7.92}$\textcolor{black}{\pm}$\textcolor{black}{ 0.06}&\textcolor{black}{ Yes}
&\textcolor{black}{\dots}&\textcolor{black}{ [18]}&&
\textcolor{black}{ 0.181}& \textcolor{black}{ 7.63}$\textcolor{black}{\pm}$\textcolor{black}{ 0.06}&\textcolor{black}{ Yes}
&\textcolor{black}{\dots}&\textcolor{black}{ [18]}\\
\textcolor{black}{ 0.189}& \textcolor{black}{ 7.61}$\textcolor{black}{\pm}$\textcolor{black}{ 0.07}&\textcolor{black}{ Yes}
&\textcolor{black}{\dots}&\textcolor{black}{ [18]}&&
\textcolor{black}{ 0.197}& \textcolor{black}{ 7.66}$\textcolor{black}{\pm}$\textcolor{black}{ 0.08}&\textcolor{black}{ Yes}
&\textcolor{black}{\dots}&\textcolor{black}{ [18]}\\
\textcolor{black}{ 0.200}& \textcolor{black}{ 7.49}$\textcolor{black}{\pm}$\textcolor{black}{ 0.04}&\textcolor{black}{ Yes} 
&\textcolor{black}{\dots}&\textcolor{black}{ [18]}&&
\textcolor{black}{ 0.204}& \textcolor{black}{ 6.83}$\textcolor{black}{\pm}$\textcolor{black}{ 0.14}&\textcolor{black}{ Yes}
&\textcolor{black}{\dots}&\textcolor{black}{ [18]}\\
\textcolor{black}{ 0.212}& \textcolor{black}{ 6.72}$\textcolor{black}{\pm}$\textcolor{black}{ 0.09}&\textcolor{black}{ Yes}
&\textcolor{black}{\dots}&\textcolor{black}{ [18]}&&
\textcolor{black}{ 0.220}& \textcolor{black}{ 6.93}$\textcolor{black}{\pm}$\textcolor{black}{ 0.15}&\textcolor{black}{ Yes}
&\textcolor{black}{\dots}&\textcolor{black}{ [18]}\\
\textcolor{black}{ 0.227}& \textcolor{black}{ 7.26}$\textcolor{black}{\pm}$\textcolor{black}{ 0.18}&\textcolor{black}{ Yes}
&\textcolor{black}{\dots}&\textcolor{black}{ [18]}&&
\textcolor{black}{ 0.327}& \textcolor{black} {7.42}$\textcolor{black}{\pm}$\textcolor{black}{ 0.34}&\textcolor{black}{ Yes}
&\textcolor{black}{\dots}&\textcolor{black}{ [19]}\\
\textcolor{black}{ 0.365}& \textcolor{black} {8.44}$\textcolor{black}{\pm}$\textcolor{black}{ 0.35}&\textcolor{black}{ Yes} 
&\textcolor{black}{\dots}&\textcolor{black}{ [22]}&&
0.408&6.58&Yes&\dots&[1] \\
0.750&6.73$\pm$0.21&Yes&\dots&[2]&&
1.400&5.99$\pm$0.14&\dots&\dots&[2]\\
1.41&5.2&Yes&\dots&[1] &&
1.46&6.648$\pm$0.047&\dots&\dots&[13][15]\\
1.5&5.2&Yes&\dots&[16] &&
\textcolor{black}{2.64}&\textcolor{black}{3.79 $-$ 5.41}&\textcolor{black}{Yes}&\textcolor{black}{8yr}&\textcolor{black}{[20][21]}\\
2.7&4.3&Yes&\dots&[1] &&
4.8&4 $-$9&Yes&\textcolor{black}{46yr}&[9]\\
4.85& 6.991$\pm0.099$&\dots&\dots&[3]&&
5.0&4.1&Yes&\dots&[1] \\
5.0&5.0&\dots&\dots&[16] &&
5.5&4.74$\pm0.13$&\dots&\dots&[15]\\
8.0&3.5 $-$ 12.5&Yes&\textcolor{black}{46yr}&[9]&&
8.1&10.5&\dots&\dots&[16] \\
8.4&6.2&\dots&\dots&[1] &&
9.0&5.002$\pm$0.050&\dots&\dots&[15]\\
14.5&2.5 $-$ 15&Yes&\textcolor{black}{46yr}&[9]&&
15 &11.0&\dots&\dots&[16]\\
22 &8.22 $-$ 11.78&\dots&\textcolor{black}{0.3yr}&[10]&&
33 &3.426$\pm$0.017&\dots&\dots&[15]\\
37 &2.25 $-$ 15.55&Yes&\textcolor{black}{25yr}&[4]&&
37 &10.51 $-$ 12.20&\dots&\textcolor{black}{0.1yr}&[10]\\
43 &8.98$\pm$1.87&\dots&\dots&[16]&&
43 &16.6$\pm$0.5&\dots&\dots&[6]\\
86 &11.2$\pm$0.3&\dots&\dots&[7]&&
\textcolor{black}{86} &\textcolor{black}{2.10 $-$ 6.11}&\textcolor{black}{Yes}&\textcolor{black}{8yr}&\textcolor{black}{[21]}\\
88 &7.8$\pm$0.5&\dots&\dots&[8]&&
90 &3 $-$ 15&\textcolor{black}{\dots}&\textcolor{black}{11yr}&[9]\\
95 &12.6$\pm$0.7&\dots&\dots&[6]&&
107&13.2$\pm$0.4&\dots&\dots&[8]\\
150&7.09$\pm$0.35&\dots&\dots&[11]&&
215&6.2$\pm$1.1&\dots&\dots&[7]\\
230&2.4 $-$ 13.0&\textcolor{black}{\dots}&\textcolor{black}{1yr}&[12]&&
230&0.857 $-$ 3.868&Yes&\textcolor{black}{16yr}&[14]\\
270&5.66$\pm0.30$ &\dots&\dots&[11]&&
345&0.604 $-$ 2.41&Yes&\textcolor{black}{15yr}&[14]\\
350&0.73 $-$ 4.08&Yes&\textcolor{black}{8yr}&[5]&&
375&4.67$\pm$0.47&\dots&\dots&[11]\\ 
\hline
\end{tabular}\\
\begin{tabular}{p{0.85\textwidth}}
{
\scriptsize 
\textcolor{black}{
Cols. 1 \& 6 -- The observing frequencies. Cols. 2 \& 7 -- The flux densities 
and 1 $\sigma$ error for the individual epochs' measurements, or the range between 
minimum and maximum of flux density determined from the monitoring programs.
Col. 3 \& 8 -- If "Yes", the mean values determined from the individual epochs' 
observations or the minimum values determined from the monitoring programs 
are used to create a low-state spectrum. These values are  marked with open circles 
in Fig. 8 and are used in the spectral fitting.
Col. 4 \& 9 -- The duration length in years are given for the flux-density monitoring
programs.
Col. 5 \& 10 -- References: $-$ } 
[1] \cite{wri90}, The Parkes Survey.  [2] \textcolor{black}{\cite{paul66}, The Green Bank 300-foot Survey.} 
[3] \cite{gri94}. 
[4] \cite{hov08}. 
[5] \cite{jen10,rob01}. [6] \cite{fal98,feng06}.
[7] \cite{kri97,feng06}. [8] \cite{reu97,feng06}.
[9] \cite{bow97,all85,all12,an13}. [10] \cite{bow97,ter92}.
[11] \cite{bow97,ste95}.
[12] \citep{bow97}. [13] \cite{kha2010}. [14] \cite{gur18}, The SMA Calibrator Catalog.
[15] This paper. The 1 $\sigma$ errors are mainly due to the core variations.
[16] \cite{vlacat18}; and {\it Test Memo \#192 of NRAO}, C. Chandler (1995).
\textcolor{black}{[17] \cite{gmrt150mhz17}, The GMRT-TGSS Survey. [18] \cite{hurley17}, The WMA-GLEAM Survey. 
[19] This paper; derived from VLA archival data AP327\_A951127.xp1 assuming the flux density of 3C 286: 26.3 Jy at 327 MHz.
[20] \cite{ange19}, Effelsberg 100-m Monitoring for the F-GAMMA program. 
[21] Personal communication (E. Angelakis) on F-GAMMA: Multi-frequency radio monitoring of Fermi blazars: https://www3.mpifr-bonn.mpg.de/div/vlbi/fgamma/Light\_Curves\_\%26\_Spectra.html.
[22] \cite{doug96}, The Texas Survey.
}}
\end{tabular}\\
\end{table*}

\section{Description of the structure of NRAO 530}

\subsection {Structure on kiloparsec scales}
Fig. 1 shows a remarkably detailed image of the radio emission structure 
related to this blazar. NRAO 530 appears to share the common radio 
emission structure  observed in radio galaxies and QSOs. From the  
high-resolution (0.5") image at 5.5 GHz made with combined A- and B-array 
data from the JVLA observations (Fig.~1) as well as the Gaussian-tapered 
image at 0.8" resolution with combined A-, B-, and C-array data 
(Fig. 2, middle), both the dominant bright core and the extended radio 
structure with an angular size of 10 arcsec show a sequence of slightly 
resolved emission knots aligned in the well-confined but curved western 
jet (W-jet). This structure bridges the hotspot (W12) in the western lobe 
(W-lobe) with the radio core. The eastern jet (E-jet) is delineated by a 
faint radiation blob (E2) and a plume of slightly curved faint emission (E3 and E4) 
connecting to the hotspot (E5) in the eastern lobe (E-lobe) 
(see the top panel in Fig. 6 for labelling, and also see Fig. 1 and
the middle panel in Fig. 2  for the detailed structure). 
In contrast to the western jet, in which 
the emission is well confined within a train of bright spots, the eastern 
jet (or jet path) is represented by a more diffuse emission 
ridge with no bright spots.  
The radio morphology is similar to those of radio galaxies and QSOs 
\citep[for example]{bri84}; the radio core of NRAO 530 dominates the 
power, with intensity ratios of 610:1 and 79,000:1 
with respect to the hotspots in the E- and W-lobes, respectively. At 5.5 GHz,
the specific power 
of the core is about 1.1$\times10^{28}$ Watt Hz$^{-1}$, dominating
the power of $\sim5.6\times10^{27}$ Watt Hz$^{-1}$ obtained by integrating 
over the extended 
components as revealed in the C-array image (see Fig. 2 and Table 3).

\begin{figure}[ht]
\centering
\includegraphics[angle=0,width=90mm]{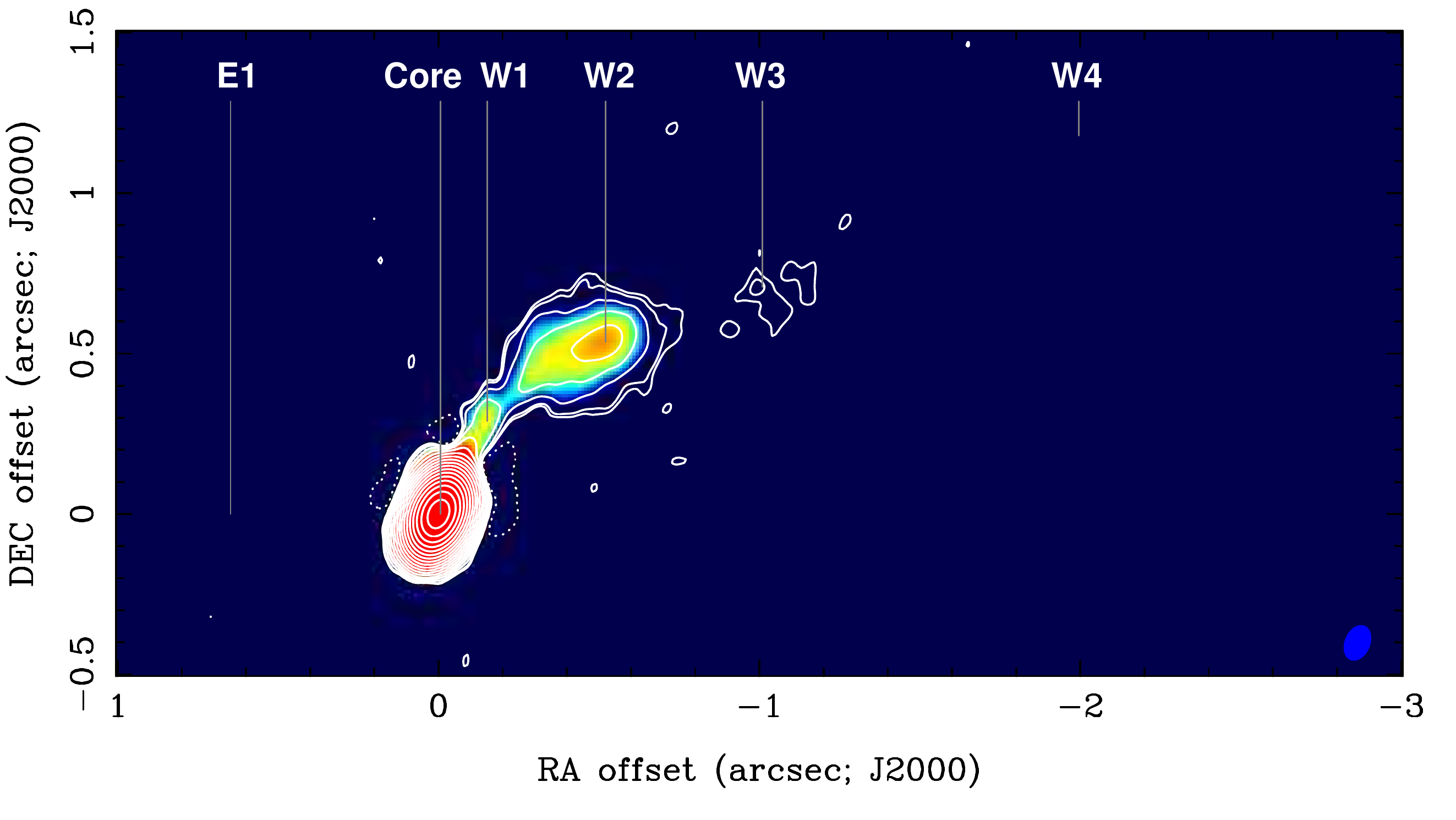}
\caption{Image of NRAO 530 at 33 GHz made with the A array data observed on 
September 11, 2015. Applying natural weighting (R=2), produces a synthesized 
beam of size 0.11"$\times$0.075" ($-$23{\degree}).  The peak intensity of the image
is 3.4 Jy beam$^{-1}$ with rms noise of $\sigma= 8 \mu$Jy beam$^{-1}$, achieving 
a dynamic range of of 420,000:1. The contours are $\sigma\times$ 
($-$4, 3, 2$^{\rm n}$), for n=2, 3, 4 ..., 18.}
\label{fig.5}
\end{figure}

\subsubsection {Eastern jet}
The 5.5 GHz images (Fig. 1, the middle panel of Fig. 2 and the top panel of Fig. 6)  
show the slightly curved  eastern jet, or jet path, with no bright spots. 
Fig. 3 shows an image at 9 GHz with angular resolution of 0.2". At this 
resolution, two  (E2 and E4) of the five features along the eastern jet 
and lobe  (E1, E2, E3, E4 and E5) are detected at 9 GHz. All
five features are observed at 5.5 GHz with angular resolution of 0.5"  
(see the top panel of Fig. 6 for labelling). The brightest component in the eastern jet, E1 
($S_{\rm p}=0.12$ mJy beam$^{-1}$), is located 0.6" 
east of the core, \textcolor{black}{and} is detected only at 9 GHz in the A-array
observations during the period between March 1 and April 28, 2014 
(see Fig. 4). The deconvolved size of E1 0.37"$\times$0.24" (PA=148{\degree}) 
has been determined from 2D Gaussian fitting; the source is slightly 
resolved at the angular resolution of 0.2". The component E1 is not 
detected at 5.5 GHz with angular resolution of 0.5", or cannot be separated
from the core owing to inadequate angular resolution. 
This component is also not detected at 33 GHz. The 
3$\sigma$ limit ($S_{\rm p}=24~\mu$Jy beam$^{-1}$) of the A-array image 
at 33 GHz is consistent with E1 \textcolor{black}{being} a newly ejected jet component with 
a steep spectral index and an intensity  below the 
detection limit at the angular resolution of 0.1".
Furthermore, the component E1 is detected in all  three epochs of the 
A-array observations at 9 GHz on 2014-3-1, 2014-4-17 and 2014-4-28. 
No significant variation in flux density, position or size
is observed during this two-month period.

\subsubsection{Hotspot of E-lobe} The component E5 appears to mark the hotspot
in the eastern lobe where the curved eastern jet terminates, showing the outer 
edge-brightened structure of the eastern lobe. The surface brightness of E5 is  
low, with peak intensities at 9 and 5.5 GHz being $S_{\rm p} =$ 30 and 
60 $\mu$Jy beam$^{-1}$ as determined from the images convolved 
to a commom beam (0.27"$\times$0.19", PA=$-17${\degree}). 
The intensity contrast to the core is $S_{\rm p }^{\rm E5}$:
$S_{\rm p}^{\rm Core}$  = 1:160,000 at 9 GHz and 1:79,000 at 5.5 GHz. At 
1.46 GHz, this ratio is less than 1:1,000, determined from the image of \cite{kha2010}
(Fig. 2 bottom). In addition, the eastern hotspot E5 is trailed by
a fainter emission plume elongated 5" towards the northwest. 
The spectral index determined from the 9 and 5.5 GHz data is 
$-1.4\pm0.7$.

\begin{figure}[t]
\centering
\includegraphics[angle=0,width=105mm]{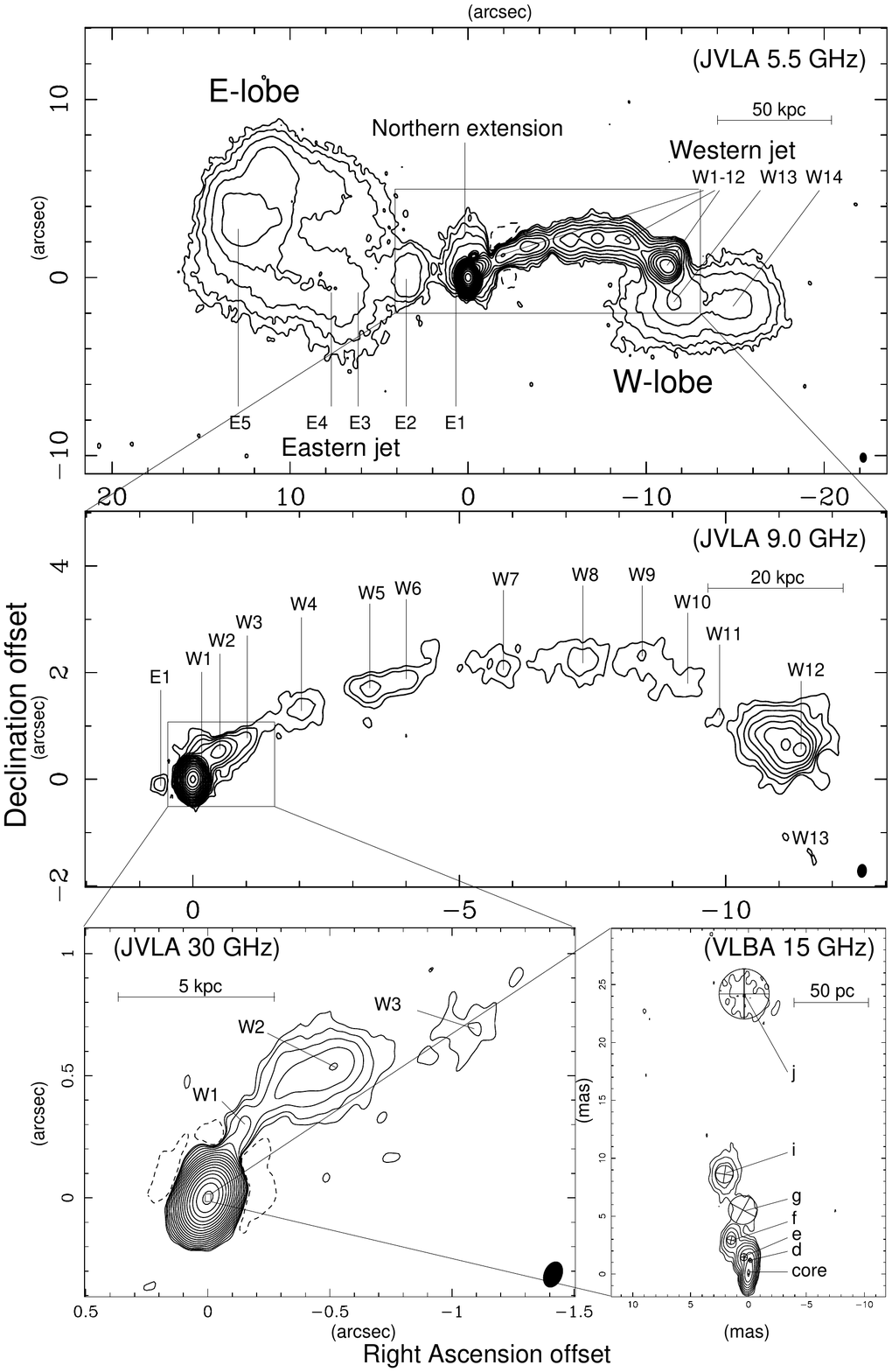}
\caption{Images of NRAO 530 made with JVLA combined  A- and B-array data 
at 5.5 GHz (top) and A-array data at  9.0 GHz (middle).
Bottom left is the Ka-band image of NRAO 530 made with JVLA A-array data at 33 GHz.
Bottom right is the VLBA image of NRAO 530 at 15 GHz from \cite{lu2011}}
\label{fig.6}
\end{figure}

\subsubsection{Western jet} Fourteen
compact components in the western jet and lobe are identified as 
W1 to W14. W1 is only detected in the high resolution image at 33 GHz,
and is not distinguishable from the core in the JVLA A-array images at both 9 
and 5.5 GHz. The components W1 to W12 are located along a curved jet 
feature, convex towards north. Table 2 lists emission quantities 
determined from the high-resolution JVLA images. The components are 
slightly resolved in the A-array images. The 9 and 5.5 GHz data at angular
resolution of 0.2" show 
that the components W2-W11 have  peak intensity $S_{\rm p}$  
ranging from 0.06 to 3.07 mJy beam$^{-1}$ at 9 GHz and 
0.13- 4.70 mJy beam$^{-1}$ at 5.5 GHz. The spectral 
index determined from these components at 9 and 5.5 GHz is about
$\alpha=-1$, consistent with synchrotron radiation.  
W3, located 0.5" northwest of  W2, 
appears to be nearly resolved out \textcolor{black}{at 33 GHz}, showing a marginal detection at the level 
of 4$\sigma$. The rest of the components in the western jet appear 
to be resolved out and not detected at 33 GHz.
\subsubsection{Hotspot of W-lobe}
After the core,  the second brightest component in NRAO 530
is  W12,  with peak intensities of 5.0 and 7.72 mJy beam$^{-1}$ 
at 9 and 5.5 GHz, respectively, marking the hotspot in the western jet/lobe 
structure. A hotspot  usually represents the termination of a jet
flow \citep[for example]{bla74}, where the jet flow impacts  
the ambient medium at the outer edge of the lobe. Similar to the {
\it northwestern} tail of the eastern hotspot E5, a {\it southwestern} 
extension (W13 and W14) appears to  be associated with
W12 (see Figs. 1, 2 and 6). We also point out a detailed 
difference of the W-lobe from the E-lobe as follows.
The {\it northwestern} tail of the hotspot E5 and the hotspot itself,
as well the eastern jet that forms the edge-brightened large lobe \textcolor{black}{
all combine to give the appearance that the jet material is flowing back 
toward the core, at leat in projection.}
The E-lobe appears to be consistent with the typical 
FR II morphology. 
The western jet, on the other hand, turns 90{\degree} to the south
after passing through the hotspot W12, and then 
stretches further out to the west.
There appears no indication of a similar continuation of the western jet 
after the hotspot W12 pointing back towards the core. 
The diffuse western  tail of the W-lobe appears to be more like the 
morphology as often observed in FR I type jets, such as the mirror-symmetric 
jets in 3C 449 \citep{perl79}.
  
\subsubsection{North-south extension to the core}

In the C-array image with low angular resolution of $\sim$4" (Fig. 2), a lower 
surface brightness structure (plume) extending 15" north and 13" south 
of the core is detected at an intensity level in the range between 0.2 and 
2 mJy beam$^{-1}$. This N-S  feature is obvious in the 1.46 GHz A-array image 
at a resolution of 1.5" from \cite{kha2010} (see the bottom panel of Fig. 2). 
In the higher angular resolution (0.8") image with combined A-, B-, and C-array 
data at 5.5 GHz, an emission plume with intensity in the range between 0.02 and 
0.3 mJy beam$^{-1}$  is located a few arcsec north of the core.
The lower surface brightness emission feature of the southern 
extension appears to be below the detection limit. 
Table 3 lists the emission quantities of 
the north-south extension for comparison with the core and eastern and western 
lobes. The 5.5 GHz flux density integrated over the northern extension is 
7$\pm$1 mJy, about 5 percent 
of the western lobe, including the jet, and the flux density of 
the southern extension is 3$\pm$1 mJy. 
The flux density integrated over the full N-S extension is about 10\%
of the total flux density integrated over the overall extended emission, 
excluding the central core emission. 
However, the total flux-density from the extended emssion in
the north-south extension is only $\sim$3\% 
of the core flux density at 5.5 GHz and about 10\% at 1.4 GHz. 

\begin{figure}[t!]
\centering
\includegraphics[angle=0,width=85mm]{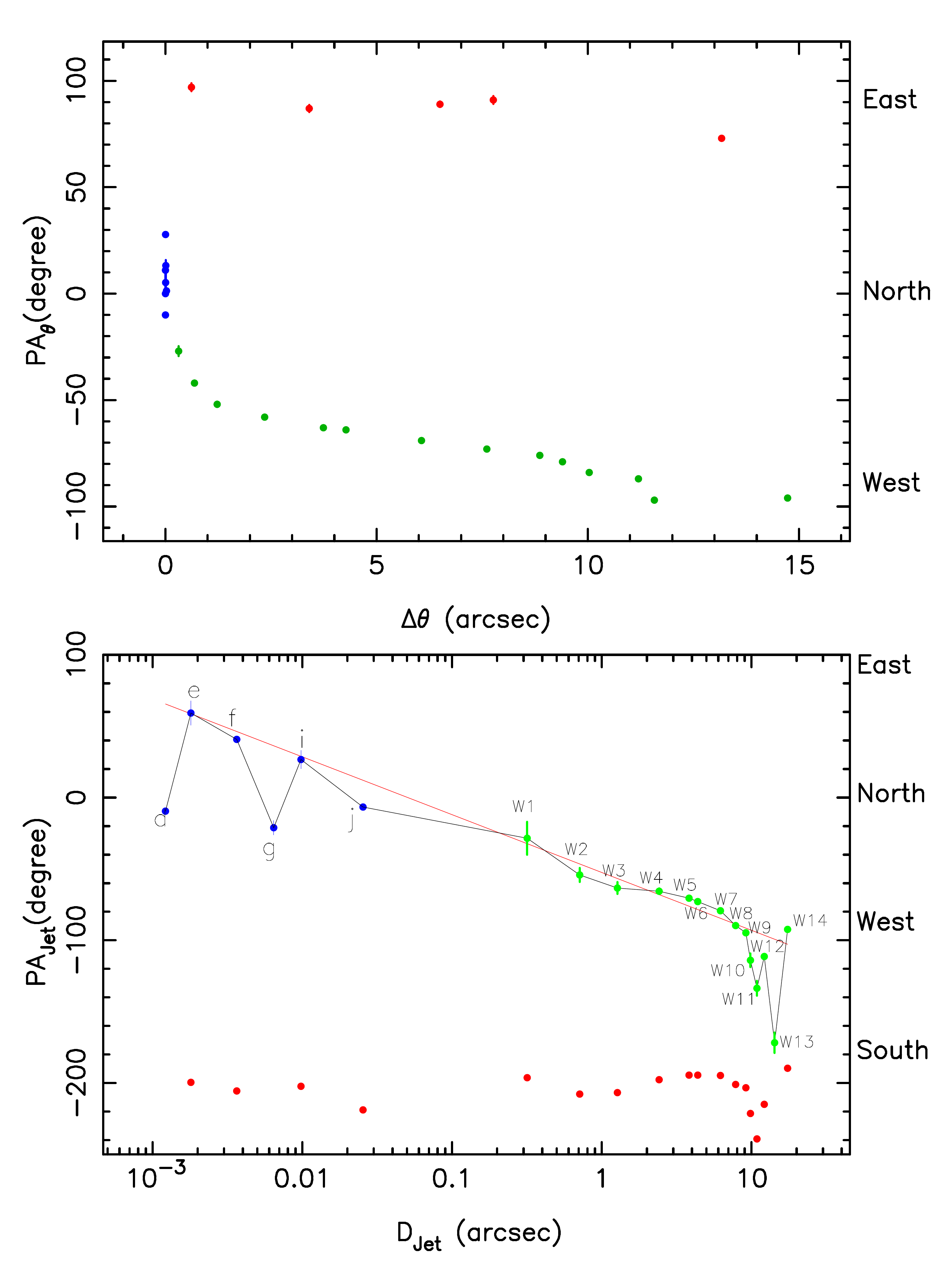}
\caption{Top: Position angle (PA$_\theta$) of the jet axis vector projected 
on the sky ($\vec{J~}$) as a function of $\Delta\theta$, where the jet axis vector 
$\vec{J~}$ for a given jet component is defined as the vector pointing to the 
component from the core with \textcolor{black}{length} $\Delta\theta$ (the angular offset  
from the core) and position angle  PA$_\theta$. The data for the western jet 
(green) and the eastern jet (red) are from this paper (Table 2). The data
for the VLBI jet components d, e, f , g, i and j (blue) are from \cite{lu2011}.
Bottom: Position angle (PA$_{\rm Jet}$) 
of the unit vector ($\vec{T}$) of the direction of the jet 
transverse velocity as a function of the distance measured along the projected
jet trajectory 
from the core to a jet component (D$_{\rm Jet}$); see section 3.2.2
for the definition of the quantities. 
The vertical bars indicate 1$\sigma$ uncertainties. 
The uncertainties for the horizontal axis are smaller than the symbols.
\textcolor{black}{The straight line (red) indicates a linear fit of
a logarithmic function of 
PA$_{\rm Jet}=-40.4{\rm Log}_{10}({\rm D}_{\rm Jet})-52.5$ to the data.
The red dots along the bottom indicate residuals of $\Delta {\rm PA}_{\rm Jet}$ (= data$-$fit) 
minus a constant of 200{\degree}. 
The mean and rms of the residuals are $-$3.6{\degree} and 1.2{\degree}
for all the data included. Eliminating VLBI and VLA components that are highly deviated from
the fitted curve, {\it i.e.} excluding the VLBI components d and g and the 
VLA components beyond W9 in the statistics,
the mean becomes 0.0{\degree} and the rms stays at a  
value $\sim1.2${\degree}. The residuals in statistics are weighted 
by the 1$\sigma$ uncertainties of the data.  
The residuals for the components d, g and W13 are out of the plot range.
}}
\label{fig.7}
\end{figure}

\subsection{The northern VLBI jet and western JVLA jet}

\subsubsection{$\vec{J}$ -- the  vector of jet axis} 
Fig. 6 shows the high-resolution images obtained from the JVLA wideband
data at 5.5, 9 and 33 GHz as compared to the VLBA image at 15 GHz.
The JVLA images trace the western jet on scales from 200 kpc (5.5 GHz)
to 700 pc (33 GHz). At the resolution of 0.1", the  components 
W1 and W2 detected at 33 GHz show an elongation to the NW. This structure 
begins at the core, linking to the large-scale western jet.
The position angles of the major axis (from 2D Gaussian fitting) 
indicate PA=146{\degree}$\pm$7{\degree}  for W1 at an angular 
offset $\Delta\theta=0.31$" from the core and  PA=111{\degree}$\pm$3{\degree} 
for W2 at $\Delta\theta=0.69$" (see Table 2). The change in PA 
between W1 and W2  suggests 
that the projected jet axis bends towards the west as the scale 
increases from a few 100 pc to a few kpc. At the size scale of  the VLBA 
structure of $<200$ pc, the axis of the jet is oriented toward the 
north \citep[for example]{lu2011}.

To quantitatively investigate the relationship between the jet on the 
VLA scale 
and the VLBI jet, we defined a jet vector  
$\vec{J}$ as the vector that connects the core to a jet component. The  
vector $\vec{J}$ can be quantitatively described by its position angle
(PA$_\theta$) and its angular offset from the core ($\Delta\theta$).
For a  jet with linear motion or no precession  involved,
its trajectory is a straight line, 
{\it i.e.}, no changes in PA$_\theta$
are expected as $\Delta\theta$ increases. Table 2 lists
$\Delta\theta$ and PA$_\theta$ in columns 7 and 8, respectively,
for all components. 

Fig. 7 (top) plots the position angle
PA$_\theta$ as function of $\Delta\theta$.
The western (eastern) jet components are represented in green
(red); the quantities PA$_\theta$ and
$\Delta\theta$ on angular scales between 0.1 to 20 arcsec, are
determined from the JVLA data (see Table 3).
The northern jet components observed with the VLBA \citep{lu2011} are in blue.
From this plot, the following obvious conclusions can be drawn:
\vskip 3pt
\noindent
(1) The JVLA western jet (green) is the continuation of 
the VLBI northern jet (blue).  The PA of the jet axis $\vec{J}$ 
is essentially to the north on the scale of VLBA component ($<50$ mas) 
and gradually changes to west as $\Delta\theta$ increase, the 
angular distance. The transition where PA$_\theta$ changes rapidly
occurs in the range between 0.05" and 1" in $\Delta\theta$.
The northern VBLI jet and the western JVLA jet belong to one
astrophysical entity, hereafter referred as \textbf{\textit{the northwestern jet}}.
\vskip 3pt
\noindent
(2) On scales $<50$ mas, the PA$_\theta$ changes by 
$>50${\degree}. According to  proper motion measurements
based on  multiple VLBA observations by \cite{lu2011}, those authors
conclude that the motion on VLBA scales is dominated by the E-W swing of 
the jet components at a rate of 3.4{\degree} yr$^{-1}$. They 
suggest that the observed angular motion is  evidence for helical motion 
of the jet component on  scales from a few mas to a few tens of mas. 
Such an E-W swing in which PA$_\theta$ changes by $>50${\degree} 
is not observed in the JVLA data. Plausibly, the rapid swing motion 
occurs only at small angular scales $<50$ mas and the angular resolution 
of the JVLA is not adequate to detect such small-scale angular motions, such
as jet component rotation around the jet axis or the inferred helical motion.
\textcolor{black}{We note that the helical motion of jet material 
occurred on the VLBI scale appears to be not
the same thing as a precession of the jet axis characterized by the jet curvature
on VLA scales. Slow precession of jet axis can also result in observed VBLI and VLA jet misalignments.}   
\vskip 3pt
\noindent
(3) It is obvious that over life time of the northwestern jet, the PA$_\theta$ changes
about $\sim90${\degree} over the scale of
$\sim$10" (or $\sim$100 kpc).

Unlike the western jet in which the bright emission spots
clearly \textcolor{black}{delineate} the projected jet trajectory,
the eastern jet components are determined from the enhanced
emission peaks along the relatively diffuse, elongated ridge that likely 
delineates the eastern jet path. The uncertainties
in the parameters of the eastern components could be larger than
the errors quoted in Table 2 concerning
the issue in identification of jet components, given
the strong likelihood that the eastern jet is  receding
and the radiation from the jet components is therefore
suppressed due to the relativistic Doppler effect, assuming 
implicitly that the jet is relativistic given that the VLBI
apparent motions are superluminal.

\subsubsection{$\vec{T}$ -- the unit vector of the jet transverse velocity}
The apparent projected direction of motion of the northwestern jet can 
also be described quantitatively.
We can determine the direction of the transverse velocity with a unit 
vector $\vec{T~}$ defined as the direction change of the vector $\vec{J~}$ 
between consecutive jet components: at the $i$-th jet component 
$\vec{T_i} = \displaystyle{ \vec{J}_{i+1}-\vec{J_i}\over|\vec{J}_{i+1}-\vec{J_i}|}$,
where $\vec{J~}$ is the vector pointing from the core to the jet component, 
defined in the previous section. 
The direction of $\vec{T_i}$ 
can be described by PA$_{\rm Jet}$, the position angle of $\vec{T_i}$.
For a curved jet, PA$_{\rm Jet}$ is a function of D$_{\rm Jet}^i$,
the distance measured along the projected jet trajectory 
from the core to the location of the $i$-th component. The quantity
D$_{\rm Jet}^i$ can be computed approximately as 
the sum of the lengths of the connecting lines between consecutive jet 
components from the core to the location of the $i$-th component:
$${\rm D}^i_{\rm Jet} = \sum\limits_{k=1}\limits^{k=i}\left|{\vec{J}}_{k+1}-{\vec{J}}_k\right|,$$
\noindent
where, $i = 1, 2, ..., n$ (see the caption of Fig. 7 for more details). 
Fig. 7 (bottom) plots PA$_{\rm Jet}$ as a function of D$_{\rm Jet}$ with 
a logarithmic scale in the D$_{\rm Jet}$ axis to facilitate visualization
of the PA$_{\rm Jet}$ 
changes in the transition from the northern jet on the VLBI scale
to the western jet on the VLA scale. 
Within a scale of 25 mas, PA$_{\rm Jet}$ varies between $-6^{\rm o}$
and 59$^{\rm o}$  based on the six brighter VLBI components 
d, e, f, g, i and j. The function PA$_{\rm Jet}$(D$_{\rm Jet}$) \textcolor{black}{indicates} 
that the overall VLBI jet \textcolor{black}{displays east-west wiggles as it moves northward.} 
The result is  consistent with VLBI observations \citep{lu2011}. In fact,
based on the VLBA observations of NRAO 530 at twenty-seven epochs between 1994 and 2011, 
the innermost jet PA can be fit to a periodic oscillation with amplitude of 20{\degree} and
a period of 9.4 yr \citep{list13}.

Then, on the scale transiting from VLBI to VLA, the abrupt northwestward bending
of the jet \textcolor{black}{occurs as PA$_{\rm Jet}$ changes $-7${\degree} (j) 
to $-54${\degree} (W2) on angular scales from 25 mas to 700 mas.}  
The corresponding  change rate is 70 deg arcsec$^{-1}$. 
Then, after the following 7" along the jet route, \textcolor{black}{the jet turns westward 
at D$_{\rm Jet}=$7.8" (PA$_{\rm Jet}=-90${\degree} at W8)} with a rate of 5 deg arcsec$^{-1}$. 
Advancing the next 3.1", at W10 before encountering
the hotspot region (W12), the jet turns southwestward at a rate 
of 14 deg arcsec$^{-1}$. Then, the jet appears to terminate 
at W12 followed by a rapid change in PA$_{\rm Jet}$, 30 deg arcsec$^{-1}$.
The rapid change in PA$_{\rm Jet}$ indicates that the jet loses its stiffness
(or collimation for straight line jets),
likely owing to the impact onto the ambient medium at the hotspot. The jet disruption
for FR I source could be also due to an unstable growing  
wave mode, such as a Kelvin-Helmholtz instability \citep[for example]{bla76,fer78, har79,hn88, nh88, zbhn92, zbns92}. 

In short, from the diagram of PA$_{\rm Jet}-$D$_{\rm Jet}$, the change in PA$_{\rm Jet}$ seems
to scale with jet kinematics, {\it i.e.} 
larger change in PA$_{\rm Jet}$ occurs in the inner region
of the core, \textcolor{black}{where the helical motion
of the jet material dominates the  PA$_{\rm Jet}$ change while
the PA$_{\rm Jet}$ change appear to be caused by a slow
precession of jet axis over the VLA scales.} On the VLBI scale, the change rate in PA$_{\rm Jet}$
drops from 83 deg mas$^{-1}$ at D$_{\rm Jet}\sim$1 mas to 2 deg mas$^{-1}$ at 25 mas.
The rapid change in PA$_{\rm Jet}$ on the VLBI mas scale is
consistent with the jet undergoing helical motion \citep{lu2011} or periodic swing \citep{list13}.
However, as compared to the rate on the VLBA scale, the rate on the VLA scale
is about three orders of magnitude smaller. 
In the course of the jet moving from the core to the hotspot W12
over its route length of 10", overall the PA$_{\rm Jet}$ 
changes about 90{\degree}. \textcolor{black}{We made a linear fit to the plotted data and
found that,
    $${\rm PA}_{\rm Jet}=-40.4\degree{\rm Log}_{10}\left({\rm D}_{\rm Jet}\over {\rm arcsec}\right)-52.5\degree, $$
\noindent where the uncertainties of the coefficients in the linear fitting are
equivalent to 1.2{\degree}, the rms of the residuals (see Fig. 7).
The fitted straight line, 
a logarithmic spiral, can be expressed  in polar coordinates if (${\rm PA}_{\rm jet}$, ${\rm D}_{\rm Jet}$) is substituted by ($\theta$, $r$):
$$r = ae^{\theta/b},$$
\noindent where the variables $\theta$ and $r$ are in the units
of degree and arcsec and the coefficients $a=0.0502''$ and $b=-17.5\degree$.} 

\textcolor{black}{
A similar logarithmic spiral function can also be fit to the 
projected trajectory that  
may serve as evidence 
for a slow precessing jet from the core in NRAO 530.}

\begin{figure}[h]
\centering
\vspace{15pt}
\includegraphics[angle=0,width=87.5mm]{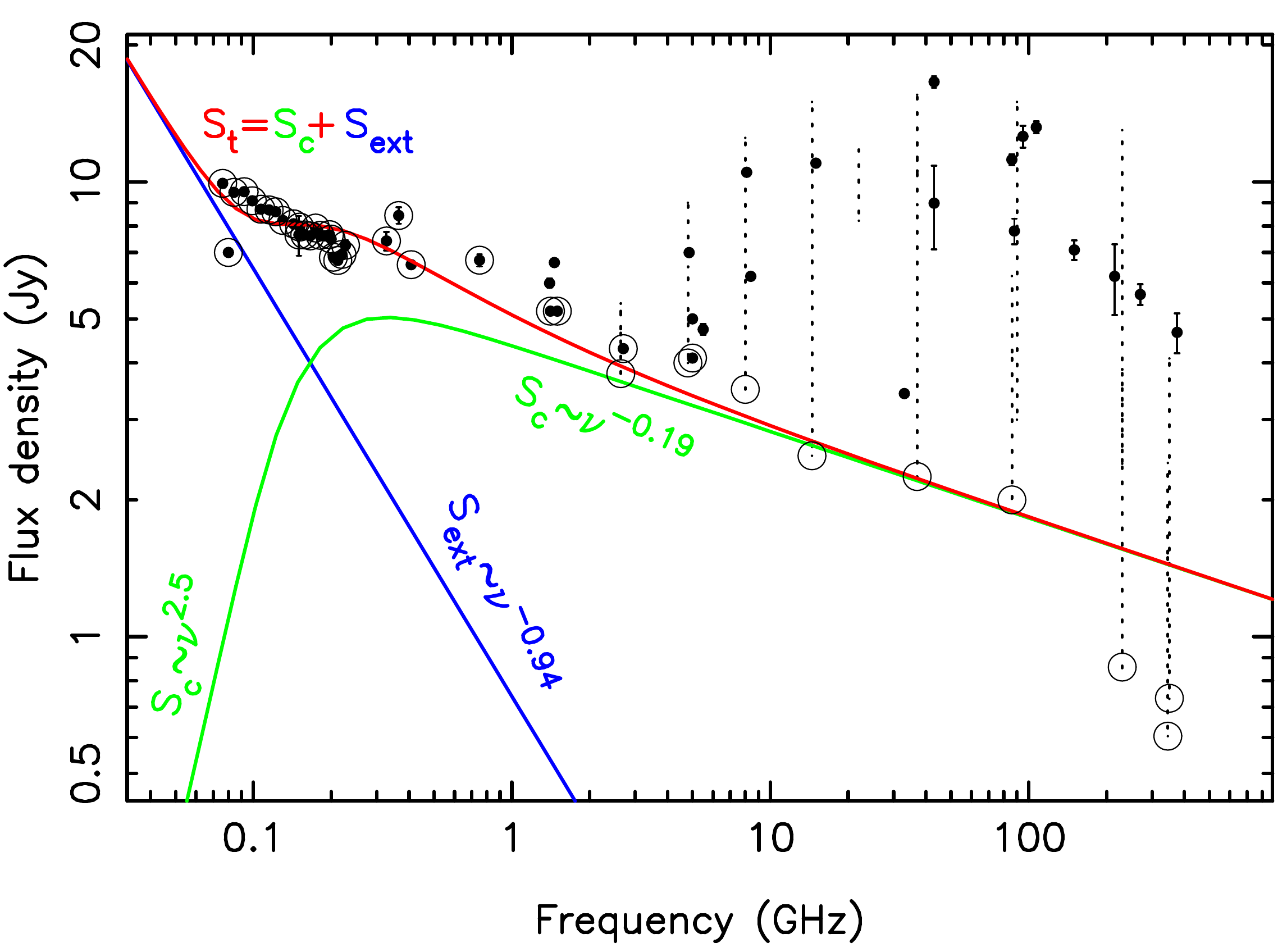}
\caption{The spectrum of NRAO 530 with the data summarized 
in \textcolor{black}{Table 4}. For the data from the monitoring programs 
\citep[\textcolor{black}{and the F-GAMMA monitoring programm$^7$}]{all85,ter92,bow97,rob01,hov08,jen10,all12}, 
the range of flux densities from minimum to maximum at the
observing frequency is scaled with the  vertical dotted line.
For individual 
observations and measurements, the mean values are marked with solid 
dots with the vertical bars for 1 $\sigma$ errors. The green solid
curve is the synchrotron self-absorption model component
\textcolor{black}{($S_{\rm C}$)} and the blue 
line is the optically thin synchrotron emission model component
\textcolor{black}{($S_{\rm ext}$)}. 
The red 
curve is \textcolor{black}{the resulant sum of the two model components 
($S_{\rm C}$+$S_{\rm ext}$), derived from
a best fitting
to the flux-density minimal values} as marked by the open circles. 
}
\label{fig.8}
\end{figure}
\subsection{Spectral fitting to the low flux-density state}
The blazar NRAO 530 exhibits a remarkable degree of flux density
variations over a wide frequency range from 5 GHz to 375 GHz
\citep[\textcolor{black}{and the F-GAMMA monitoring program\footnote{\textcolor{black}{https://www3.mpifr-bonn.mpg.de/div/vlbi/fgamma/fgamma.html}}}]
{all85,ter92,bow97,rob01,feng06,hov08,jen10,all12,an13}.
Superluminal motions of the northern jet components within the VLBI core
have been detected with apparent transverse velocities
ranging from 2.3c to 26.5c \citep{lu2011} which agree with
the more accurate determination of maximum 27.37$\pm$0.97c 
and median 13.74$\pm$0.40c based on the inner six components
by the MOJAVE project\footnote{https://www.physics.purdue.edu/MOJAVE/sourcepages/1730-130.shtml} 
\citep{list13,list16,list18}.
Thus, the core-dominated emission is highly enhanced via the relativistic 
Doppler-boosting effect. Table 4 lists the total flux densities of NRAO 530.
Fig. 8 plots the spectrum of NRAO 530, implying two discernible states:
(1) the high state characterized by highly fluctuating flux densities
at the wavelengths from short centimeter to sub-millimeter;
\textcolor{black}{(2) the low state that appears to be delineated by
the minimal values of the flux densities determined at each of the observing
frequencies $\nu>1$ GHz. The minimal flux density in each of monitoring programs 
appears to depend on the length of monitoring time.} 
A model composed of  a synchrotron self-absorption 
component  and an optically thin component 
is used to fit the low-state spectrum.
Allowing for the emission from the core, the 
synchrotron self-absorption can be described as,
$$ S_{\rm c} = S_0 \left(\nu \over \nu_0\right)^{5/2}\left(1- {\rm exp}
\left[-\left(\nu\over \nu_0\right)^{\alpha_{\rm c}-5/2}\right]\right),$$
\noindent
where $\nu_0$ is the turnover frequency at which the synchrotron
optical depth becomes \textcolor{black}{unity}, $S_0$ is the flux density at $\nu_0$
multiplied by a factor of $\displaystyle {e\over e-1} = 1.58$, and $\alpha_{\rm c}$
is the spectral index when the core emission becomes optically thin. 

The emission from the extended
lobes and jets is assumed to be optically thin synchrotron radiation with
spectral index $\alpha_{\rm e}$ and the flux density of the extended emission
at 1 GHz of $S_{1}$ for $\nu$ in units of GHz,
$$S_{\rm ext}=S_{1}\nu^{\alpha_{\rm e}}.$$
\noindent \textcolor{black}{
There are five parameters ($S_0$, $\nu_0$, $\alpha_{\rm c}$, $S_1$ and $\alpha_{e}$)
to fit. The spectral index  $\alpha_{e}$ can be determined from
the extended emission. With $\alpha_{\rm e}=-0.94$, determined from
the VLA images at 1.46 GHz \citep{kha2010} and 5.5 GHz (this paper, see Table 3),
the number of free parameters is then reduced to four.}
We fit the low-state flux-density data  in Fig. 8 (open
circles) to derive the parameters.  The turnover frequency is \textcolor{black}{$\nu_0=0.16$} GHz, 
\textcolor{black}{$S_0 =6.2$} Jy or \textcolor{black}{$S_{\rm c} = 3.9$} Jy at 
$\nu =\nu_0$, and \textcolor{black}{$\alpha_{\rm c}=-0.19$
which is consistent with the value determined from the core. The extended 
flux density at 1 GHz, $S_{1} =$ 0.739 Jy,  is also consistent with 
the value extrapolated from the VLA measurement for the extended flux density at 1.46 GHz
\citep{kha2010}}. 
We note that the core in this paper refers to the region within 50 mas from the source 
center and the emission from the outside of the region is included in the extended component.
The spectral index of \textcolor{black}{$\alpha_{\rm c}=-0.19$} appears
to be consistent with the result derived from the VLBI core \citep{lu2011}.
A few notes on the spectral fitting follow:
\vskip 3pt
\noindent
(1) at \textcolor{black}{$\nu=142$} MHz, the contributions in flux density 
 from the core and extended emission are comparable. At 
\textcolor{black}{$\nu<$ 142} MHz,
the extended emission becomes prominent, {\it e.g.} at 80 MHz the flux
density from the extended emission is \textcolor{black}{88\%} of the total flux density.
At \textcolor{black}{ $\nu>>$ 142} MHz,
the core becomes dominont, having \textcolor{black}{88\% and 99\%} 
of the total flux densities
at 1.4 and 90 GHz, respectively. The
corresponding  total radio power in the low state is 
\textcolor{black}{$1.0\times10^{28}$ and $4.5\times10^{27}$} 
W Hz$^{-1}$ at 1.4 and 90 GHz, respectively.
\vskip 3pt
\noindent
(2) At most times, the core appears to be in the high flux-density state.
Higher variability in flux density tends to occur towards 
high frequencies. \textcolor{black}{This result is 
consistent with the VLBI determinations of the variation in superluminal 
velocity at different angular scales, suggesting that the high-state flux 
density at  high frequencies is relativistically Doppler boosted
\citep[{\it e.g.}\rm ][]{lu2011}.}

\section{Implication and discussion}
Extragalactic radio sources with jets and lobes can be classified
into two morphological classes, FR I and FR II \citep{FR1974}.
The FR I jets appear to get weak at large distances from the radio core,
{\it e.g.}, 3C 449 \citep{perl79} while the FR II sources, 
{\it e.g.}, 3C 47 \citep{bri94},  show edge-brightened lobes that are 
characterized by a hot spot. These two distinct classes in apparent radio 
\textcolor{black}{morphology} appear to be related to the powers of the 
radio sources and absolute isophotal magnitude \citep{led96}. In the 
logarithmic diagram of the radio power at 1.4 GHz ($P_{\rm 1.4~GHz}$) 
and absolute isophotal magnitude (M), the authors find that a non-zero 
slope line divides FR I from FR II sources, which corresponds to 
$L_{\rm radio}\propto~L_{\rm opt}^{1.8}$. Based on their nearby sample 
($z<0.5$), the FR I/II division appears to occur over a range in 
$P_{\rm 1.4~GHz}\sim10^{24-26.6}$ Watt Hz$^{-1}$.


On the other hand,  studies of the correlation between radio power at 1.4 GHz 
and optical luminosity \citep{led96} lead to the conclusion that all 
radio galaxies live in similar environments in that  optical luminosity 
and other properties of the host galaxy are the most important parameters 
affecting radio source formation and evolution. Relating the absolute optical 
R-band magnitude $M_{\rm R}$ of the host galaxy to the central black hole mass 
$M_{\rm BH}$ \citep{mcl01} and the radio power at 1.4 GHz, $P_{\rm 1.4GHz}$, to 
the kinetic luminosity of the jet,  $L_{\rm jet}$ \citep{will99}, \cite{ghi2001} 
find that the dividing line of the radio power at 1.4 GHz between
FR I and FR II corresponds to a constant ratio, $\displaystyle{ 
L_{\rm jet}/ M_{\rm BH}}$. The statistical analysis of the FR I/II dividing 
line suggests that for radio sources situated on the FR I/II boundary, 
higher radio powers imply a larger mass of black holes harbored in their cores 
\citep[{\it e.g.}\rm][]{ghi2001, will99}.

\subsection{FR II or FR I type source?}
Based on  1.4 GHz VLA observations of the extended radio structures 
associated with the blazars in the complete flux-density-limited 
MOJAVE sample, \cite{kha2010} find that a substantial fraction of 
MOJAVE sources fall into FR I/II category according to their 
luminosities and morphologies. 
Either the total power of \textcolor{black}{$P_{\rm 1.4 GHz}= 1.0 \times10^{28}$ W Hz$^{-1}$}
determined from the spectral fitting to the low flux-density state or
the power from the extended emission in kpc scale 
\textcolor{black}{$P_{\rm ext, 1.4 GHz}= 2.1 \times10^{27}$ W Hz$^{-1}$} suggests
that NRAO 530 has a high optical luminosity if it falls on the FR I/II dividing line 
\citep{owe94,led96,ghi2001}. Therefore, the energetic jets in NRAO 530 might imply 
a high accretion rate \citep{ghi14} onto a very massive black hole 
 \citep{raw91} at the blazar core.

However, the radio power at 1.4 GHz from the core is a factor of 5 greater than that 
of the extended emission even in the low state. The flux density from the core is  
boosted due to the relativistic  Doppler effect as superluminal motions have been 
observed with the VLBA \citep{lu2011,list13,list16}, becoming highly variable.
Taking the median value determined from the inner core components with the VLBA,
the apparent transverse velocity v$_{\rm a}$ in units of $c$ (the speed of light)
is $\beta_{\rm a} =14$, giving a lower limit on the Lorentz factor of 
$\gamma_{\rm min} = \sqrt{1+\beta_{\rm a}^2} \approx 14$  \textcolor{black}{, 
assuming the jet to be viewed at a critical angle $\theta_{\rm c}$. Then,  
the corresponding view angle $\theta_{\rm c}$, the angle between the direction 
of jet velocity and the line of sight to the observer, can be determined
as ${\theta_{\rm c}=\displaystyle{{\rm arcsin}\left(1\over\gamma_{\rm min}\right)}\approx4}${\degree} \citep{lu2011}}. 
\textcolor{black}{Thus, the minimum Doppler factor,
${\displaystyle \delta\equiv{1\over\gamma}{1\over\left(1-\beta{\rm cos}\theta\right)}}$,  
would be $\delta_{\rm min}=\gamma_{\rm min}\approx 14$.} 
\textcolor{black}{ 
The enhanced factor of observed luminosity for a relativisitic jet is: 
$\delta^p$ with $p=3-\alpha_{\rm c}$ for a spherical core and $p=2-\alpha_{\rm Jet}$
for an optically thin jet \citep[{\it e.g.}\rm][]{cawt91,urry1995}.} 
Therefore, the radiation from the core is enhanced due to the Doppler boosting by
an amount of \textcolor{black}{ 
$\gamma^{\alpha_{\rm c}-3}(1-\beta {\rm cos}\theta)^{\alpha_{\rm c}-3}$
\citep{kell88,cawt91,urry1995}}. 
In the case of $\gamma=\gamma_{\rm min}=14$, the flux densities from the inner 
components of the core can be boosted by $\sim\gamma_{\rm min}^3$, about 3 orders 
of magnitude, \textcolor{black}{ assuming $\alpha_{\rm c}=0$}. The Doppler 
boosting effect explains why only the northern jet components (approaching the 
observer) are detected if the emission from the receding southern jet is suppressed 
by $\sim\gamma_{\rm min}^{-3}$ and the bipolar jet is intrinsically symmetric.  

Taking account of the Doppler boosting, we find that the actual intrinsic 
radio power at 1.4 GHz of NRAO 530 is likely on the order of the power from 
the extended emission, or a few times 10$^{27}$ W Hz$^{-1}$. Although the 
edge-brightened eastern lobe with barely visible jet is very consistent with 
the FR II morphology, the relatively bright western jet with a faint emission 
lobe  stretching further out appears to be more like an FR I-type radio source.
Some of the extremely misaligned MOJAVE blazar jets could be  hybrid morphology 
sources, with an FR I jet on one side and an FR II jet on the other \citep{kha2010}. 
Our deep images \textcolor{black}{may} provide further evidence for NRAO 530 
falling into the FR I/II category. \textcolor{black}{However, the classification 
based on edge-brightening appears to be dependent on the viewing direction 
with respect to the orientation of jet-flow. From a certain azimuth and 
elevation angle, the western lobe of NRAO 530 could also be viewd as an edge 
brightened lobe, {\it i.e.}, FR II morphology.}  

\subsection{Mildly relativistic jets on the VLA scale?}
On the VLBI scale, the jet axis lies essentially  north of the core on angular 
scales 1-25 mas. The jet axis of the northern jet components in the inner core  
rapidly swings at a rate of 3.4{\degree} yr$^{-1}$ in change of the position 
angle \citep{lu2011}, that \textcolor{black}{can be modelled} with a periodic 
oscillation of the jet position angle for the innermost component (amplitude 
of 20{\degree} and period of 9.4 yr) \citep{list13}. On VLA scale, over the 
anglar scale 0.1-10 arcsec, the position angle of the northwestern jet drifts 
towards the west. In addition, neglecting the contrast in the brightness of 
the western and eastern jets, the curvature of the eastern jet appears to be 
antisymmetric with respect to the western jet. The twin jet in NRAO 530 can 
be characterized as an S-shaped structure.

The S-shaped morphology differs from the U-shaped narrow-angle-tailed and 
wide-angle-tailed twin-jet sources that are often observed in the environment 
of rich clusters of galaxies  \citep{odea86,odon93}. Such mirror-symmetric 
jet structures are usually associated with FR I radio sources, {\it e.g.}, 
3C 449 \citep{perl79}. The ram-pressure owing to the motion of the jet-parent 
galaxies through the intracluster medium is suggested to be responsible for the 
observed, symmetrically curved radio trails \citep{BRB1979}. Unlike the collinear 
twin jet observed in many FR II sources in which the view angle of the jet axis 
keeps constant, the projected S-shaped jet indicates that the jet axis in 
those cases likely drifts on the sky slowly over the jet lifetime. 
Consequently, the view angle of the jet axis also changes over  time. 
\textcolor{black}{The S-shaped antisymmetric jet is a feature predicted 
by precessing jet models, discussed in detail by \cite{gower82} via their 
extensive numerical simulations and used for the interpretation of the VLBI features 
observed in AGNs, such as 4C+12.50 \citep{list03}.}  

For the case of NRAO 530, if the \textcolor{black}{mean} intensity ratio 
of the western jet hotspot (W12) to the eastern one (E5) \textcolor{black}{at 
5.5 and 9 GHz, $R_{\rm hspt}$= 150} is owing to the relativistic Doppler effect, 
then we can estimate the bulk Lorentz factor $\gamma$ \textcolor{black}
{\citep[\it e.g.\rm][]{laing15}.} The emission from the western jet, plausibly 
moving towards the observer, \textcolor{black}{is enhanced by $\delta^{2.94}$ 
for a coninuous jet with spectral index $\alpha_{\rm Jet}\approx\alpha_{\rm e}=-0.94$. 
If we further assume that the western jet is viewed at a critical angle and 
the eastern jet is in antisymmetric to the western jet, the enhancement factor 
is $\displaystyle{\left[\gamma_{\rm min}(1-\beta_{\rm a}^2)\right]^{-2.94}}$
for the approaching western jet while the receding eastern jet is suppressed by 
a factor of $\displaystyle{\left[\gamma_{\rm min}(1+\beta_{\rm a}^2)\right]^{-2.94}}$.} 
From the intensity ratio \textcolor{black}{$R_{\rm hspt}$ for W12 and E5, the bulk 
Lorentz factor of $\gamma_{\rm min}\sim \sqrt{{R_{\rm hspt}^{1/2.94}+1\over2}} = 1.8$ 
can be inferred.} Then, the view angle of the hotspot W12 can be estimated as \textcolor{black}{
$\theta_{\rm c}={\rm sin}^{-1}\left(\displaystyle{1\over \gamma_{\rm min}}\right)=34${\degree}}. 
Compared to \textcolor{black}{$\theta_{\rm c}=4${\degree}}, determined for the inner jet 
in the VLBI core, the large view angle of the hot spot and the curvature of the western jet 
imply that the jet axis has precessed, presumably in concert with the rotation axis of 
an accretion disk, changing the \textcolor{black}{critical} view angle from 
\textcolor{black}{34{\degree}} at the earlier epoch to 4{\degree} at the recent epoch. 
The slow precession of the jet axis could occur if the jet nozzle has been subject 
to a torque as has been postulated in many theoretical models of accreting black 
holes \citep[{\it e.g.}\rm][]{ghi14}. The simple model adopted in this paper, 
which assumes an intrinsic bi-directional symmetry in a mildly relativistic 
bipolar jet coupled with a slow precession of the jet axis, acoounts for 
the observed antisymmetry of the curved, large-scale twin jet morphology 
and the intensity ratio of the oppositely directed jets. The observed properties of 
the northwestern jet seem to also fit into the theoretical model for 
the hybrid FR I/II sources,  in which the jets become transonic 
in the core and decelerate to mildly relativistic velocities 
from moderately relativistic or ultrarelativistic jets \citep{bic95}.

\section{Conclusion}
With a procedure developed within the CASA software package, we show that the 
time-variable residual delay in the JVLA data can be effectively eliminated 
for deep imaging of faint, extended radio structure associated with dominant 
cores. \textcolor{black}{ Applying the technique that is described in the Appendix,
sensitivities or dynamic ranges of images made with wideband interferometric array
data can be improved by a factor of several tens to over a hundred.} As an 
example for demonstration, the data for NRAO 530, a gain calibrator used in 
the JVLA observations of Sgr A at 5.5, 9 and 33 GHz during the period between 
March 29, 2012 and September 11, 2015, have been processed following the 
procedure of residual-error correction and high dynamic-range imaging discussed 
in this paper. Deep images of NRAO 530 at 5.5, 9, and 33 GHz at angular 
resolutions from 0.1" to 4"  have been achieved with dynamic ranges in the order 
of 1,000,000 : 1. The detailed emission structure revealed on scales of 1-100 kpc
suggests that the edge-brightened eastern lobe/jet structure
appears to fit with the FR II category.  \textcolor{black}{The observed 
morphology of the western jet, with the western extension, 
seems to more likely fall into the FR I category. The classification for the western jet 
perhaps is owing to a specific viewing angle.} 
The apparent antisymmetric, curved 
twin-jet morphology and the observed contrast
of their radiation intensity suggest that the twin jets in NRAO 530 on VLA scales
is mildly relativistic and that the view-angle of the jet axis likely evolves over time.

\acknowledgments
{We are grateful to  Kumar Golap for advice on the use of CASA. We also thank 
Ken I. Kellermann and Matt L. Lister for their valuable comments. 
\textcolor{black}{We want to thank the anonymous referee for providing 
useful input and suggestions, especially for instructive suggestions 
on the low-frequency spectrum fitting  as well as fitting a logarithmic 
spiral function to the northwestern jet data.} The Very Large
Array (VLA) is operated by the National Radio Astronomy Observatory
(NRAO). The NRAO is a facility of the National Science Foundation
operated under cooperative agreement by Associated Universities, Inc.
The research has made use of NASA's Astrophysics Data System.
}
\appendix

\section{A procedure for High dynamic range imaging}

Wideband capability enabled by recent technology advancements has enhanced the power
of interferometer arrays at radio, millimeter and submillimeter wavelengths.
Sensitivities of radio telescopes have been dramatically improved in recent years.
Source structure and variability of core-dominant radio objects such as blazars
and quasars have become noticeable issues in calibration and imaging. Less attention
has been paid to the effects of these radio properties in handling radio data
when the image dynamic range (DR) is limited by the telescope's sensitivity.
For example, the image of NRAO 530 observed with the VLA in 1997 (see Fig. A1)
indicates that the blazar is a point-like source at an angular resolution of $\sim$1
arcsec when the DR is below 10,000:1. In addition, at sub-arcsec resolution,
residual delays ($\delta\tau_{\rm A}$), varying with time, become a noticeable
issue in high-dynamic range imaging with wideband data taken with modern
telescopes. The corresponding phase gradient across a wide frequency band causes
sidelobe smearing of a dirty beam which is difficult to clean with the algorithms
that have been widely used in radio astronomy \citep{hogb74, clar80, steer84} as
well as the newer algorithms developed for multi-frequency-synthesis (MFS) and
multiple-scale (MS) imaging \citep{rau11}. The effect becomes aggravated and
significantly degrades the image quality in the field near a strong compact
source. For example, Fig A2 shows residual bandpass solutions in phase, solved
from JVLA A-array data of NRAO 530 at 9 GHz after the standard delay corrections
derived from 3C 286 are applied. A phase slope of 2{\degree} across a sub-band of
BW 128 MHz is present, suggesting that a typical residual delay of 0.04 nsec
or 1/3 $\lambda$ remains in the data. A time variation of the residual delays is
also evident from the time-dependent bandpass solutions. A larger phase slope of
4{\degree} per 128 MHz is present occasionally on some antennas (see Antenna='ea08'
Time = '11:12:35.1' of Fig. A2, bottom-left). As a consequence, the presence of 
residual delays in deep image
has several implications.

\renewcommand{\thefigure}{A1}
\begin{figure}[ht!]
\centering
\includegraphics[angle=0,width=43mm]{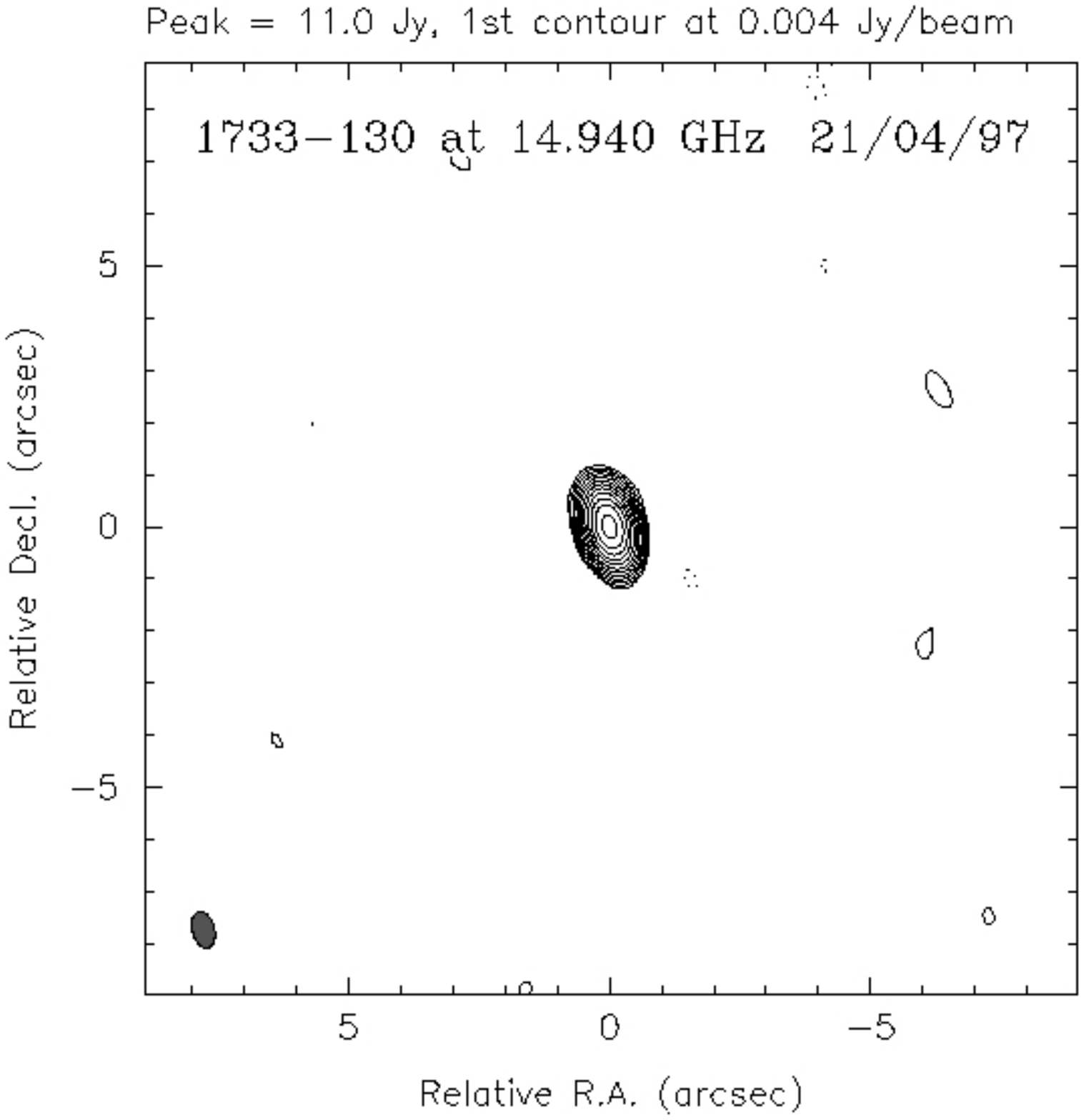}
\includegraphics[angle=0,width=43mm]{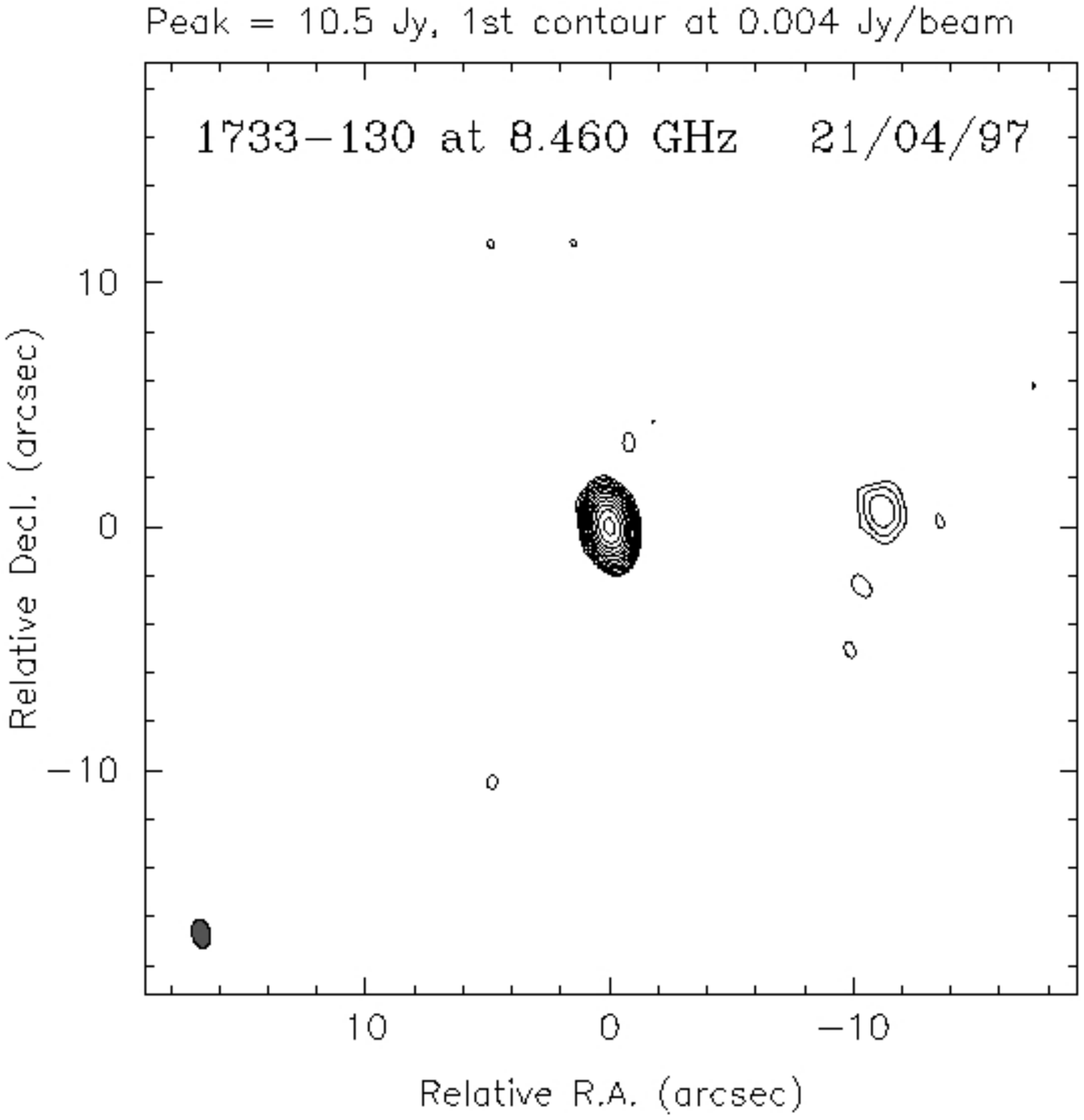}
\caption{The images of NRAO 530 in 1997 from the VLA catalog,
https://science.nrao.edu/facilities/vla/observing/callist,
observed in the B-array at 14.9 GHz (left) and 8.46 GHz (right),
indicating an image DR of $\sim$8,000:1.}
\label{figA1}
\end{figure}

\vskip 10pt
{
\noindent
1. Residual phase delays generate errors in the position of a radio core that
is supposed to be placed at the phase center or the delay center. For example,
phase errors ($\Delta\phi$) caused by residual delays
($\delta\tau_{\rm A}$) across a sub-band (${\rm BW}=\Delta\nu$)
could be due to pointing errors or can propagate to positional uncertainty
of the radio core at the sampling channel frequencies in synthesis imaging
with MFS algorithms. For example,  given fixed baselines,
positional errors ($\Delta\theta$) on the radio core can be related
to the quality of data described by the  residual delays ($\delta\tau_{\rm A}$)
remaining after the calibration process,
$\Delta\theta = {\displaystyle{c\delta\tau_{\rm A}\over B}}$, 
where $c$ is the speed of light
and $B$ is  the separation between two antenna elements, 
or the baseline length.
In the case of JVLA A-array observations at 9 GHz, corresponding to the longest baselines,
the positional errors of the core caused by residual errors are $\sim$0.07\arcsec, about
a quarter of the synthesized beamwidth, $\theta_{\rm syn}$. The main sidelobe pattern in
the dirty image will be smeared owing to the small deviation in phase between the channels,
corresponding to the positional errors caused by residual delays. The consequence of the
residual delays will limit the benefits of deep cleaning in the process of
multiple-frequency synthesis imaging. Given a fixed residual delay error
$\delta\tau_{\rm A}$, the positional error  $\Delta\theta$ is inversely proportional
to the baseline length. The shorter baselines of the small array configurations
produce larger uncertainties in source positions. On the other hand, the sythesized
beam of the array is also inversely proportional to the baseline
length, $\theta_{\rm syn} = {\displaystyle {\lambda \over B}}$, where $\lambda$ is the observing wavelength.
The baseline term is cancelled in the ratio  of the positional
error to the size of the sythesized beam, 
namely,
${\displaystyle {\Delta\theta\over\theta_{\rm syn}}} = {\displaystyle{c\delta\tau_{\rm A} \over \lambda}}$.
Therefore, the issue in deep cleaning of wideband imaging caused by dirty
beam smearing due to the residual delay is not dependent on the size of the
array configuration, given a fixed observing wavelength. However, for the same
array configuration, the issues in deep imaging due to the residual delay become
worse for the data observed at shorter wavelengths due to the difficulty in 
achieving a reliable model from the contaminated visibility data.

\vskip 10pt
\noindent
2. Vector-averaging sub-band data along with self-calibration can eliminate 
the residual delay issue but only at the cost of increasing the noise in 
the data. This approach has been used in handling earlier generation, narrow 
bandwidth VLA continuum data for each of the IFs, or by creating "ch0" data 
to form the continuum data from spectral line observations. The "ch0" is 
simply produced from vector-averaging the channels across a spectral band. 
Indeed, the frequency-dependent phase errors due to $\delta\tau_{\rm A}$ can 
be eliminated completely. The remaining phase errors related to source position 
can be further removed by applying additional self-calibration solutions derived 
from the "ch0" data. However, the trade-off using the vector-averaging technique 
for sub-bands is that the phase variation is effectively turned into signal noise, 
producing a coherence loss. For a visibility function $V(\nu)=A(\nu){\rm exp}[i\phi(\nu)]$, 
the ratio of the vector-averaged amplitude to the actual amplitude drops with the amount of 
phase changes in phase across the sub-band ($\Delta\phi$) due to a residual delay:
${\displaystyle {<A>\over A}=1-{\Delta\phi^2\over24}}$.
This ratio can be used for quantitative assessment of the coherence loss.
The maximum loss in fringe intensity owing to a residual delay of 0.04 nsec is 
$\sim$0.06\% of the core flux-density ($F_{\rm core}$). The uncertainty of the 
vector-averaged amplitude is ${\displaystyle\sigma=A{\Delta\phi\over2\sqrt{6n}}}$,
where $n$ is the number of independent data channels in a given sub-band. Thus, 
the process of vector-averaging produces an uncertainty in the calibration modeling 
and generates additional noise in the data as well.
\vskip 10pt
\noindent
3. The method of vector-averaging sub-band data is subject to bandwidth smearing if 
a residual delay is present. The resultant fringes of vector-averaged sub-band data 
are modulated by a sinc function $\displaystyle \left[ {\rm sin}(\pi\Delta\nu_{\rm BW}
\tau_{rs})\over \pi\Delta\nu_{\rm BW}\tau_{\rm rs}\right]$, the delay beam or the 
bandwidth pattern \citep{thom86}. The residual delay here ($\tau_{\rm rs}$) is considered 
from two aspects: (1) The residual delay $\delta\tau_{A}$; as discussed above, this delay 
term is referred to the time delay from unknown source when the target source is placed 
at the phase center that is presumed to be accurately tracked. (2) The geometrical delay, 
$\delta\tau_{\rm pnt}=\displaystyle{B\over c}\Delta\theta_{\rm pnt}$, for a source offset 
by $\Delta\theta_{\rm pnt}$ from the phase tracking center or the array pointing center. 
Hereafter, we refer to the geometrical delay, $\delta\tau_{\rm pnt}$, as the source delay.
Both delay terms, $\delta\tau_{A}$ and $\delta\tau_{\rm pnt}$, are time variable.
Given the unclear origins, the behavior of $\delta\tau_{A}$ appears unpredictable in time. 
The term $\delta\tau_{\rm pnt}$ is traceable. The phase compensation of the source delay 
at each of the spectral frequencies can be computed with a high precision interferometer 
model if no sub-band averaging is applied. The MFS algorithms have incorporated such 
corrections in the imaging process.

However, the shortcut with sub-band averaging will lead to a loss of valuable 
frequency-dependent information and therefore the image fidelity will be
substantially degraded. For sub-band width of $\Delta\nu_{\rm BW}=128$ MHz and the 
maximum baseline of the JVLA, at angular distance of 5" from the delay center or 
phase center, the fringe amplitude will drop to 75\% of the value in the spectral data
without sub-band averaging. In addition, the structure of the source will be distorted 
due to the bandwidth-smearing effect. At the radial distance of $\Delta\theta_{\rm pnt}$ 
away from the phase center, a point source will be smeared to an angular extent of
$\Delta\theta_{\rm pnt} \displaystyle{\left(\Delta\nu_{\rm BW}\over \nu\right)}$
\citep{thom86}. For example, at  $\Delta\theta_{\rm pnt}=10$" from the phase center,
the angular distortion of a point source can be $\sim$0.14", more than  half of the A-array
synthesized beam, if sub-band averaging is applied to a 128 MHz sub-band data 
at 9 GHz.
\vskip 10pt
\noindent
4. The residual delay, $\delta\tau_{\rm A}$, varies with time in an unpredictable way.
The unclear origin of $\delta\tau_{\rm A} $ causes
difficulties in eliminating such frequency-dependent phase errors from
the data of target sources.
The phase issue caused by $\delta\tau_{\rm A} $ can be corrected partially by applying
the frequency-dependent phase corrections interpolated from the time-dependent bandpass
solutions derived from a reference calibrator close to the target. A more sophisticated
technique using modern computing resources is needed that can extract the requisite information
from the data of the target source itself. Development of new technique for data calibration
and deep imaging with wideband data appears to be imperative.
}
\vskip 15pt
In addition to the phase errors caused by residual delays,
the variation in visibility amplitude as a function of both time
and frequency leads to detrimental effects in deep imaging with
wideband data if this variability is not accounted for. The source
activity as function of time and the distribution of the radiative
energy as a function of frequency or source spectrum are not unusual
among astronomical sources. The issues related to the amplitude
variation in DR imaging with wideband data include the following:
\vskip 10pt
{
\noindent
1. Source spectrum. For a given spectral energy distribution, the visibility 
amplitude varies as a function of frequency across a broad frequency band.
For a source having a steep power-law spectrum ($\sim\nu^{\alpha}$) with $\alpha=-1$,
the fractional amplitude changes by $\Delta S/S=\Delta\nu/\nu \sim36\%$ across
the 2-GHz JVLA wideband  at 5.5 GHz. Unlike the former VLA continuum data
with narrow IF bandwidth, the amplitude variation across such a wide band challenges
traditional imaging techniques. In order to cope with the spectral issues, a new
algorithm for imaging the first two terms in the Taylor expansion of the visibility
spectra has been developed and has been used in the \textbf{\textit {clean}} program within
CASA \citep{rau11}. This technique has been successfully employed for
deep imaging of the first term, the intensity distribution.
\vskip 10pt
\noindent
2. Time variation. Fig. A3 shows the issue in deep imaging owing to the time-variation 
of core flux-density. A variation of 10\% in flux density of a 1-Jy core can reduce the 
dynamic range by a factor of 6-7, limiting deep imaging. In addition, artifacts around 
the core can be generated in the "cleaned" image if the variability during the observation
is not removed from the data \citep{zhao91}. A procedure for removal of the time
variation of a compact radio core has been described using the software tools
in CASA \citep{zmg2016}.
 }

\vskip 15pt
The rest of the content in this appendix concerning the method of RE-correcting 
and DR-imaging is organized as follow. Section A.1 formulates a general model of 
the radiation structure and the variable components for given a source. Section 
A.2 describes a set of fomulas for transforming extended source structure into 
a point-source model. Section A.3. discusses the JVLA standard data reduction procedure. 
Based on the modeling discussed in these three sections, we develop a procedure 
for residual-error (RE) correction and dynamic-range (DR) imaging utilizing CASA 
software. Section A.4 outlines five major steps involved in the RE-correcting and 
DR-imaging process, with a demonstration using JVLA Ka-band data from the A-array
observation of NRAO 530 at 33 GHz. Section A.5 introduces a case involving a more  
sophisticated procedure for handling the JVLA C-array data of NRAO 530 observed 
at 5.5 GHz. Section A.6 discusses the application of residual-delay corrections
derived from gain calibrators to target sources. A summary is given in Section A.7. 
The techniques discussed here can be applied to general interferometer array data,
although the suggested procedures are based on JVLA data and CASA algorithms.


\renewcommand{\thefigure}{A2}
\begin{figure*}[t]
\centering
\includegraphics[angle=0,width=169mm]{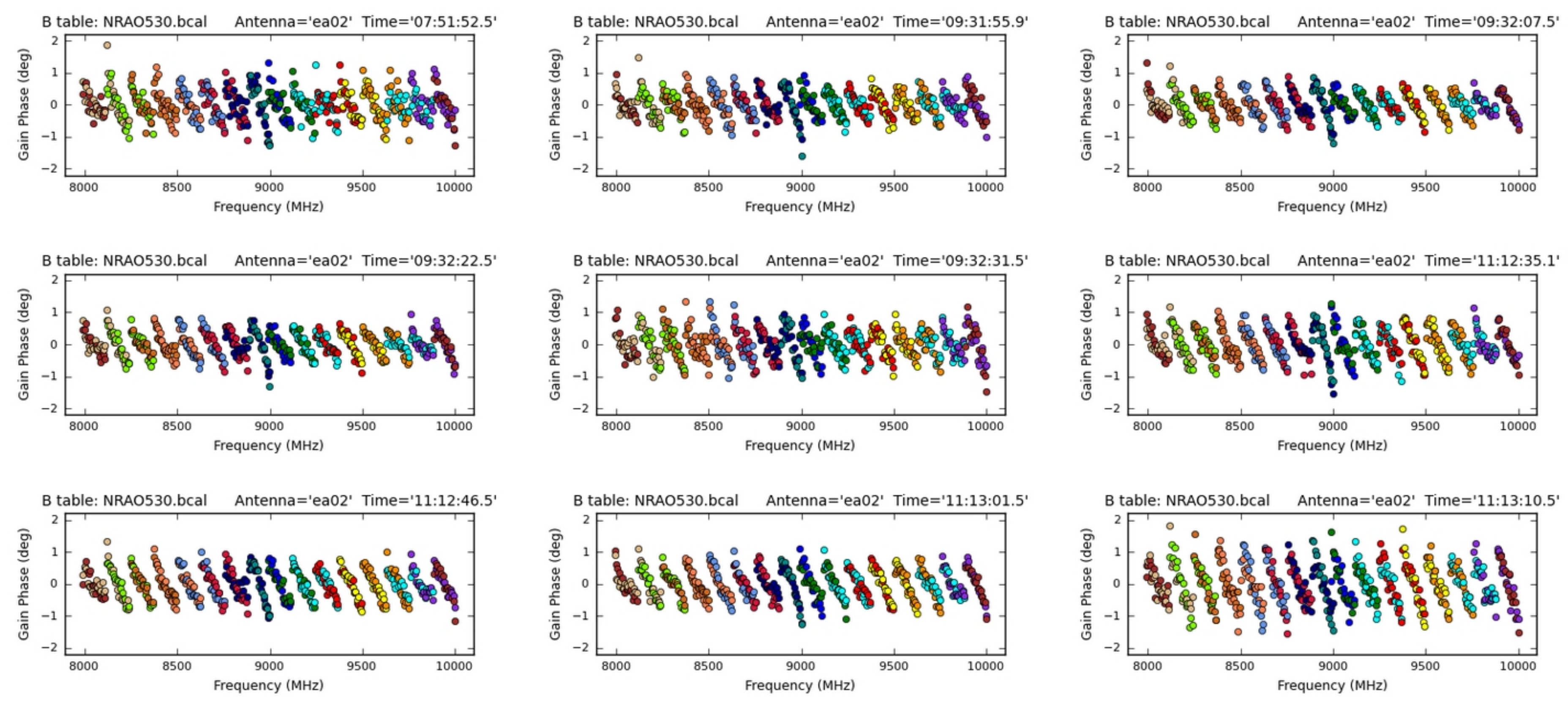}
\includegraphics[angle=0,width=169mm]{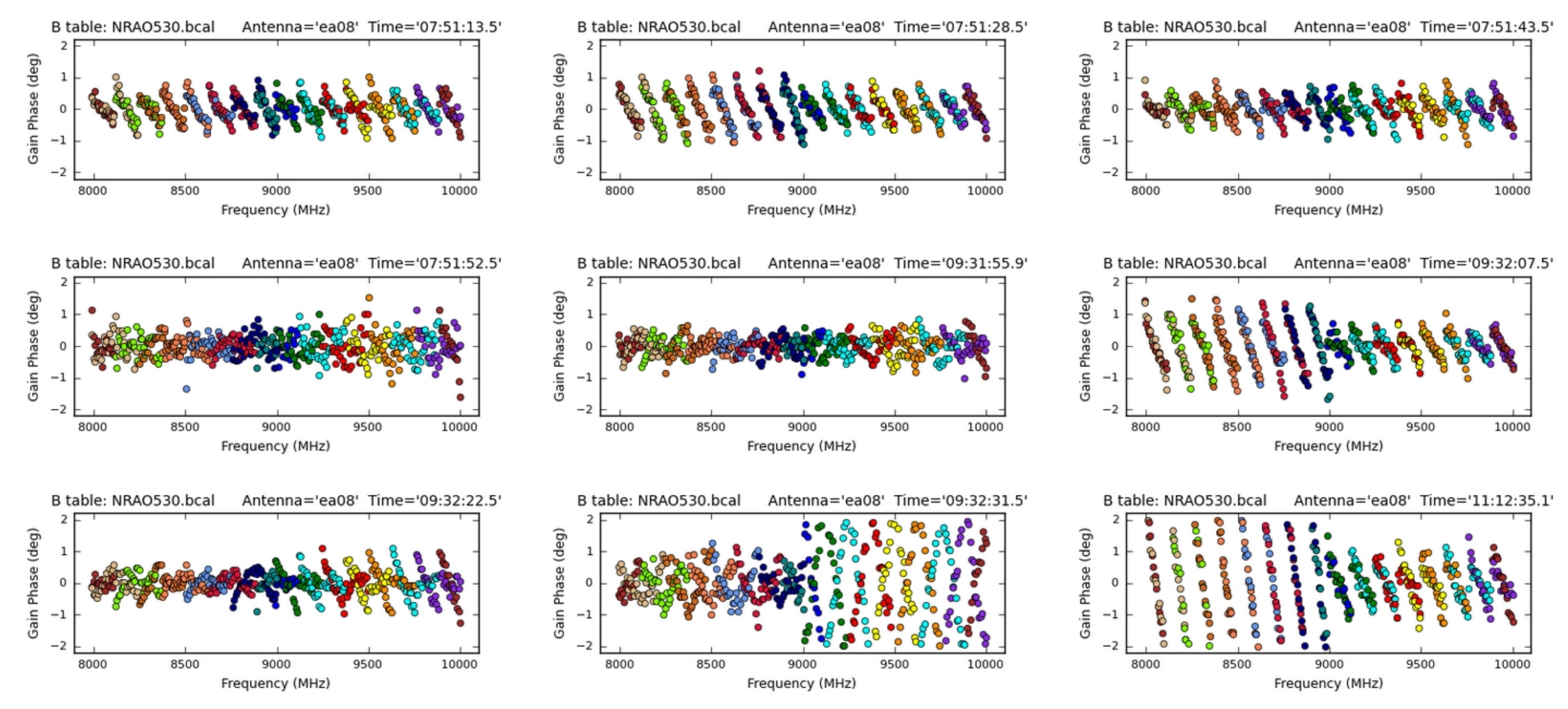}
\caption{The residual bandpass solutions solved from NRAO 530 data after
standard JVLA calibration, showing slopes of the
phase as function of frequency. The typical residual phase slope of
$\sim2${\degree} across a sub-band
of 128 MHz, which corresponds to a residual delay of 0.04 nsec or 1/3 $\lambda$,
remains in the data. The JVLA data were observed
on April 17, 2014 in the A-array configuration at 9 GHz.
The top group of nine sub-plot panels shows
a part of the phase solutions related to antenna "ea02" solved
with the task \textbf{\textit{bandpass}} in CASA using an interval
of 15 sec while the bottom group is for the solutions of antenna "ea08".
The antenna ID and the specific time of the solutions are indicated on top-left of each subplot.}
\label{figA2}
\end{figure*}

\renewcommand{\thefigure}{A3}
\begin{figure*}[h]
\centering
\includegraphics[angle=0,width=60mm]{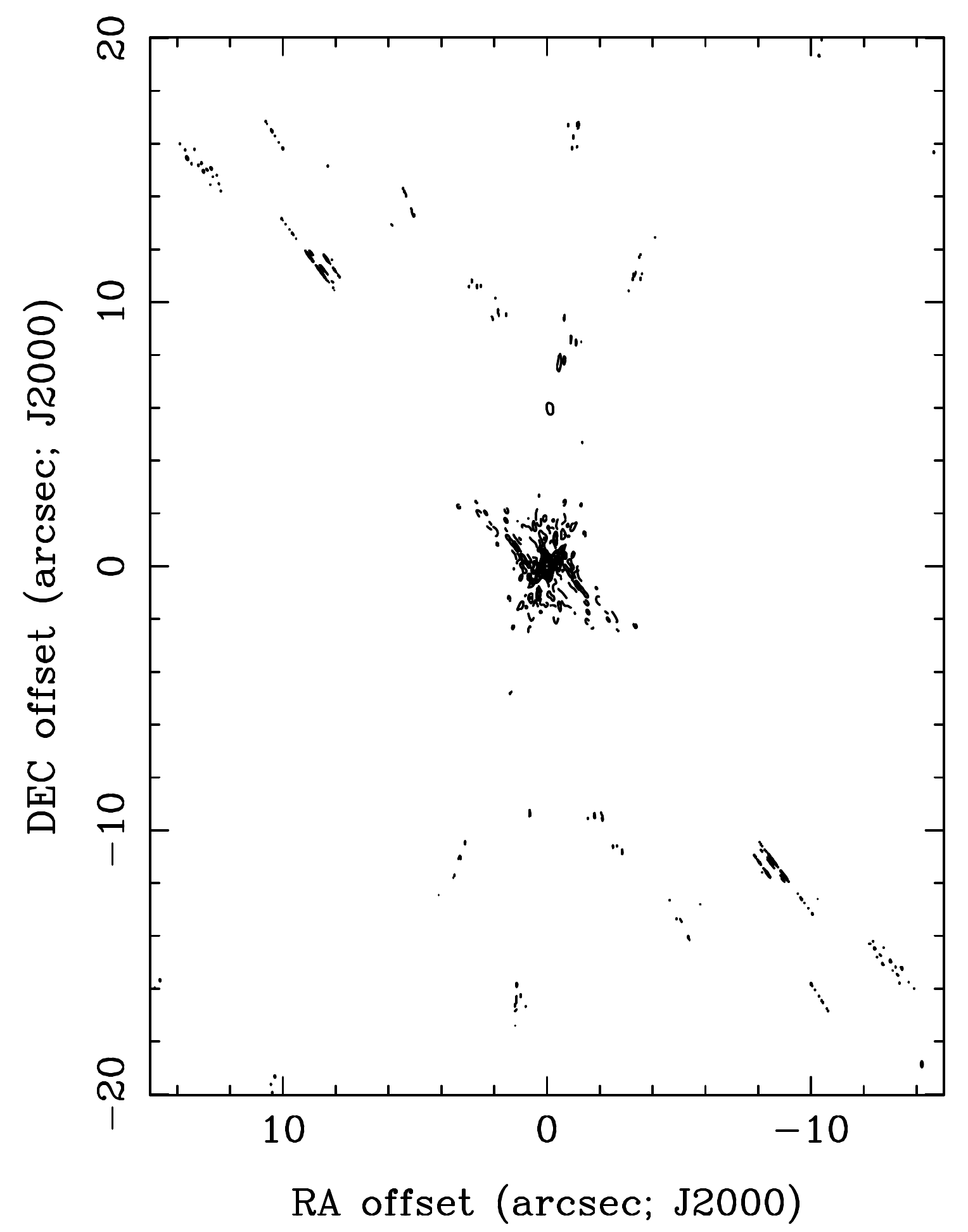}
\includegraphics[angle=0,width=60mm]{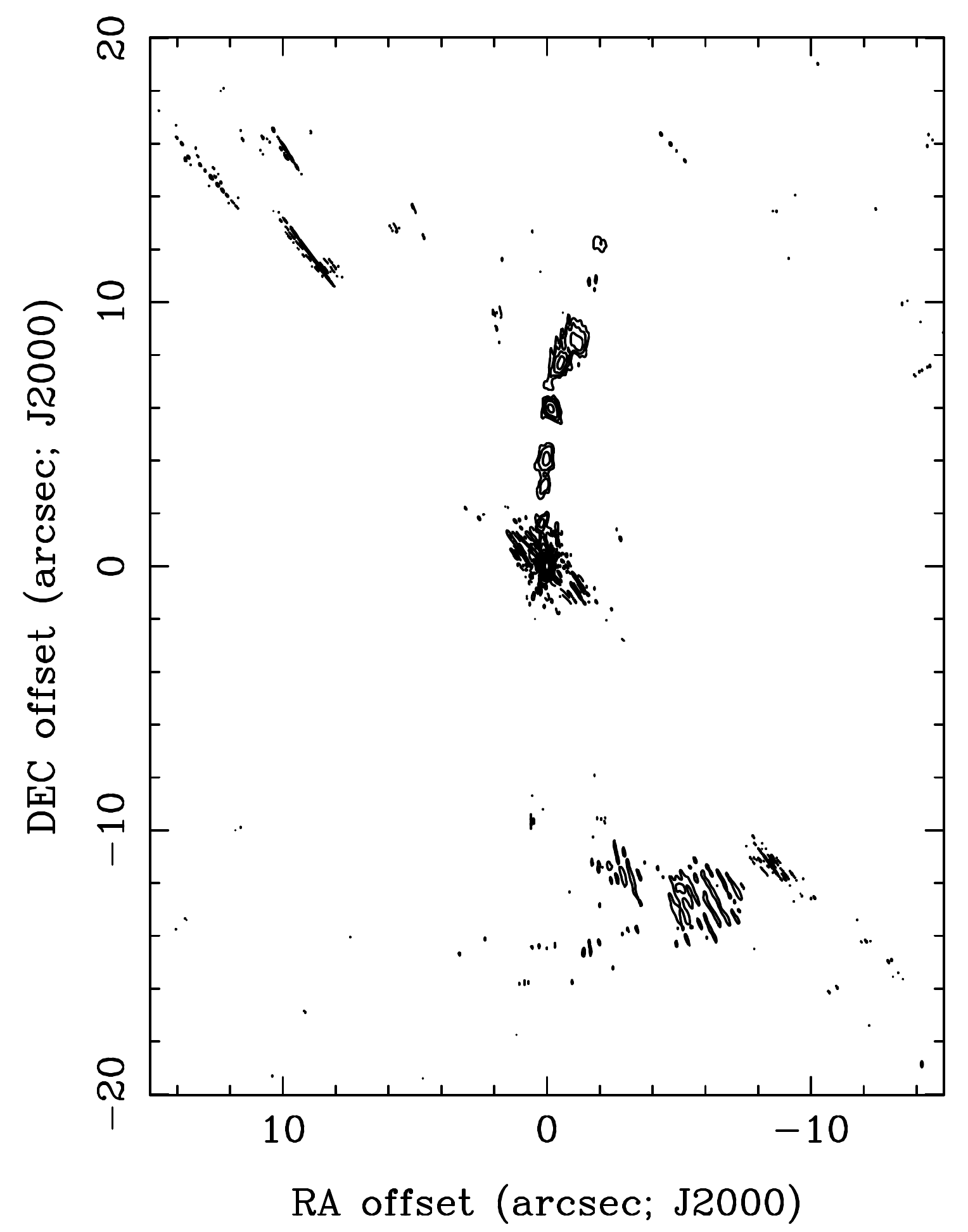}
\includegraphics[angle=0,width=60mm]{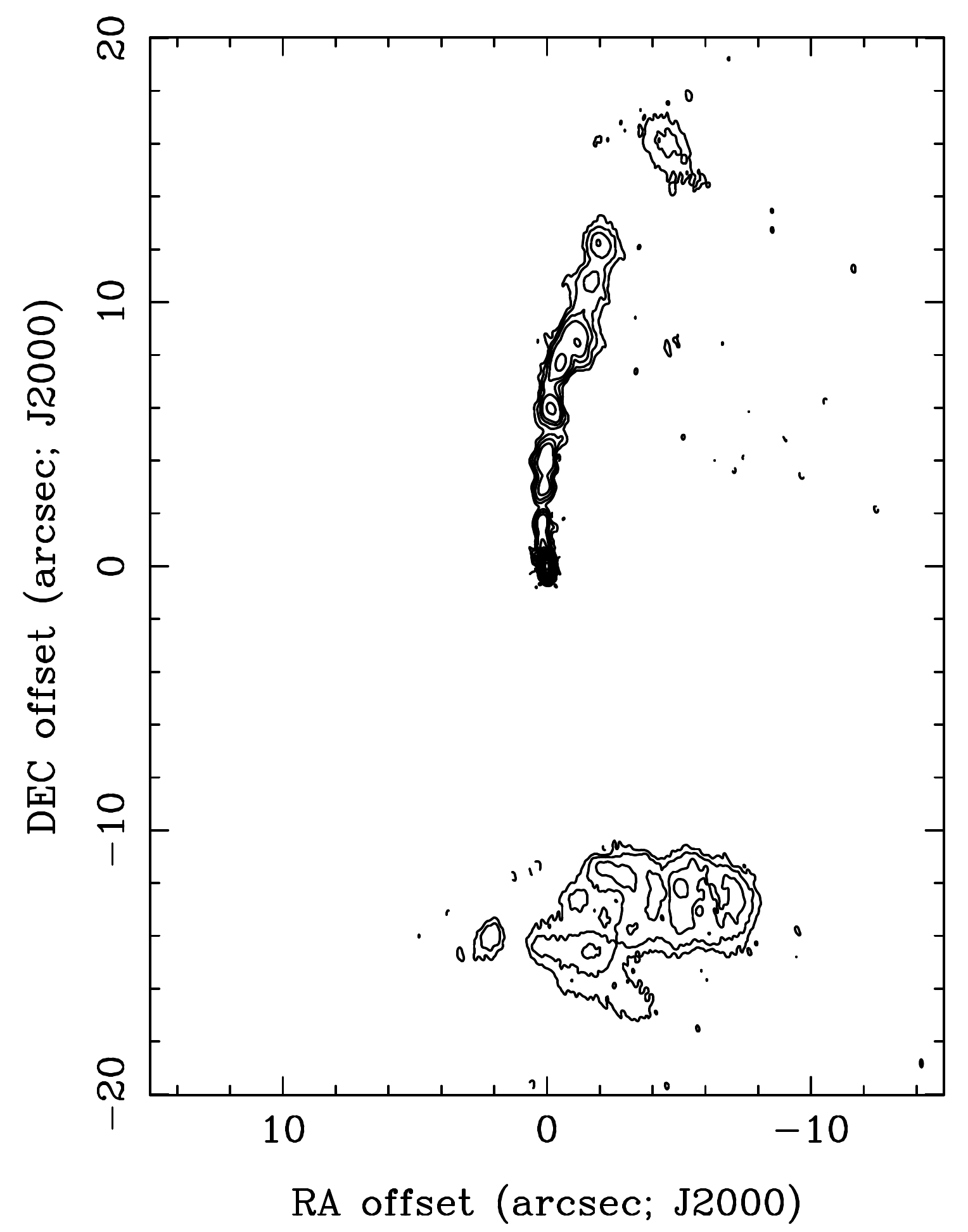}
\caption{Images of J1744-3116 constructed with a combination of two sets of JVLA data 
observed on 2014-3-1 (Day 1)  and 2014-4-17 (Day 2) at 9.0 GHz, in the A array. Left: 
Image made from a combination of the two data sets which have been self-calibrated with 
independently determined models, including corrections for residual delay and intra-day 
variability of the core, but with no correction for the variation from 0.71 Jy on Day 1 
to 0.77 Jy on Day 2. A high noise level is present owing to the 10\% variation in flux 
density of the radio core. The rms noise level in the region far from the core is 
$\sigma_{\rm far}\sim$45 $\mu$Jy beam$^{-1}$. Middle: The image made from the same 
uncorrected data but adding a value of 0.06 Jy at the core to the data of Day 1,
establishing a constant flux density of the core in both data sets. The rms noise of 
$\sigma_{\rm far}$ is improved to the level of $\sigma_{\rm far}\sim$6 $\mu$Jy beam$^{-1}$.
Right: The image made with both data sets with further corrections for the residual errors
using the improved model. The rms noise $\sigma_{\rm far}$ is $\sim$2 $\mu$Jy beam$^{-1}$.
The contours are 2$\sigma_{\rm far}\times$($-2,2^{\rm n}$) with n = 1, 2, 3, until 
reaching the peak. The synthesized beam is 0.20"x0.39" (-31{\degree}).}
\label{figA3}
\end{figure*}
\subsection{Modeling structure and variability of a radio source}
The visibility function on the $\left[u, v\right]$-plane of a radio
source sampled with a wide-band interferometer
at observing time $t$ and frequency $\nu$ can be written as:

\begin{equation} 
{V({\rm u}, {\rm v}, \nu, t) = V^{\rm S}({\rm u}, {\rm v}, \nu, t)  
e^{2\pi i {\rm \nu\tau_A(t)}}} G_A(\nu, t),
\end{equation}
\noindent where
$\tau_A(t)$ and $G_A(\nu, t)$ correspond to time delays and time-dependent complex gains,
respectively. Both terms are antenna-based, containing instrumental errors
and atmospheric effects. The source function, $V^{\rm S}(u, v, \nu, t)$,
composed of numerous emission components, can be described as:
\begin{equation}
V^{\rm S}({\rm u}, {\rm v}, \nu, t) = \sum_{i=0}^{m-1}V_i^{\rm Ext}({\rm u}, {\rm v}, \nu) + 
\sum_{i=0}^{n-1}V_i^{\rm C}({\rm u}, {\rm v}, \nu, t) 
\end{equation}
\noindent where
$V_i^{\rm Ext}({\rm u}, {\rm v}, \nu)$ is the source visibility function of the $i$-th extended
component, and $V_i^{\rm C}({\rm u}, {\rm v}, \nu, t)$ is the visibility function of the $i$-th
unresolved compact or point source. The integers {\it m} and {\it n} are assumed the
total numbers of extended and unresolved components in the field of view,
respectively. The extended components are presumably stable during the observing 
period while the flux-densities of the compact components can vary with time. In
practice, for a QSO or a blazar in which a dominant radio core is present, the source
function can be simplified as a strong point-source function
$V^{\rm C}({\rm u}, {\rm v}, \nu, t)$ and an arbitrary extended emission 
$V^{\rm Ext}({\rm u}, {\rm v}, \nu)$, so that,
\begin{equation}
V^{\rm S}({\rm u}, {\rm v}, \nu, t) = V^{\rm C}({\rm u}, {\rm v}, \nu, t) + 
V^{\rm Ext}({\rm u}, {\rm v}, \nu).
\end{equation}
\noindent
For the MOJAVE sources at 1.4 GHz, the ratio of the integrated flux density from 
the extended components to that of the core is distributed across a very wide range 
between a few thousands to less than a thousandth \citep{kha2010}. For the two gain 
calibrators NRAO 530 and J1744-3116 at 5.5 and 9 GHz, the integrated flux density 
of the extended portion is only a few percent that of the core. In core-dominant 
sources such as most blazars and QSOs at high frequencies, 
$V^{\rm S}({\rm u}, {\rm v}, \nu, t) \sim V^{\rm C}({\rm u}, {\rm v}, \nu, t)$ 
can be used in the initial approach. 
\vskip 15pt
As interferometer arrays improve in both sensitivity and angular resolution,
the previous assumption of a calibrator (a QSO or a small planet) as a point source
is no longer suitable. The extended structure and variability of a calibrator appear
to generate significant effects, limiting  both the calibration precision
and the image quality as well as fidelity. Given a celestial object, a time-invariant
structure of the source emission can be constructed, as was shown in the analysis
of the radio observations and wideband imaging of the complex region Sgr A after 
fixing the variable component of Sgr A* to a constant value \citep{zmg2016}. 
Then, a model for the structure of stable emission from a calibrator or a radio 
source can be expressed as:
\begin{equation} 
V^{\rm S}({\rm u}, {\rm v}, \nu) = V^{\rm C}({\rm u}, {\rm v}, \nu) 
+V^{\rm Ext}({\rm u}, {\rm v}, \nu)
\end{equation}

\subsection{Building a point source model}

\noindent  Dividing the observed complex visibility $V({\rm u}, {\rm v}, \nu, t)$
of Eq(A1) by the source model $V^{\rm S}({\rm u}, {\rm v}, \nu)$ of Eq(A4), 
one can approach to point-like visibility function:
\begin{equation}
V^{\rm P}({\rm u}, {\rm v}, \nu, t) = 
V({\rm u}, {\rm v}, \nu, t) / V^{\rm S}({\rm u}, {\rm v}, \nu),
\end{equation}
\noindent
where the normalized source function $V^{\rm P}({\rm u}, {\rm v}, \nu, t)$  
is an {\it initial calibration model} in which the deviations from a point source 
are due to the atmospheric effects, such as amplitude attenuation and phase 
distortion, and the antenna-based instrumental errors in both amplitude and 
phase, as well as error due to imperfections of the initial source model.
The process of calibrations for various effects will eventually converge 
the normalized source function to a point source function. For example, 
the linear changes in phase as a function of frequency across the sampling 
band can be fit by time delays and then be removed from the data. In addition, 
for QSOs or blazars, time variations of the flux densities from their 
radio cores often become significant. Special attention is needed in 
the calibration and imaging processes \citep{zhao91,zmg2016}. Thus, when 
a variable core is present, the source function can be separated into two 
components, a component, $V_0^C({\rm u}, {\rm v}, \nu) = 
V^{\rm C}({\rm u}, {\rm v}, \nu, t_0)$,  the flux-density  at 
the beginning  of the observation, $t_0$, and a time-variable term giving
the deviations from that initial value,
$\Delta V^C({\rm u}, {\rm v}, \nu, t)$:
\begin{equation}
V^{\rm C}({\rm u}, {\rm v}, \nu, t) = V_0^C({\rm u}, {\rm v}, \nu) + \Delta V^C({\rm u}, {\rm v}, \nu, t).
\end{equation}
\noindent For calibration concerns, the radio-dominant core is placed at the phase 
tracking center of the interferometer. Given a variable intensity $I_0 +\Delta I$ of 
the radiation from its core, the complex finction (A6) of the compact source can be 
simplified as a real function,
 $V^{\rm C}({\rm u}, {\rm v}, \nu, t) = I_0 +\Delta I(t)$.
If the extended component $V^{\rm Ext}({\rm u}, {\rm v}, \nu)$ is not time variable,
then the normalized source function of Eq(A5) can be expressed as,
\begin{equation}
V^{\rm P}({\rm u}, {\rm v}, \nu, t)= \left[1 + {\Delta I(t) \over I_0 + 
V^{\rm Ext}({\rm u}, {\rm v}, \nu) }\right] e^{2\pi i {\rm \nu\tau_A}} G_A(\nu, t).
\end{equation}
\noindent
\noindent
where the spectrum of the core is further assumed to be flat for simplicity.
Removing the time-variable term is necessary prior to further calibration 
of complex gains \citep{zmg2016}. For a calibrator with no time-variations 
and the source structure is simple enough, the normalized source function
$V^{\rm P}({\rm u}, {\rm v}, \nu, t)$ 
is essentially a complex function tracing the antenna gains multiplied by the 
antenna-based delay term, {\it i.e.}, 
\begin{equation}
V^{\rm P}({\rm u}, {\rm v}, \nu, t)= e^{2\pi i {\rm \nu\tau_{\rm A}(t)}} G_{\rm A}(\nu, t).
\end{equation}
\textcolor{black}{
the right of Eq(A8) can be defined as the overall antenna-based complex gain 
$
\mathscr G_{\rm A}({\rm u}, {\rm v}, \nu, t) \equiv e^{2\pi i {\rm \nu\tau_{\rm A}(t)}} G_{\rm A}(\nu, t).
$
}
Furthermore, the antenna-based delay can be further separated into two terms,
a constant delay $\tau_0$ and a time-dependent residual delay:
\begin{equation}
    \tau_{\rm A}(t) = \tau_0 + \delta\tau_{\rm A}(t)
\end{equation}
For data from the JVLA, $\tau_0$ can be removed by the standard delay
correction in the beginning of data calibration or in a pipeline process, and
$\delta\tau_{\rm A }(t)$ is the residual delay that appears to be dependent 
on time. Furthermore, the frequency-dependent and time-dependent functions 
in the complex gain can be decoupled if the instrumental bandpass is stable 
enough in time, \textcolor{black}{ {\it i.e.}, $G_{\rm A}(\nu, t) = G_{\rm A}(t)B_{\rm A}(\nu)$,
 the overall antenna-based gains becomes}
\begin{equation}
\textcolor{black}{
\mathscr G_{\rm A}(\nu, t) = G_{\rm A}(t)B_{\rm A}(\nu)e^{2\pi i {\rm \nu\delta\tau_{\rm A}}(t)},
}
\end{equation}
\noindent 
where $G_A(t)$ is the antenna-based complex gain, a function of time;
with a scaling factor of the amplitude $\mathscr{F}$,  the antenna gain can be expressed
as,
\begin{equation}
G_{\rm A}(t)  = \mathscr{F} g_{\rm A} (t) e^{i\phi_{g_{\rm A}}(t)},
\end{equation}
\noindent and
$B_{\rm A}(\nu)$ is the antenna-based, instrumental complex bandpass, a
function of frequency,
\begin{equation}
B_{\rm A}(\nu)  = b_{\rm A} (\nu) e^{i\phi_{b_{\rm A}(\nu)}}.
\end{equation}

\subsection{The standard data reduction procedure for JVLA data}
In principle, the calibrations of delay, bandpass, complex gain and
flux scale with the JVLA standard data reduction procedure or in VLA
pipeline\footnote{see https://science.nrao.edu/facilities/vla/data-processing/}
are essentially to correct for the parameter terms
$\tau_0$, $b_{\rm A}(\nu)$, $\phi_{b_{\rm A}}(\nu)$, $g_{\rm A}(t)$, 
$\phi_{g_{\rm A}}(t)$ and $\mathscr{F}$,
respectively (Fig. A4, top box). After going through 
the standard data reduction procedure,  various errors discussed
above are corrected, neglecting the extended emission structure.  
\textcolor{black}{
Then, residual errors remained in $G_{\rm A}(t)$ and $B_{\rm A}(\nu)$, or Eq(A11) and Eq(A12), 
can be assessed by the normalized gains,  
$g_{\rm A} (t) = 1+O[g_{\rm A}(t)]$ and $ b_{\rm A} (\nu) = 1+O[b_{\rm A}(\nu)]$ for amplitude, 
as well as
$e^{i\epsilon_{\phi_{g_{\rm A}}}(t)}$ and $e^{i\epsilon_{\phi_{b_{\rm A}}}(\nu)}$ for phase.
The residual errors are small, {\it i.e.}, the residual amplitude errors 
$O[g_{\rm A}(t)]<<1$ and $O[b_{\rm A}(\nu)]<<1$ and the residual phase error $\epsilon_{\phi}\equiv\epsilon_{\phi_{g_{\rm A}}}(t)+ \epsilon_{\phi_{b_{\rm A}}}(\nu)$ fluctuates 
around zero as a function of time and frequency.
}  The point source model
as described in  Eq(A8) is reduced to be a function of residual errors:
\begin{equation}
V^{\rm P}({\rm u}, {\rm v}, \nu, t)\approx 
e^{2\pi i {\rm \nu\delta\tau_{\rm A}(t)} +i\epsilon_{\phi}} \left(1+ 
O\left[g_{\rm A}(t)\right] + O\left[b_{\rm A}(\nu)\right]\right),
\end{equation}
\noindent \textcolor{black}{ neglecting the terms with high order of residual errors in amplitude.} 
If the residual
errors $O\left[b_{\rm A}(\nu)\right]$ in frequency-dependent gains are much
smaller than the residual errors $O\left[g_{\rm A}(t)\right]$ in the
time-dependent gains and the  phase term owing to residual delay  dominates frequency-dependent 
residual errors in phase, ignoring $O\left[b_{\rm A}(\nu)\right]$ and $\epsilon_{\phi}$ in Eq(A13),
the first order of the residual errors is reduced to:
\begin{equation}
V^{\rm P}(u, v, \nu, t)\approx e^{2\pi i {\rm \nu\delta\tau_{\rm A}(t)}} \left(1+ 
O\left[g_{\rm A}(t)\right]\right).
\end{equation}
\noindent 
The remaining residual errors can be separated, namely the residual delay 
$\delta\tau_{\rm A}(t)$ and the residual gain $O\left[g_{\rm A}(t)\right]$. 
Both are time dependent. The residual gains can be corrected with self-calibration, 
the well-known technique  \citep{schw80} that has been successfully applied 
in continuum imaging for radio interferometer arrays. The residual delay 
$\delta\tau_{\rm A}(t)$ can be fit to time-dependent antenna-based phase solutions
for the bandpass. Then, the residual delay can be essentially removed in the process of
constructing a high-precision point-source model.
    
\renewcommand{\thefigure}{A4}
\begin{figure}[t]
\centering
\includegraphics[angle=0,width=91mm]{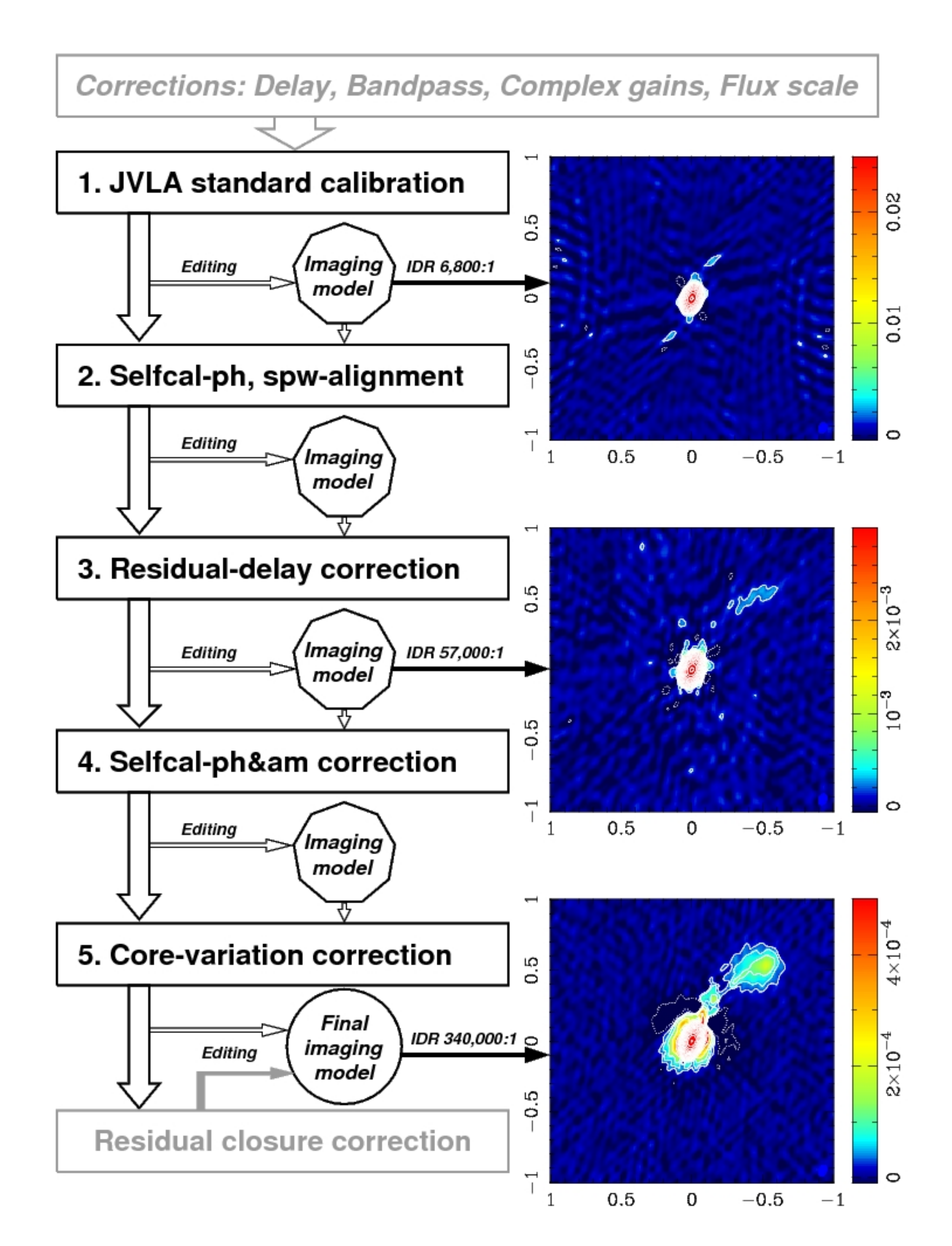}
\caption{A flow chart illustrating the procedure of corrections and
calibrations for the instrumental and atmospheric issues in the process
of HDR imaging. The imaging models (on the right column) were produced from
a 10-mininute test data of NRAO530 observed on 2015-9-11
using JVLA in the A-array at 33 GHz in the Ka-band with a total bandwidth
of 8 GHz. \textcolor{black}{The images are scaled in units of arcsec.}
{\bf Step 1} is the JVLA standard calibrations, including the corrections
for instrumental delay $\tau_0$ and bandpass $G_{\rm A}(\nu)$,
calibrations for complex gains $G_{\rm A}(t)$ of interferometer data,
and boostrap flux density scale $\mathscr{F}$ from a primary calibrator.
Top-right image shows the model image produced from this step with a 
DR of 6,800:1. {\bf Step 2} is further corrections for phase errors 
and alignment of the sub-band offsets in both phase and amplitude, 
utilizing the self-calibration technique and multiple-spw bandpass 
solutions, improving the imaging model. {\bf Step 3} is to further 
remove the residual delays in each of sub-bands; an improved model 
image (Middle-right) with an improved image DR of 57,000:1 is produced 
from this step.
{\bf Step 4} is to correct for the residual gain errors in both phase 
and amplitude with solution intervals from self-calibration as short 
as possible.
{\bf Step5} is to remove the time variation in flux density from 
the compact core, reaching the final imaging model if no significant 
residual closures present. Otherwise, corrections for closure errors 
are needed. The bottom-right image shows the final model image with
image DR 340,000:1 limited by sensitivity.}
\label{figA4}
\end{figure}
\subsection{Corrections for residual delays}
In processing broad bandwidth interferometer data with high sensitivity, 
both the residual delay and extended emission struction prevent from  
building  a high precision point-source model. In this section, we discuss 
the procedure of RE-correcting and DR-imaging based on the JVLA data. 
Iterations of deliberately designed steps in processing wideband data 
appears to be a practical way on correction  for the residual errors. 

For the JVLA data at 9 GHz, the typical residual delays $\delta\tau_{\rm A}(t)$ 
produces a phase change at a level of 2{\degree} across a sub-band of BW 128 
MHz. The origin of the residual errors is not ascertained, but they likely 
arise from a combination of the instrumental issues related to the changes 
in the telescope backend electronics and the telescope frontend. The later 
includes pointing drifting owing to weather conditions such as a strong wind 
as well as the gravitational deformation of telescope beams as function of 
sky position as the telescope tracking a celestial source. The changes are 
dependent on the conditions of the individual telescope elements across the 
array. Based on the analysis discussed in A.1, A.2 and A.3, a procedure for 
RE-correcting and DR-imaging can be developed with the tasks and modules in 
the CASA software. Fig. A4 outlines the basic steps in corrections for the 
residual delays, residual gains and time-variable issues. The method for 
removing the time-variable portion of the core flux-density has been discussed 
\citep{zmg2016}. The critical issue of the procedure is to provide a way of 
approaching a high-precision source model with the steps of corrections for 
residual errors. We use the JVLA Ka-band data of NRAO 530 as an example to 
illustrate the process. A description of the steps is outlined in the caption 
of Fig. A4.

The image created after applying JVLA standard calibration ({\bf step 1})
is a point-like source with DR of 6,800:1 (top image in Fig. A4). With
this model, the sub-band or spectral window (spw) based residual errors
of antenna gains in phase (time-dependent) are corrected with self-calibration.
Then, alignment of the time-dependent offsets in both phase and amplitude
between the sub-bands are carried out with the solution (time-dependent)
derived from the CASA task \textbf{\textit {bandpass}} by averaging all
the channels in each of the sub-bands.  Solutions for both gains and
bandpass are derived from time intervals equal to each of the observing
scans. The \textbf{\textit {bandpass}} solutions are equivalent to
the scan-based offsets in ampltiude and phase between the "ch0" data of
each sub-band. Correcting for the residual phase errors from self-calibration
with the task \textbf{\textit {gaincal}}  along with applying the
\textbf{\textit {bandpass}} solutions of spectral window alignment for
each of the 64 sub-bands ({\bf step 2}), a better image model is produced. 
With the updated model, further \textbf{\textit {bandpass}} solutions are 
solved in a shorter time interval 30 sec by averaging every 4 channels in 
each of the sub-bands. The phase slopes across each of the sub-bands present 
in the \textbf{\textit {bandpass}} solutions suggests the residual delays 
that have not been fully corrected in the data. Applying the new 
\textbf{\textit {bandpass}} solutions to the data ({\bf step 3}), the 
improved image is constructed, showing an image DR $\sim$57,000:1 
(the middle image in Fig. A4). The reference model is converging to 
the true source model. With the updated image model produced from the 
previous iteration, the corrections for the residual gains in both phase 
and amplitude can be solved with \textbf{\textit {gaincal}} in the 
integration time, the shortest time interval of the data. Applying the new 
gain solution ({\bf step 4}), the further corrected dataset is generated.
Then, the variability of the core is inspected with a light curve showing 
the varibility index in flux-density
${\displaystyle 2\left(S_{max}-S_{min}\over S_{max}+S_{min}\right)=0.1}$\%
within the observing period of 2.7 hr. A model for the variable component 
is built on the scan-averaged basis with the value derived from the  first 
time scan as the reference flux density. The offsets from the reference flux 
density at the following time scans are computed. This variability model is 
then subtracted from the data with the method discussed by \cite{zmg2016}. 
After {\bf step 5}, the antenna-based issues as well as flux-density 
variation have been corrected. A new model image can be constructed. In 
the region a few arcsec from the core, the rms noise in the new model image 
appears to be at the level of the thermal noise. One may go back to {\bf step 2} 
and redo the procedure if the image DR is not limited by telescope sensitivity. 
If the baseline-based residual closure errors\footnote{The {\it closure} 
errors referred here are baseline-based errors from the possible origins as 
discussed in \cite{corn86} and \cite{clar81} for the former VLA. Unlike the 
antenna-based errors that are vanished in the well-known closure relationships, 
{\it e.g.} Eqn. (9.104) and Eqn. (10.44) of \cite{tms2017} for phase and 
amplitude, and can be corrected with self-calibration technique
\citep{schw80,corn86}, the baseline-based errors remain in the closure
relationships and the self-calibration will fail to correct them.
Baseline-based corrections for the {\it closure} errors can be done,
however \citep{perl86}.} present, the baseline-based corrections may need to 
apply to the data. For the case of NRAO 530 Ka-band data, we redo the procedure 
from {\bf step 2} and apply baseline-based corrections with \textbf{\textit {blcal}}. 
The resulting image is consistent with the image made in the {\bf step 5}. 
The rms noise of the final image made with robustness weight (R=0) is 
$\sim$10 $\mu$Jy beam$^{-1}$, reaching the sensitivity limit of the JVLA. 
The image DR of the final image (the bottom-left image in Fig. A4) is 340,000:1.

\renewcommand{\thefigure}{A5}
\begin{figure}[t]
\centering
\includegraphics[angle=0,width=94mm]{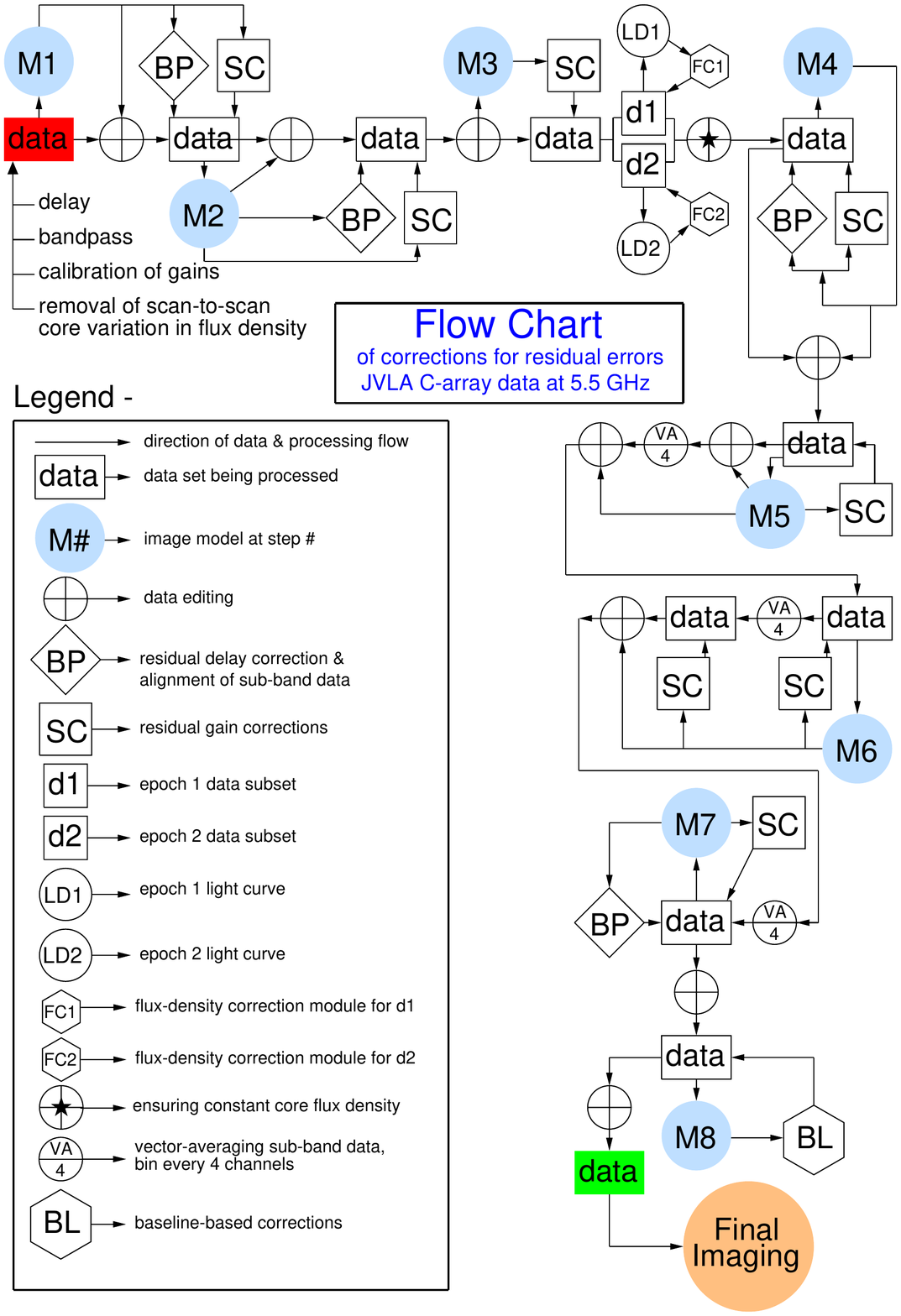}
\includegraphics[angle=0,width=42mm]{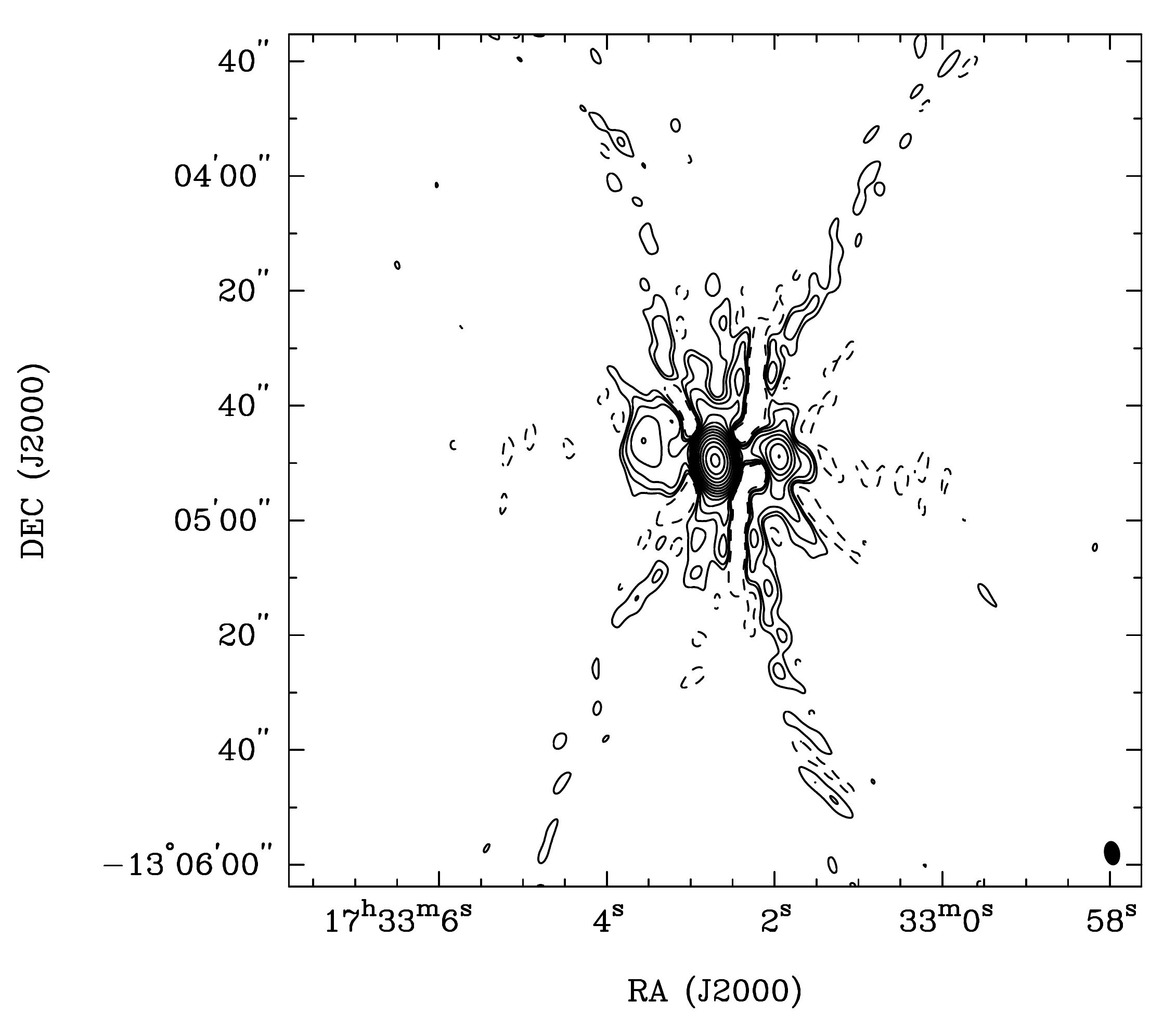}
\includegraphics[angle=0,width=42mm]{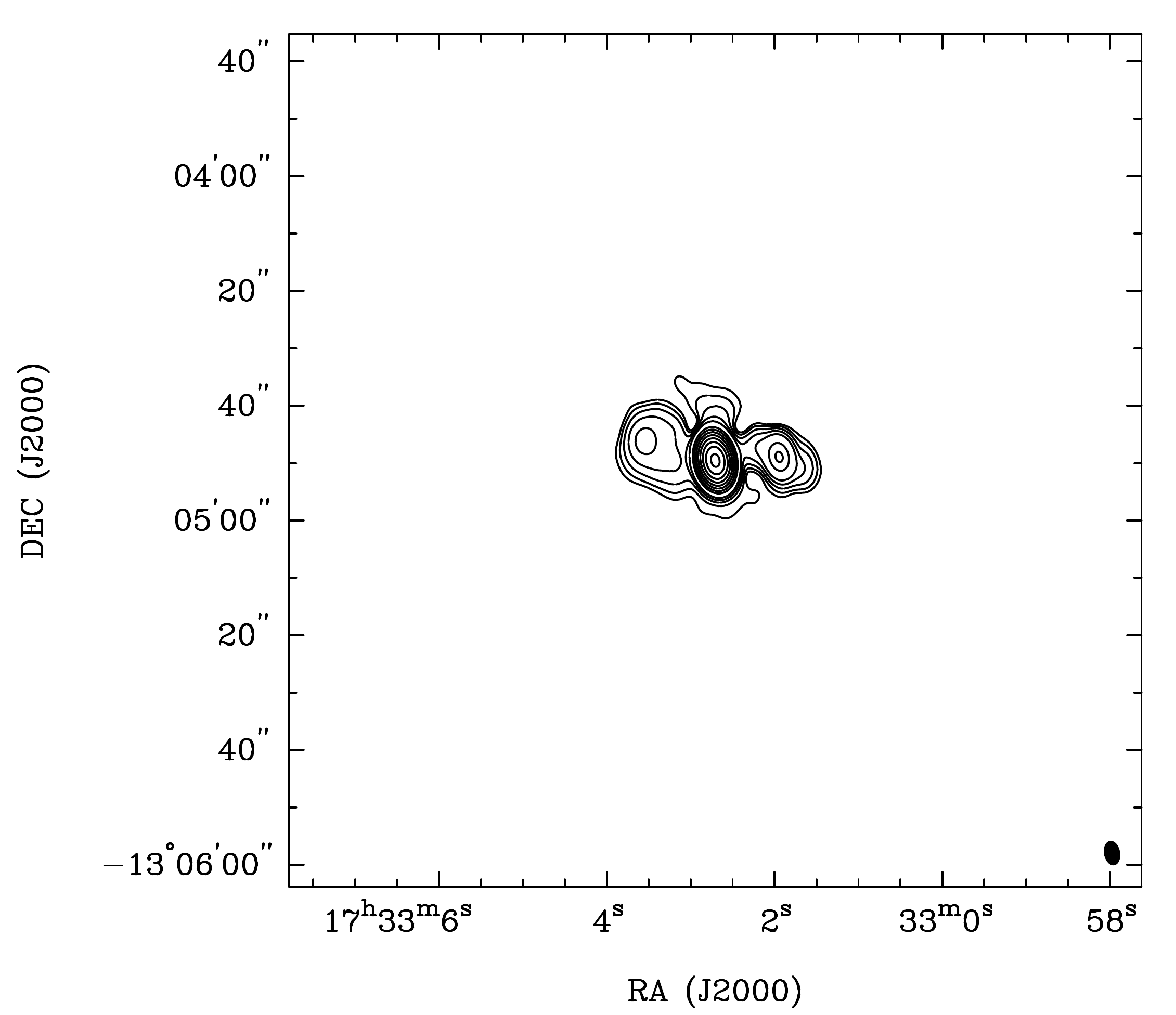}
\caption{Top: A flow chart of the procedure for imaging and
correcting residual errors; this particular procedure was developed
based on the JVLA C-array data observed in the spring of 2012.
The two epoch data sets appear to be contaminated by lower RFI
and other errors from various origins.
Bottom: Images from the data (redbox) with residual errors  and
the corrected data (greenbox) with the procedure outlined in the flow chart
above. Left panel is for M1 imaging model and right panel is the final image.
Contours are  4.45 Jy beam$^{-1} \times$ (-0.0001, 0.0001$\times2^{\rm n}$),
where n=0, 1, 2, ..., 13.
}
\label{figA5}
\end{figure}

\subsection{A note to the 5.5-GHz C-array data}
The above procedure appears to be quickly converging for the high-resolution data
sets used in this paper. Owing to large positional uncertainty generated from
the residual delays in the shorter baselines or lower-resolution data as well as
contamination by low-level RFI, the initial model is poorly determined from
our JVLA C-array data taken at 5.5 GHz in the spring of 2012. The process of 
RE-correcting and DR-imaging is difficult to converge following the steps 
outlined in Fig. A4. Consequently, a more elaborate procedure has been developed 
based on the C-array data, adding sub-steps to complement the major
correction steps of the procedure.

Fig. A5 (top) shows a flow chart of the RE-correcting and DR-imaging
procedure. The details appear to be necessary to converge the imaging 
process while fixing data issues using software in CASA. In the data 
correcting flow, eight progressive imaging models, marked with a symbol 
of capital M followed by a serial number (the filled light-blue circles 
in Fig. A5), are created. The imaging model is needed for two purposes: 
(1) to be utilized as an input imaging model to generate solutions from 
the data processed in previous step; (2) to make statistics with the
output image as compared with the model in previous step to judge if the 
process converging; if failed in converging, adjusting the input parameters 
and redo the step; if converging, then comparing the data with the improved 
imaging model to further edit and flag the data to eliminate the lower-level 
RFI and the erroneous data owing to incorrigible telescope issues. 
Thus, the details of each step might be different for different data sets. 
For the C-array data at 5.5 GHz used in this paper, we summarize the 
eight sub-steps marked in the flow chart after the initial process that 
includes the JVLA standard calibrations and fixing core flux-density to 
a constant. The output data  from the initial calibration are averaged 
to 15s in integration and the frequency configuration remains as the
original, namely 64 spectral channels with a uniform width in each of 
16 sub-bands. A brief description of the sub-steps in the process, 
identified with the serial number of the imaging model, is provided below:
\vskip 15pt
\noindent \hangindent 0pt
{\bf M1 -} A model created from the data calibrated initially. Using this 
model, the data are edited; then using the edited data, solutions for 
bandpass correction are solved by averaging the 64 channels in the basis 
of sub-band with a time interval of 60 sec; the solutions are applied to
data to align the sub-bands in both phase and amplitude; self-calibration 
is applied to the aligned data for residual gain corrections.
\vskip 5pt
\noindent \hangindent 0pt
{\bf M2 -} A model produced from the data processed in step M1;
the process in this step is similar to  step M1 except for the
bandpass solutions computed by averaging every 8 channels in each
of the sub-bands. A shorter interval of 30 sec is used in the averaging.
The bandpass solutions are applied to the data for correcting the residual 
errors in delay.
\vskip 5pt
\noindent \hangindent 0pt
{\bf M3 -} A new model created from the data post the steps M1 and M2. 
Using this model, the residual gain errors are further corrected with 
self-calibration in the interval of 15s. Then, light curves for each 
of the two epochs are produced to fix the residual variations in 
flux-density with the  module composed of
\textit{\textbf{add.component}, \textbf{ft}} and \textit{\textbf {uvsub}} 
in CASA.
\vskip 5pt
\noindent \hangindent 0pt
{\bf M4 -} This sub-step is similar to the sub-steps M1 and M2, except for
shorter time interval (15s) used for solving bandpass solutions in
a spectral bin of every two channels. The remaining residual delays are
further corrected.

\vskip 5pt
\noindent \hangindent 0pt
{\bf M5-M6-M7 -} The three sub-steps to average 64 channels into single channel
by vector averaging the data in each of sub-bands, producing  "ch0"
data in each of the sub-bands. Every four channels are averaged in each 
of the three sub-steps; consequently, the S/N of the new channel data in each
of the sub-steps increases by a factor of 2, which helps to discern the 
lower-level RFI. Data editing is applied to reject the erroneous data 
that are hidden in the noise. Further self-calibration is carried out,
ensuring the previous gain corrections to be not affected by bad data 
and lower S/N issues. Post M7, the output data becomes single channel 
(continuum) in each of the sub-bands. We note that one needs to be cautious 
for possible distortions of the source structure owing to the bandwidth smearing effect 
at a large distance from the phase center when applying the vector averaging across each
of the sub-bands. 
\vskip 5pt
\noindent \hangindent 0pt
{\bf M8 -} a new imaging model created from the data corrected for 
antenna-based error. This model is used to correct for baseline-based 
errors. The scan-averaged solutions are applied to the data. After 
flagging the data with large deviation from the baseline-based
fringe curve, the output is the final data.
\vskip 15pt
Fig. A5 (bottom) shows a comparison between the image of M1 (left panel) 
made from the initially input data (red box) and the image (right panel)  
from the final output data (green box). The process of RE-correcting and 
DR-imaging is converged. The residual error correction appears to be the 
key process in the restoration of image from radio interferometer array 
data in addition to FFT and deconvolution or cleaning process.

\subsection{Applying residual-delay corrections to target sources}
The time-dependant bandpass solutions for the residual delays derived from 
calibrators can be applied to target sources. In \cite{mzg2017}, we 
demonstrated that the residual delay corrections derived from NRAO 530 
at 5.5 GHz and J1744-3116 at 9 GHz are applied to Sgr A* producing HDR 
images. However, the resultant images of Sgr A* suggest that the corrected 
data at 9 GHz produces a better image than the one at 5.5 GHz. We note 
that both NRAO 530 and J1744-3116 were frequently scheduled as a gain 
calibrator for Sgr A* in the observations at 5.5 GHz and 9 GHz, 
respectively. However, J1744-3116 is 2.5{\degree} away from Sgr A* 
while NRAO 530 is 16{\degree} away. The difference between 9 GHz and 
5.5 GHz images might suggest that the accuracy in the correction for 
the residual delay is sensitive to the amount of positional offsets 
between target and calibrator.

\subsection{Summary}
In short, implementing the algorithms and methods for processing wideband data in addition
to the fundamental technology developed during the former, narrow bandwidth VLA
operation, the CASA has successfully provided users a software platform for handling 
JVLA data. With the procedures as discussed in this Appendix, we have demonstrated that, 
within CASA, it is possible to achieve HDR images, reaching the telescope sensitivity limits. 
However, the approach is tedious and time consuming. Various input parameters for the 
CASA programs used in the procedures require specifically adjusting in the RE-correcting and DR-imaging 
process. With the modern computing power available, it becomes possible to have a better 
way using more advanced algorithms to cope with these issues that we have confronted 
in the operation of wideband telescopes and handling big data. Therefore, improvements 
in programming are needed to arrange and manipulate the computer units in a more efficient 
way for processing wideband data in a combination of the advanced statistical algorithms 
with the knowledge that radio astronomers have developed over the past decades.

\end{document}